\documentclass[runningheads]{llncs}
\usepackage[T1]{fontenc}
\usepackage[english]{babel}
\usepackage{setspace}
\usepackage{lineno}
\usepackage{color}
\usepackage[T1]{fontenc}
\usepackage[section]{placeins}
\usepackage[pdftex]{graphicx}
\usepackage{floatflt}
\usepackage{float}
\usepackage{amsmath}
\usepackage{orcidlink}
\usepackage[
backend=biber,
style=numeric, maxbibnames=20,
sorting=nty
]{biblatex}
\newcommand{\mc}{\mathcal}

\usepackage{mathtools}
\DeclarePairedDelimiter\abs{\lvert}{\rvert}

\usepackage{array}
\newcolumntype{M}[1]{>{\centering\arraybackslash}m{#1}}
\newcolumntype{N}{@{}m{0pt}@{}}

\begin{document}

\title{Workshop on Information Theory and Related Fields --
In Memory of Ning Cai}
\titlerunning{Workshop on Information Theory --
In Memory of Ning Cai}

\author{Christian Deppe\inst{2,3}\orcidlink{0000-0002-2265-4887} \and
Holger Boche\inst{1,3,4,5}\orcidlink{0000-0002-8375-8946} \and \\ Rami Ezzine \inst{1,3}\orcidlink{0000-0002-3432-4447}\and Wafa Labidi\inst{1,3}\orcidlink{0000-0001-5704-1725}}
\authorrunning{C. Deppe, H. Boche, R. Ezzine, W. Labidi}
%
\institute{Technical University of Munich, TUM School of Computation, Information and Technology, Munich, Germany \and
Technical University of Braunschweig, Institute for Communications Technology, Braunschweig, Germany \and 6G-life, 6G research hub, Germany
\and  {{Munich Center for Quantum Science and Technology, Munich, Germany}} \and {{Munich Quantum Valley, Munich, Germany}} \\
\email{christian.deppe@tu-bs.de, boche@tum.de, rami.ezzine@tum.de, wafa.labidi@tum.de}}

\maketitle

\begin{abstract}
Sadly, our esteemed colleague and friend Ning Cai passed away on 25th May, 2023. In his memory, Ingo Althöfer, Holger Boche, Christian Deppe, Jens Stoye, Ulrich Tamm, Andreas Winter, and Raymond Yeung organized the “Workshop on Information Theory and Related Fields” at the Bielefeld ZiF (Center for Interdisciplinary Research). This special event was held from 24th November to 26th November, 2023.
The workshop celebrated Ning Cai's remarkable contributions to the field of information theory and provided a platform for discussing current research in related areas. Ning Cai's work has had a significant impact on many domains, and this gathering brought together colleagues, collaborators, and young researchers who were influenced by his pioneering efforts.
\end{abstract}
\sloppy
\section{Introduction}

It was a profound shock when the news spread at the IEEE International Conference on Communications 2023 in Rome at the end of May that one of our most important researchers and friends, Ning Cai, had passed away. At the ICC, Holger Boche, Christian Deppe, and Frank Fitzek fondly reminisced about the results, work, and shared experiences with Ning. In recent years, Holger Boche had repeatedly invited Ning to the TUM as a guest scientist. Christian Deppe had known him since 1996, when he had shared an office with Ning as a doctoral student in Bielefeld. Ning Cai was a mentor to him during his doctoral studies.

Holger Boche and Christian Deppe soon decided to organize a memorial workshop for Ning. Since Ning had lived in Bielefeld for a long time, the memorial event took place there. The Center for Interdisciplinary Research (ZiF) at Bielefeld University was an ideal location. Ning had conducted enthusiastic research there from 2002 to 2006 as part of the "General Theory of Information Transfer" project. Another participant in the project was Jens Stoye, who is now the director of ZiF. When Christian Deppe approached him with the idea, he immediately agreed and became a co-organizer. Ning’s closest former colleagues and friends—Ingo Althöfer, Ulrich Tamm, Andreas Winter, and \mbox{Raymond} \mbox{Yeung}—also agreed to help with the organization.

Ning would certainly have been very pleased with the composition of the workshop participants. In addition to experienced colleagues and old companions, many young scientists also attended the workshop. The topics covered included information theory, post-Shannon theory, network coding, quantum information theory, and combinatorics—all areas in which Ning had conducted significant research and achieved fundamental results.

\section{Poster Sessions}
Most of the posters were directly related to Ning's scientific fields of work. A vast field of classical information theory, quantum information theory, combinatorics, communication theory, optimization theory, and computing theory was covered. The posters have shown that all of Ning's work, and these areas in general, greatly influence the future development of communication systems and information processing. These applications range from 6G, quantum communication, molecular communication, network optimization, resilience and latency optimization, post-Shannon communication, secure communication to privacy protection. This mixture of theory and relevant practical open problems was a special quality in the poster session and in the presentations at the workshop. It is particularly noteworthy that many new papers and even patents were created as a result of the discussions at the poster session and during the lecture breaks.
Since Ning would certainly have appreciated it, we present here the research results that were shared during the workshop. We begin with the poster sessions:
\newpage
\subsection{Andrea Grigorescu: "Algorithmic Computability of the Capacity of Additive Colored Gaussian Noise Channels"}
\begin{floatingfigure}[r]{6cm}
\mbox{\includegraphics[width=5.5cm]{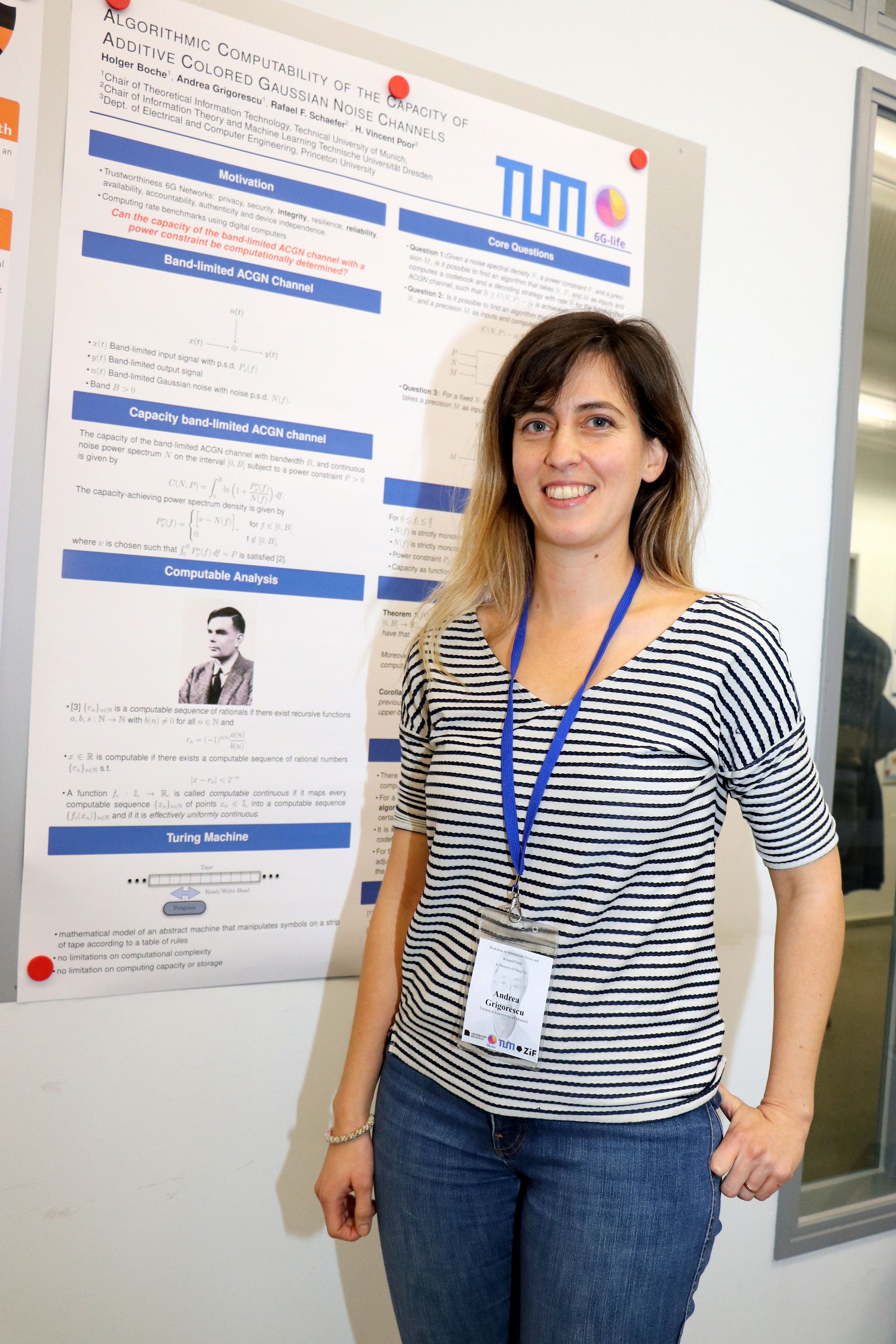}}
\caption{Andrea Grigorescu}
\end{floatingfigure}
Andrea Grigorescu's poster addressed the topic of the algorithmic computability of the capacity of additive colored Gaussian noise (ACGN) channels. Designing capacity-achieving coding schemes for band-limited ACGN channels remains a significant challenge. The poster explored the computability of the capacity of these channels from an algorithmic perspective, specifically investigating whether it is possible to compute the capacity algorithmically.

To examine this, the study uses the concept of Turing machines, which set fundamental performance limits for digital computers. The findings demonstrate that there are some band-limited ACGN channels with computable continuous spectral densities where the capacity is a non-computable number. Additionally, it is shown that for these channels, it is not possible to determine computable sequences of asymptotically sharp upper bounds for their capacities \cite{PosterAndrea}.

The poster also had some {\color{red} of} Andrea's questions about when capacity can be calculated. A simple sufficient condition for computability was found in a recent publication \cite{CapacitycoloredNoise}. However, the capacity calculation shows a complexity blowup even for additive Gaussian noise with strictly positive smooth spectral noise powers that can be calculated in polynomial time \cite{CapacitycoloredNoise}. Some participants then asked about the practical consequences of this complexity blowup. This question was very interesting and was taken up by the authors of \cite{CapacitycoloredNoise}. The finite blocklength behavior for Gaussian channels was investigated, and the results were presented at the recent results session ISIT 2024. An extended abstract can be found in \cite{FiniteBL}.

\vspace*{0.5cm}

\newpage
\subsection{Ark Modi: "Quantum and Quantum-Inspired Stereographic kNN Clustering"}
\begin{floatingfigure}[r]{6cm}
\mbox{\includegraphics[width=5.5cm]{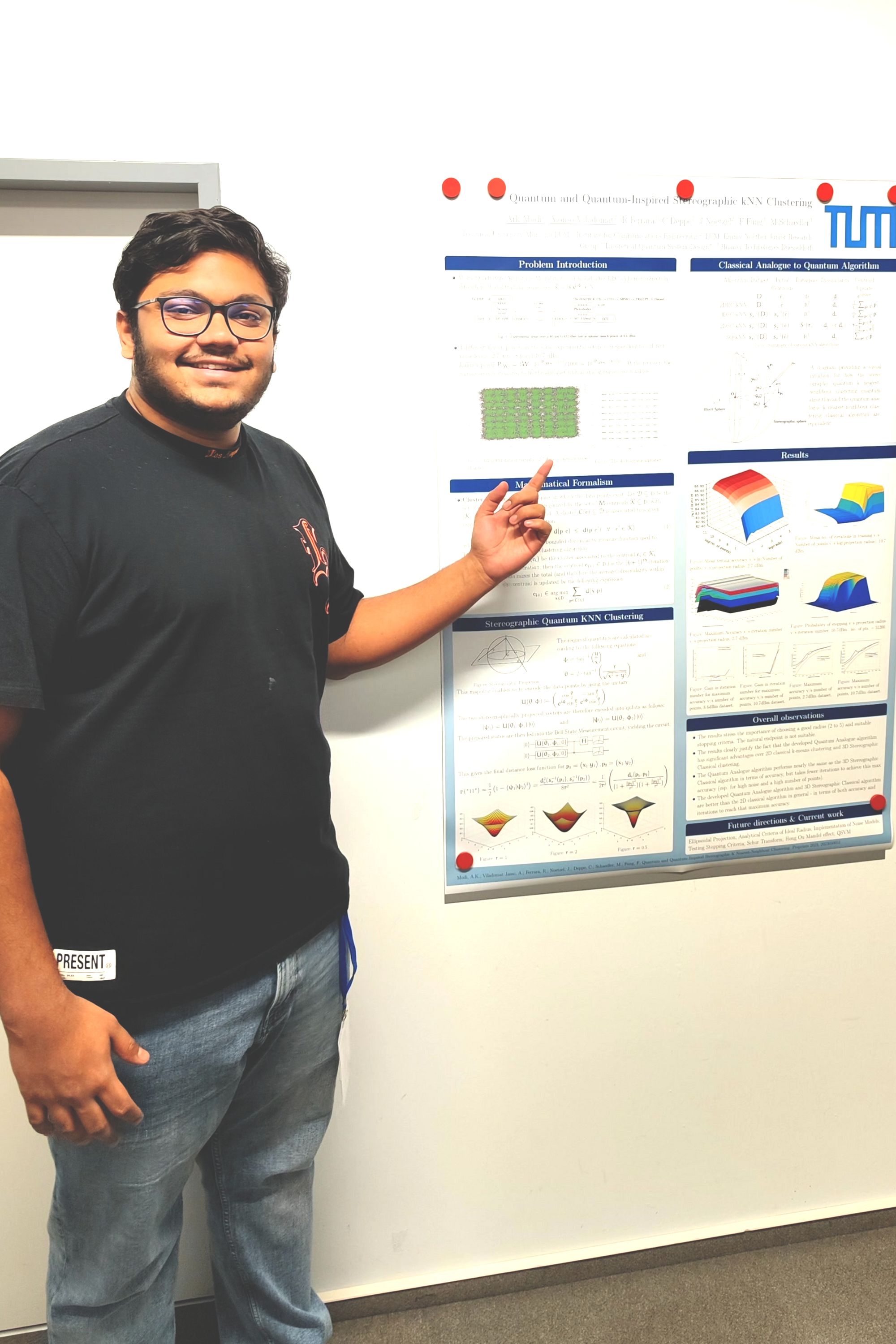}}
\caption{Ark Modi}
\end{floatingfigure}
Ark Modi presented on his poster quantum and quantum-inspired stereographic k-means nearest-neighbour (kNN) clustering. Nearest-neighbour clustering is a straightforward yet effective machine learning algorithm commonly used for decoding signals in classical optical-fibre communication systems. While quantum k-means clustering offers potential speed-up over classical k-means, no advantage for decoding optical-fiber 
 signals has been demonstrated yet because of the inaccuracies and slowdowns introduced by embedding classical data. Current Noisy Intermediate-Scale Quantum (NISQ) implementations do not achieve exponential speed-up, so this work proposes the generalized inverse stereographic projection as an improved method for embedding into the Bloch sphere for quantum distance estimation in the k-nearest-neighbour clustering. This improvement brings the performance closer to classical algorithms. Additionally, the generalized inverse stereographic projection is used to develop a classical clustering algorithm, which is then benchmarked for accuracy, runtime, and convergence in decoding real-world experimental optical-fiber communication data. This `quantum-inspired' algorithm enhances both the accuracy and the convergence rate compared to the classical k-means algorithm. Therefore, this work presents two main contributions. First, it introduces the general inverse stereographic projection into the Bloch sphere as a superior embedding technique for quantum machine learning algorithms, using the clustering of quadrature amplitude-modulated optical-fiber signals as an example. Second, it proposes and benchmarks a classical clustering algorithm, inspired by the quantum approach, that utilizes the general inverse stereographic projection and spherical centroid for optical-fiber signal clustering, demonstrating that optimizing the radius consistently improves the accuracy and the convergence rate. The results have been published in \cite{PosterArk}.

\newpage

\subsection{Benedikt Baier: "Simulation-based analysis of delay-reducing quantum repeater protocols"}
\begin{floatingfigure}[r]{6cm}
\mbox{\includegraphics[width=5.5cm]{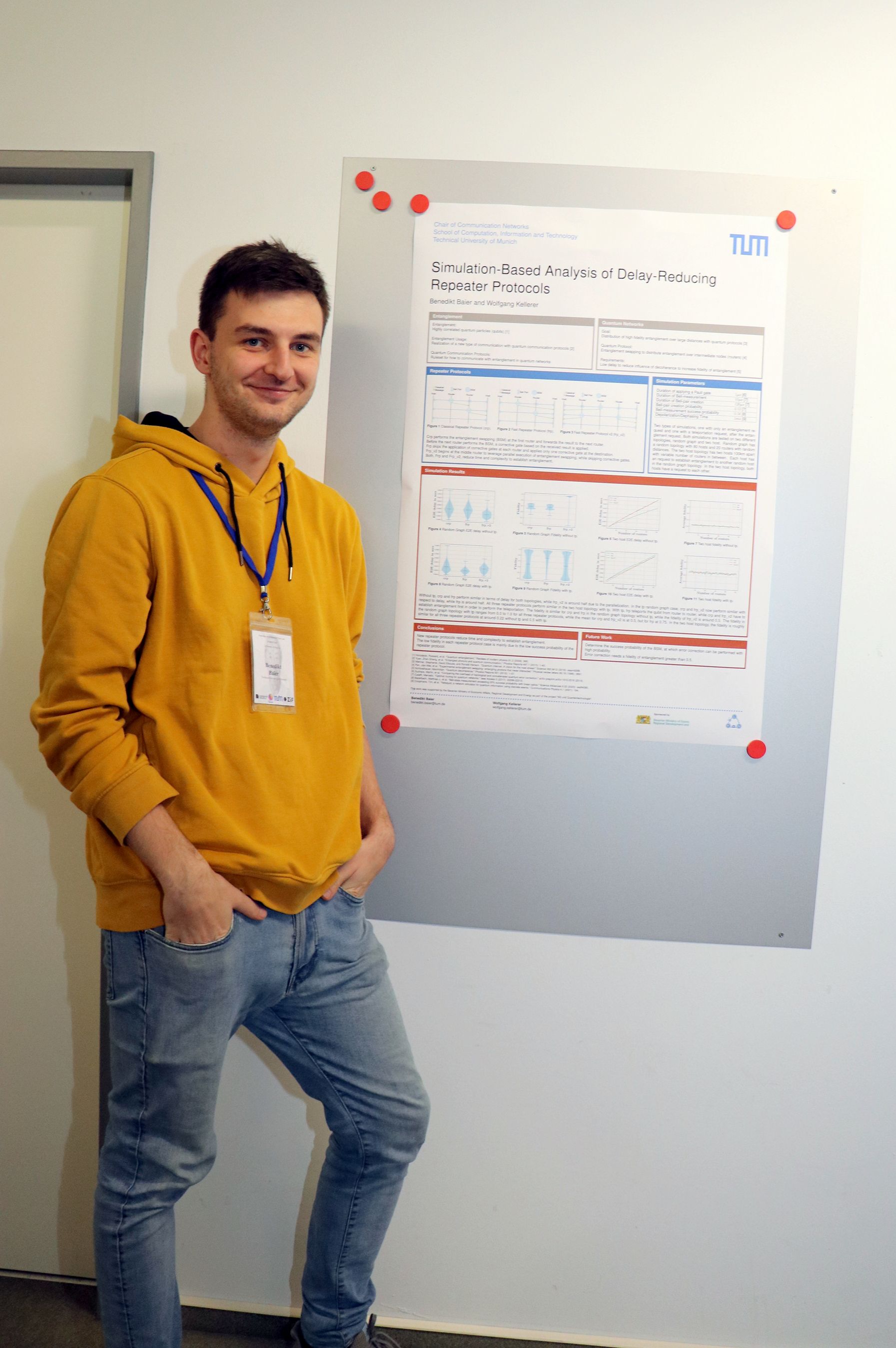}}
\caption{Benedikt Baier}
\end{floatingfigure}
Benedikt Baier dealt with the topic of quantum repeaters in his poster. In quantum networks, entanglement enables a new form of communication through quantum protocols. Achieving high-fidelity entanglement distribution over large distances is a primary goal, necessitating the use of quantum repeaters. These repeaters employ entanglement swapping to extend entanglement across intermediate nodes (routers). Minimizing delay is crucial to mitigate decoherence and enhance entanglement fidelity.
We examine two types of simulations: one involving an entanglement request only and another that includes a teleportation request following the entanglement request. Both simulations are tested on two different topologies: a random graph and a two-host configuration. The random graph topology consists of 80 hosts and 20 routers with random distances between them, while the two-host topology involves two hosts 100 km apart with a variable number of routers in between. In the random graph topology, each host requests entanglement with another random host. In the two-host topology, both hosts request entanglement with each other.
Benedikt's results demonstrate that new repeater protocols significantly reduce the time and complexity required to establish entanglement. However, the primary factor contributing to low fidelity in these protocols is the low success probability of the repeater mechanisms.

\vspace*{0.5cm}

\newpage

\subsection{Caspar von Lengerke: Beyond the Bound: A New Performance Perspective for Randomized Noiseless Identification Codes}
\begin{floatingfigure}[r]{6cm}
\mbox{\includegraphics[width=5.5cm]{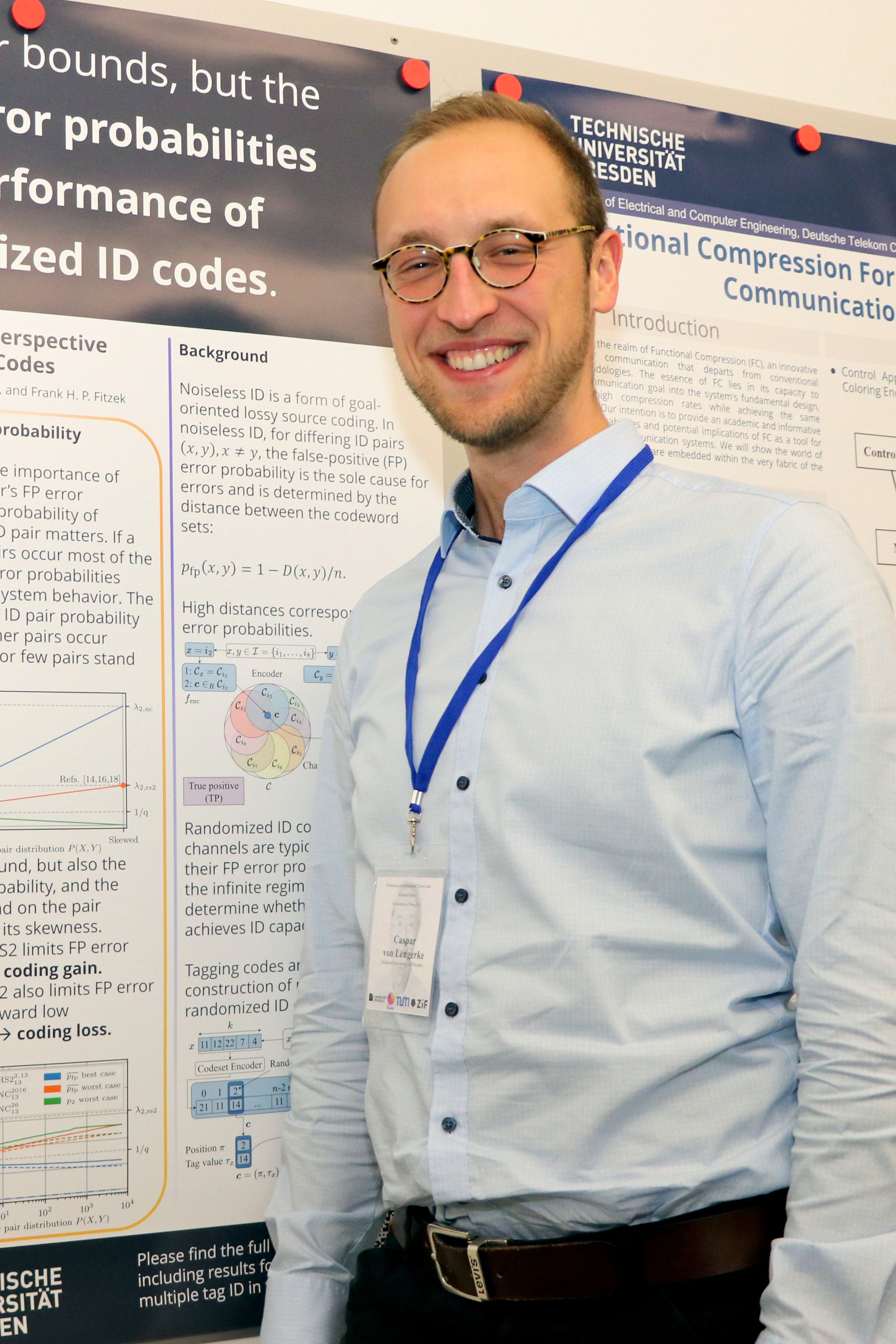}}
\caption{Caspar von Lengerke}
\end{floatingfigure}
Caspar von Lengerke presented his poster on new perspectives for randomized noiseless identification codes. Identification via channels (ID) is a goal-oriented communication paradigm that verifies the matching of message (identity) pairs at the source and sink, representing a shift beyond traditional Shannon-based communication models. Traditionally, ID research has focused on the upper bound $\lambda$ for the probability of a false-positive (FP) identity match, primarily using ID tagging codes that represent identities through sets of position-tag tuples.

This work expands the scope of ID research by introducing new performance metrics. These include the expected FP-error probability, which considers the distance properties of ID codeword sets along with the probability of selecting ID pairs; threshold probabilities, which characterize quantiles of FP probabilities; and the distance tail uplift ratio (DiTUR), which measures the fraction of ID pairs whose distance is increased beyond the minimum distance corresponding to $\lambda$.

A No-Code (NC) approach is introduced as a baseline for ID, where identification is conducted directly with the messages (identities) without additional coding. Additionally, a concatenated Reed-Solomon ID code and a Reed-Muller ID code are investigated, finding that they do not consistently offer advantages over the NC approach. The analytical exploration of reducing error-prone ID pairs by sending multiple tags is also presented.

The findings suggest the importance of examining the distance distribution of ID codes and incorporating the ID pair distributions of real ID systems in future research. The results presented have been published in \cite{PosterCaspar}.

\vspace*{0.5cm}
\newpage
\subsection{Evagoras Stylianou: "Minimal Trellises for Decoding Quantum Stabilizer Codes"}
\begin{floatingfigure}[r]{6cm}
\mbox{\includegraphics[width=5.5cm]{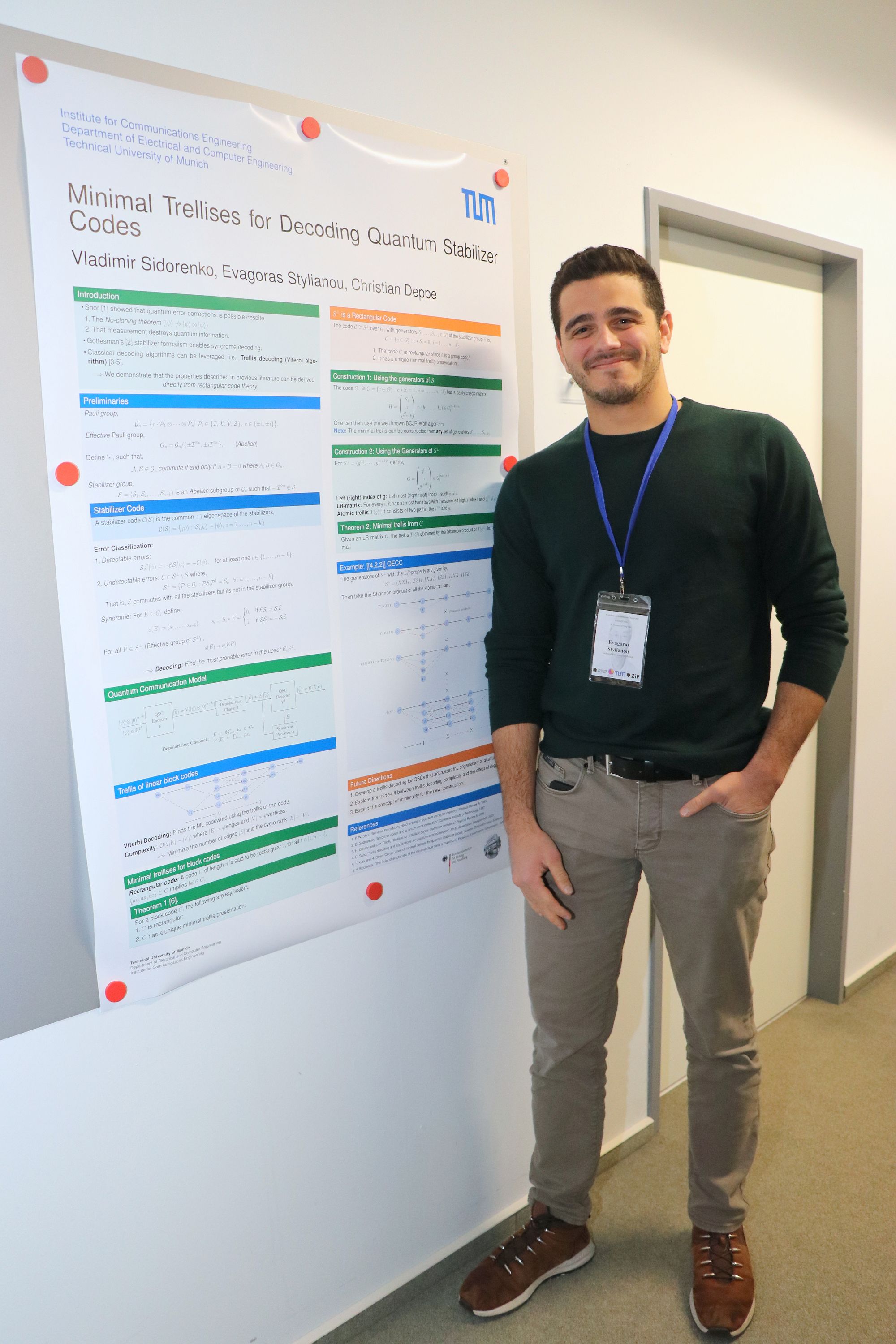}}
\caption{Evagoras Stylianou}
\end{floatingfigure}
Evagoras Stylianou presented an in-depth exploration of his work on \textit{Minimal Trellises for Decoding Quantum Stabilizer Codes}. This result was first published in \cite{PosterEvagoras}. The foundation for this work dates back to 2006, when Ollivier and Tillich introduced a trellis representation for quantum codes associated with a stabilizer group $S$. This representation facilitates efficient maximum likelihood error estimation by decoding the coset of the code $S^\perp$, which is the normalizer of $S$. 
In Stylianou's presentation, he elaborated on how this decoding process can be carried out effectively using a trellis constructed for the coset of the code $S^\perp$, which is determined by the syndrome obtained from quantum measurements. The semi-tutorial poster he presented delves into several methodologies for designing a minimal trellis for $S^\perp$ and its associated cosets. The primary goal is to streamline the complexity involved in maximum likelihood error estimation using the Viterbi algorithm.
A significant portion of the findings presented in his poster is grounded in the theory of rectangular codes, which is particularly relevant since the code $S^\perp$ exhibits a rectangular structure. Stylianou's results contribute to advancing the understanding and practical implementation of efficient quantum error correction.
Additionally, these findings are included in the survey \cite{LNCSQEC}.

\vspace*{0.5cm}

\newpage
\subsection{Farzin Salek: "Multi-User Distillation of Common Randomness and Entanglement from Quantum States"}
\begin{floatingfigure}[r]{6cm}
\mbox{\includegraphics[width=5.5cm]{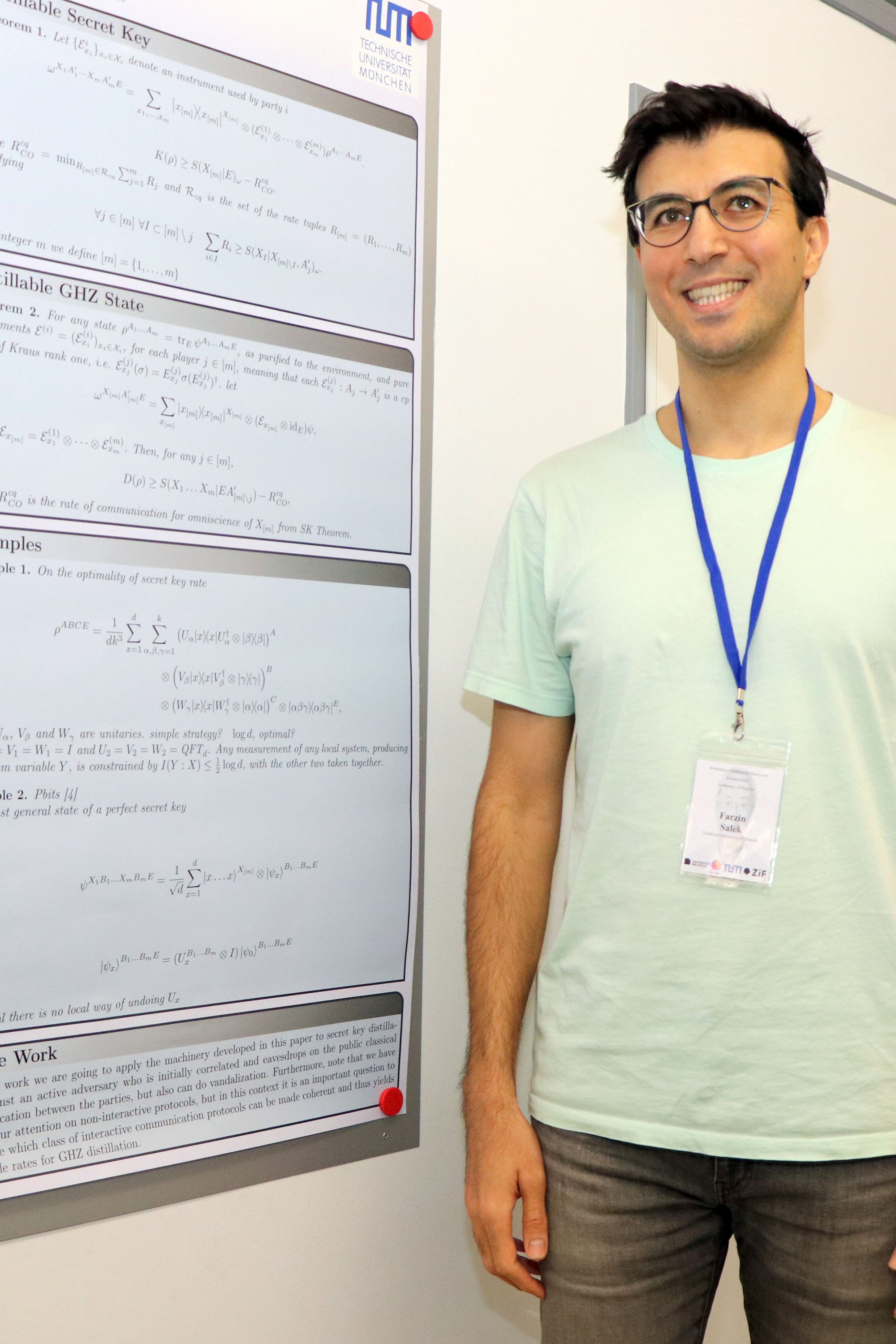}}
\caption{Farzin Salek}
\end{floatingfigure}
In Farzin Salek's poster, new protocols for converting noisy multipartite quantum correlations into noiseless classical and quantum correlations using local operations and classical communications (LOCC) are developed. For common randomness (CR) distillation, two new lower bounds on the ``distillable common randomness,'' an operational measure of the total genuine classical correlations in a quantum state, are introduced. This proof extends the communication for the omniscience (CO) concept from Csiszár and Narayan's work \cite{Csiszar2004}. A key contribution is the introduction of a novel simultaneous decoder for compressing correlated classical sources through random binning with quantum side information at the decoder.

Additionally, two new lower bounds on the rate at which Greenberger-Horne-Zeilinger (GHZ) states can be asymptotically distilled from any given pure state under LOCC are established. This approach involves ``making coherent'' the proposed CR distillation protocols and resource recycling, building on the work of Devetak et al. \cite{Devetak2008}. The first lower bound matches a recent result by Vrana and Christandl, achieved through a combinatorial method \cite{Vrana2019}. The second lower bound generalizes and improves upon this result, unifying several other known lower bounds on GHZ distillation. The detailed results can be found in the work \cite{Salek2022}.

\vspace*{1cm}
\newpage

\subsection{Holger Boche: "Wiener Prediction Theory is not Effective"}
\begin{floatingfigure}[r]{6cm}
\mbox{\includegraphics[width=5.5cm]{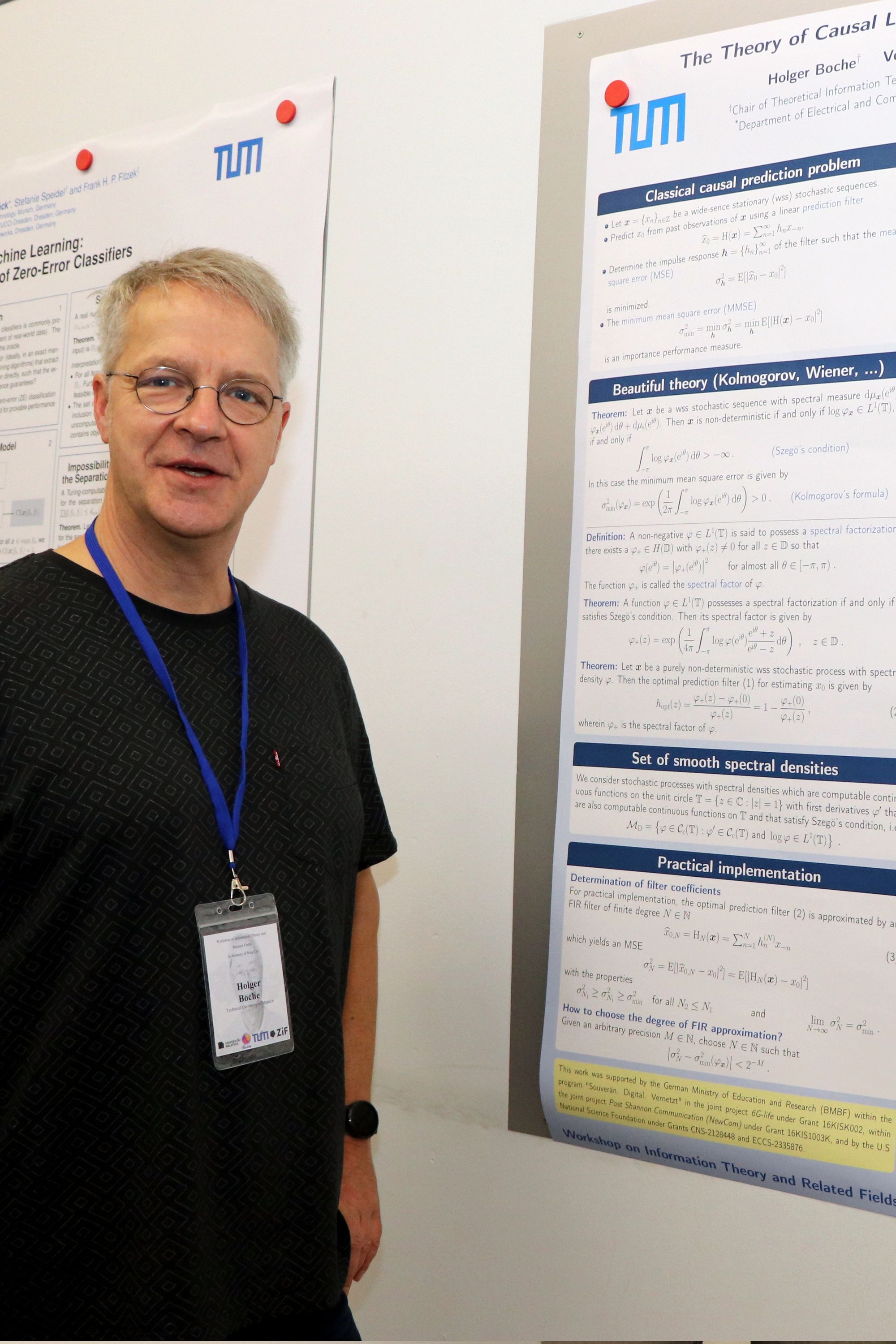}}
\caption{Holger Boche}
\end{floatingfigure}
Holger Boche demonstrated in his poster that the minimum mean square error (MMSE) for predicting a stationary stochastic time series from its past observations is not generally Turing computable, even when the spectral density of the stochastic process is differentiable with a computable first derivative. This indicates that there is no algorithmic stopping criterion for any approximation sequence converging to the MMSE that can ensure that the computed approximation is sufficiently close to the true MMSE value. Additionally, it was shown that under the same conditions on the spectral density, the coefficients of the optimal prediction filter are also not generally Turing computable \cite{PosterHolger}. There were also highly intensive discussions at the poster session, where Holger received numerous intriguing questions. In general, the optimal Wiener
prediction estimators are always filters with an infinite impulse response. Many questions were about whether these optimal estimators can also be approximated by finite impulse response filters. The poster
showed that this is generally not possible under the control of the
approximation error. Moreover, for processes with polynomially computable spectral power densities, the sequence of finite impulse response approximations cannot be computed in polynomial time.

\newpage

\subsection{Johannes Rosenberger: Functional Communication and $K$-Identification}
\begin{floatingfigure}[r]{6cm}
\mbox{\includegraphics[width=5.5cm]{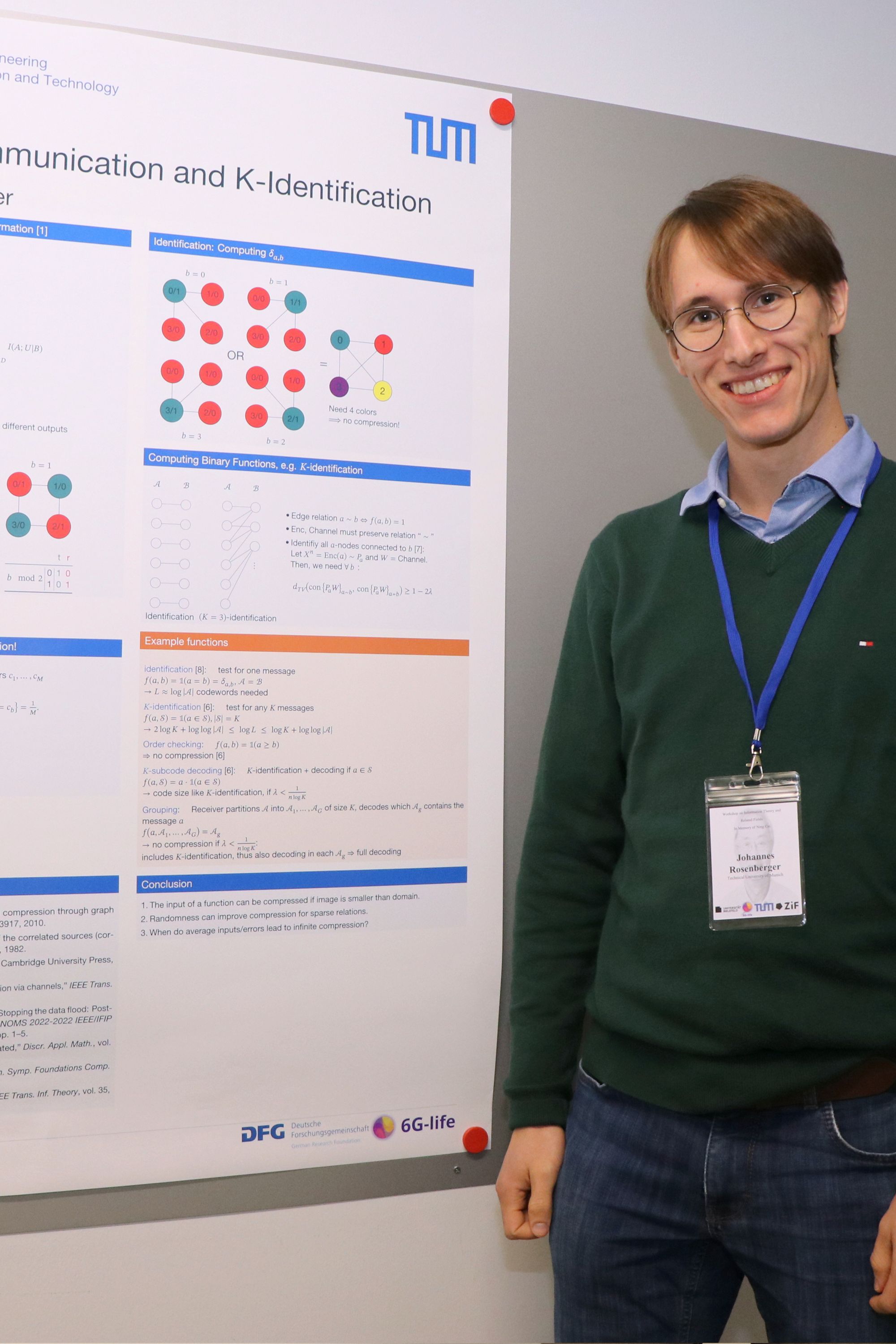}}
\caption{Johannes Rosenberger}
\end{floatingfigure}
Johannes Rosenberger introduced the concept of a locally homomorphic channel in his poster and established an approximate equivalence between these channels and codes for computing functions. Additionally, decomposition properties of locally homomorphic channels were derived, which were used to analyze and construct codes where two messages must be encoded independently. This resulted in new findings for identification and $K$-identification, demonstrating significant rate improvements over naive code constructions.

Despite these advancements, the optimality of these constructions in many cases remains uncertain. A deeper understanding of the input hypergraphs to locally homomorphic channels is necessary to evaluate this. Specifically, it is unclear if the vertex set of the target hypergraph of a bipartite encoder must be strictly rectangular, and how to optimize these, as they are also the input hypergraphs to the actual channel. Ensuring that the vertex sets are rectangular and the channels are locally homomorphic and edge-bijective is crucial. Resolving these questions would fully elucidate the tradeoff between the rates of the two messages. The detailed results can be found in \cite{PaperJohannes}.

\vspace*{0.5cm}
\newpage

\subsection{Juan Cabrera: "Network Coding in Molecular Communications"}
\begin{floatingfigure}[r]{6cm}
\mbox{\includegraphics[width=5.5cm]{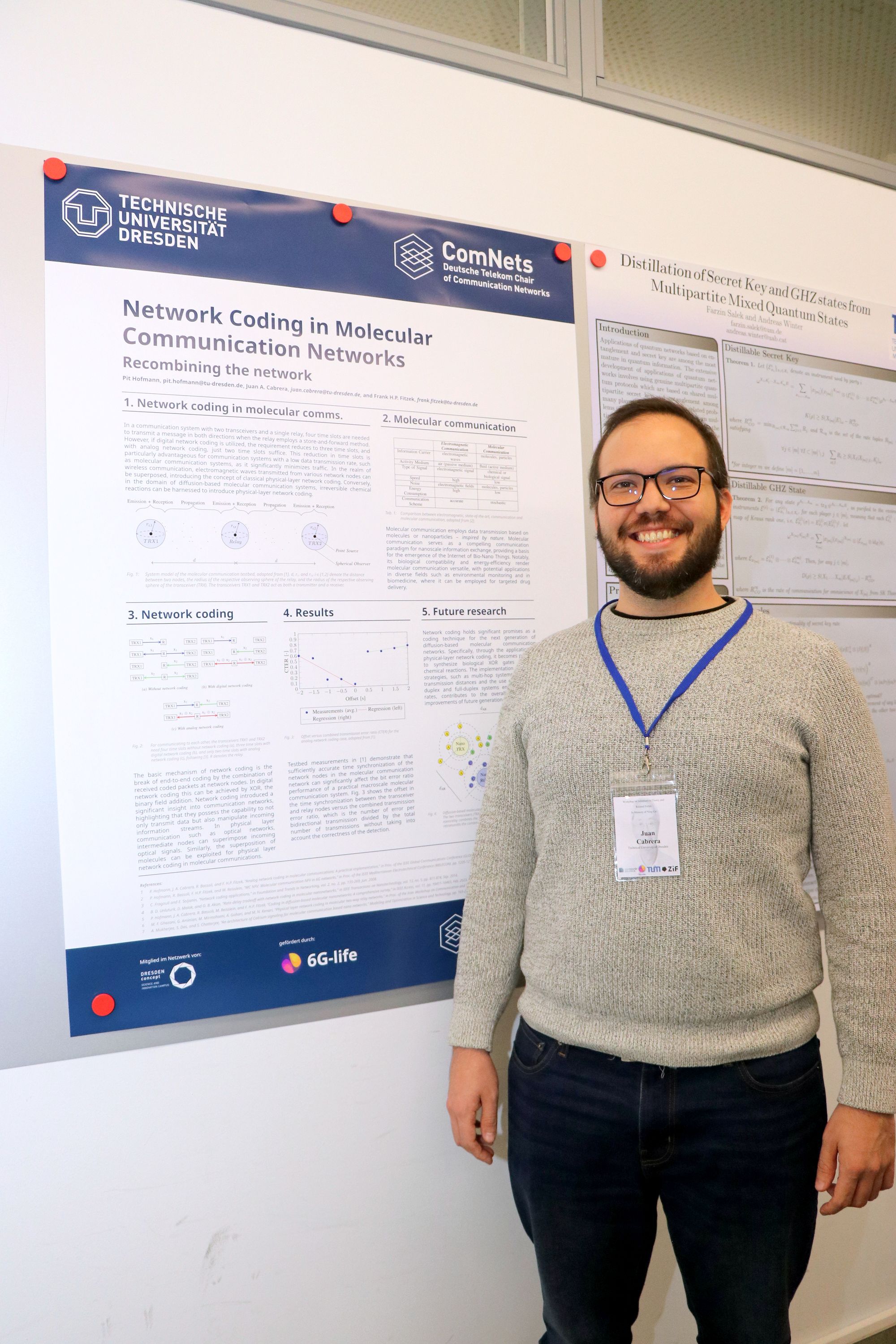}}
\caption{Juan Cabrera}
\end{floatingfigure}
Juan Cabrera presented results on network coding in molecular communication in his poster. In a typical communication scenario involving two transceivers (TRXs) and a relay, four time slots are generally required to exchange one packet in each direction if the relay employs a store-and-forward approach. Digital network coding (NC) reduces this requirement to three time slots, while analog network coding (ANC) further reduces it to just two time slots. This reduction in time slots is particularly beneficial for systems with low transmission rates, such as molecular communication (MC) systems.

Previous studies have largely concentrated on theoretical aspects of NC applications in MC nanonetworks. The poster introduces the first macroscale MC testbed for ANC, marking a significant advancement from theoretical to practical implementation. The testbed demonstrates the impact of time synchronization on the bit error ratio (BER) performance of a network-coded MC system. It also investigates how the length of the bit sequence influences the error ratio.

Additionally, duplex network coding was implemented, revealing that while error ratios for half-duplex and full-duplex NC are similar, full-duplex NC offers a time gain of up to \( (n-1) \) time slots for a bit sequence length of \( n \). The detailed results have been published in \cite{PosterJuan}.

\vspace*{0.5cm}
\newpage
\subsection{Julia Kunzelmann: "Scalability of the Repeater Rate in Multipartite Quantum Networks"}
\begin{floatingfigure}[r]{6cm}
\mbox{\includegraphics[width=5.5cm]{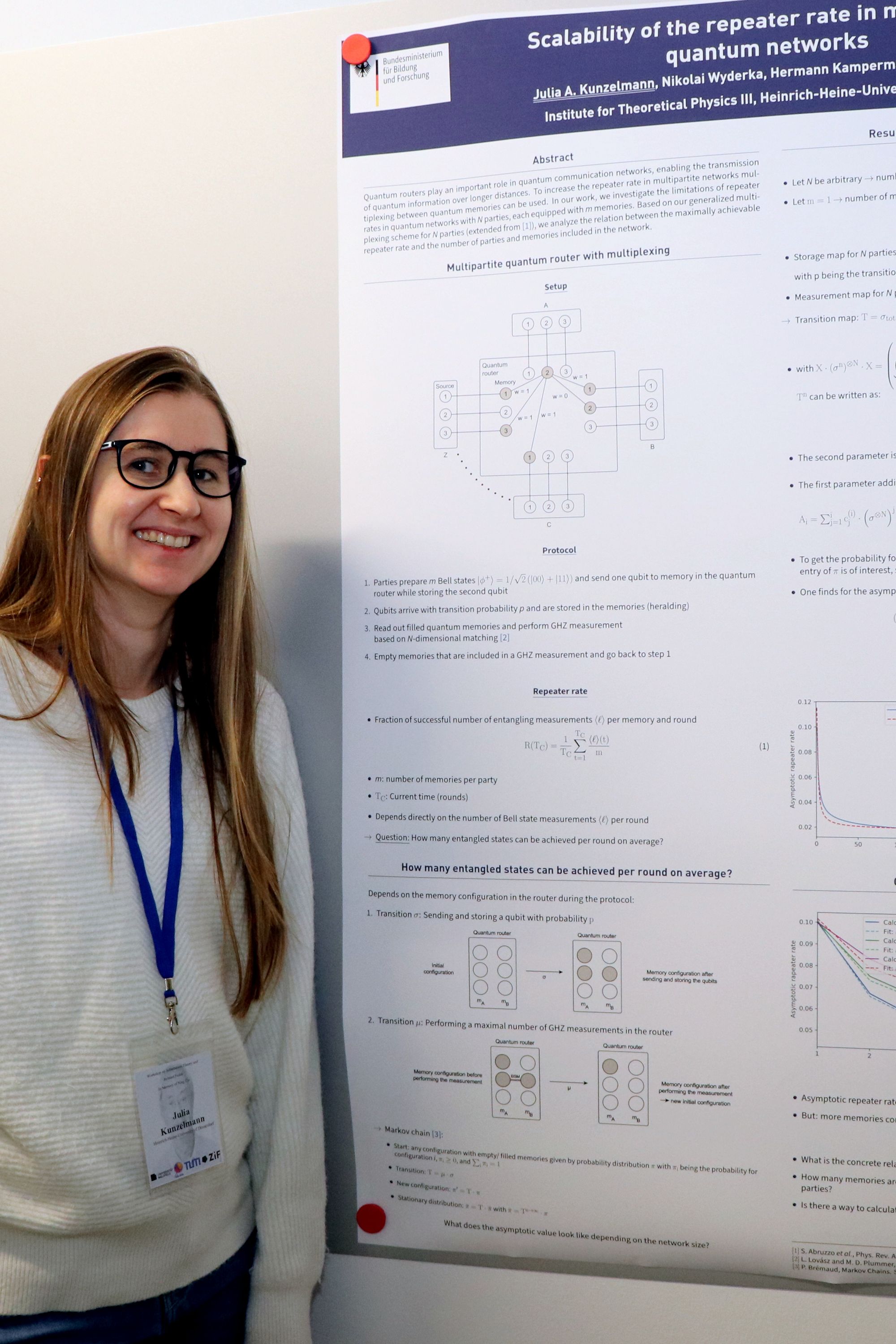}}
\caption{Julia Kunzelmann}
\end{floatingfigure}
Julia Kunzelmann presented her research on the scalability of the repeater rate in multipartite quantum networks in her poster. Long-distance communication in quantum networks remains challenging due to the interaction between qubits and their environment. Quantum repeaters are used to address this challenge, and their performance can be optimized through multiplexing. In multipartite networks, multiplexing between quantum memories can improve the repeater rate.

The poster explored the limitations of repeater rates in quantum networks with \( N \) parties, each equipped with \( m \) memories. Using a generalized multiplexing scheme for \( N \) parties, the study analyzed the relationship between the maximally achievable repeater rate and the number of parties and memories in the network. The findings indicate that while the asymptotic repeater rate tends to decrease as the number of parties increases, this decrease can be mitigated by increasing the number of memories. An open question remains concerning the exact relationship between the asymptotic repeater rate and the network size. The detailed results are published in \cite{PaperJulia}.

\vspace*{0.5cm}
\newpage

\subsection{Marcel Mross and Jyun-Sian Wu: "From Early Decoding to Asynchronous Unsourced MAC"}
\begin{floatingfigure}[r]{6cm}
\mbox{\includegraphics[width=5.5cm]{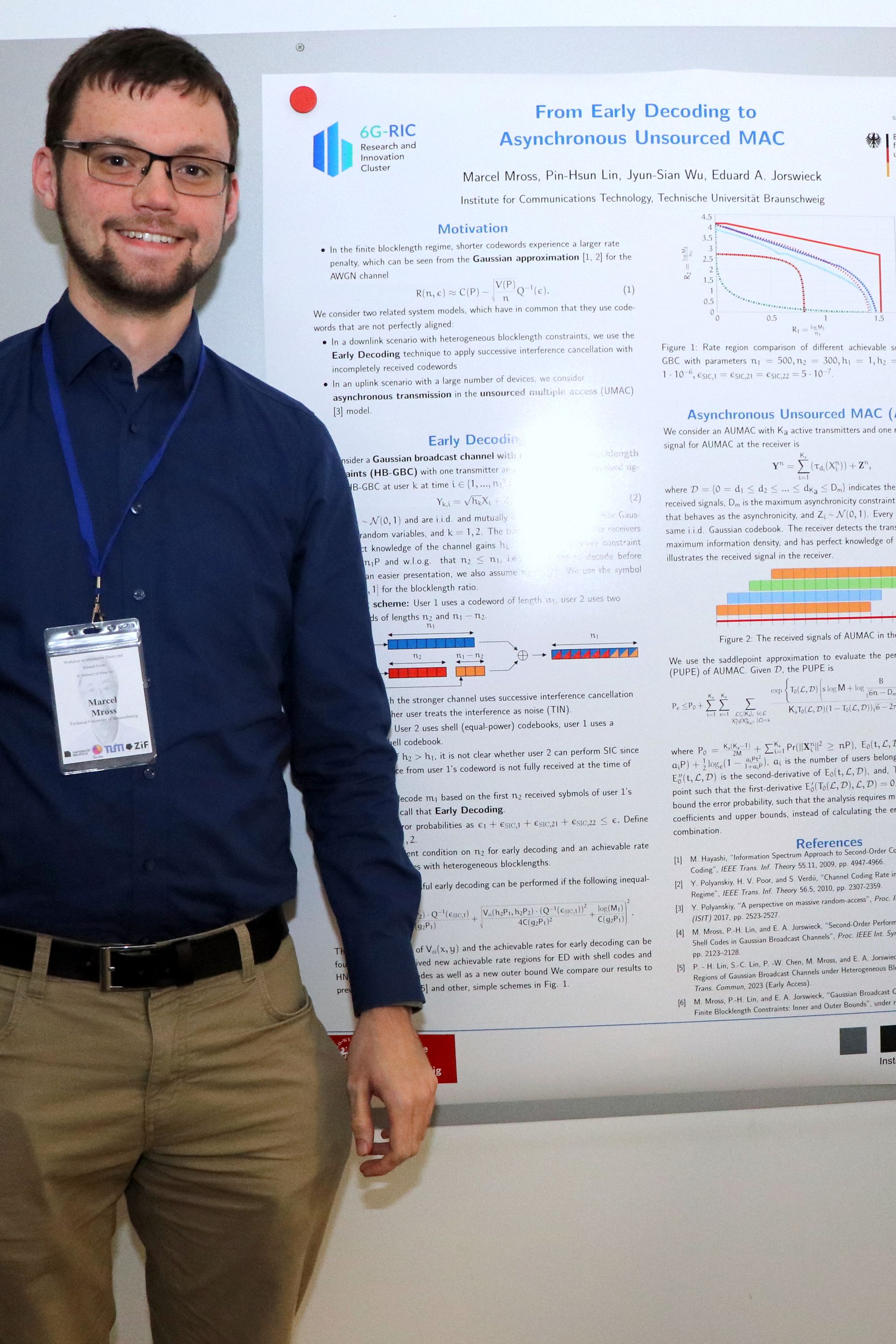}}
\caption{Marcel Mross}
\end{floatingfigure}
In their poster, Marcel Mross and Jyun-Sian Wu examined two connected topics. First, they investiagted a Gaussian Broadcast Channel with two receivers and heterogeneous blocklength constraints (HB-GBC), i.e., the two receivers have different latency requirements, which is modeled by using codewords of different blocklengths. This makes the usage of successive interference cancellation (SIC) difficult, since the interference may not be completely received at the time of decoding. They analyze the rate region of the \textit{early decoding} technique, where the user with the stricter latency requirement can perform SIC under certain conditions on the channel gains and the power allocation. They analyzed second-order rate expressions for early decoding using an adapted version of equal-power codebooks (or \textit{shell codebooks}). The results are published in \cite{PosterMarcel1}.

They also examined an asynchronous $K_a$-active-user unsourced multiple access channel (AUMAC) under worst-case asynchronicity conditions. The challenge is to decode transmitted messages within $n$ channel uses, despite some codewords being only partially received due to asynchronicities. A constraint is imposed on the maximum allowable transmission delay. Unlike the synchronous UMAC, the AUMAC does not possess the permutation-invariant property because different permutations of the same codewords with fixed asynchronicity are distinguishable. 
To address this issue, a uniform bound on the per-user probability of error (PUPE) is derived by examining the worst-case asynchronous patterns under the delay constraint. The results were later extended and published in \cite{PosterMarcel}.

\vspace*{0.5cm}

\newpage

\subsection{Moritz Wiese: "Upper and Lower Bounds for the Security Performance of Wiretap Channels”}
\begin{floatingfigure}[r]{6cm}
\mbox{\includegraphics[width=5.5cm]{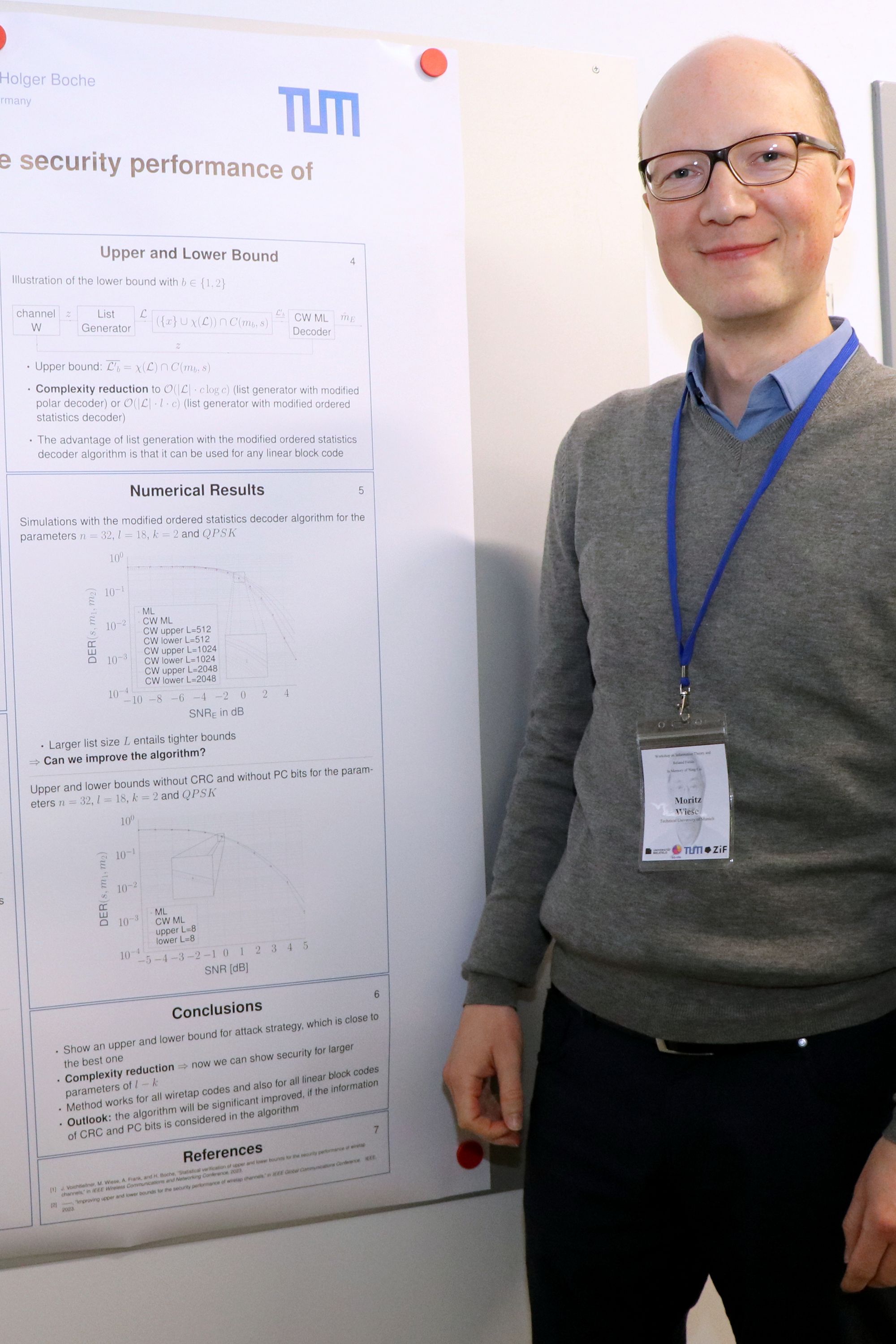}}
\caption{Moritz Wiese}
\end{floatingfigure}
Moritz Wiese compared various algorithms for evaluating semantic security on AWGN wiretap channels in his poster. Each algorithm provides upper and lower bounds on the performance of an attack strategy that approximates the best possible attack strategy. The key advantage of these algorithms is their reduced computational complexity compared to the optimal attack strategy. Additionally, it was demonstrated that the proposed algorithms can be further enhanced by incorporating cyclic redundancy check bits and parity check bits, such as those generated when using polar codes or LDPC codes in accordance with the 5G standard. Furthermore, the compatibility of these algorithms with both polar codes and LDPC codes is illustrated \cite{PosterMoritz}. 

The results have just been published in \cite{Collisionflat} and show an interesting connection to combinatorics. Relevant open questions in combinatorics have already been addressed \cite{SmallMosaics}.

\vspace*{0.5cm}
\newpage

\subsection{Olaf Gröscho: “Source Identification”}
\begin{floatingfigure}[r]{6cm}
\mbox{\includegraphics[width=5.5cm]{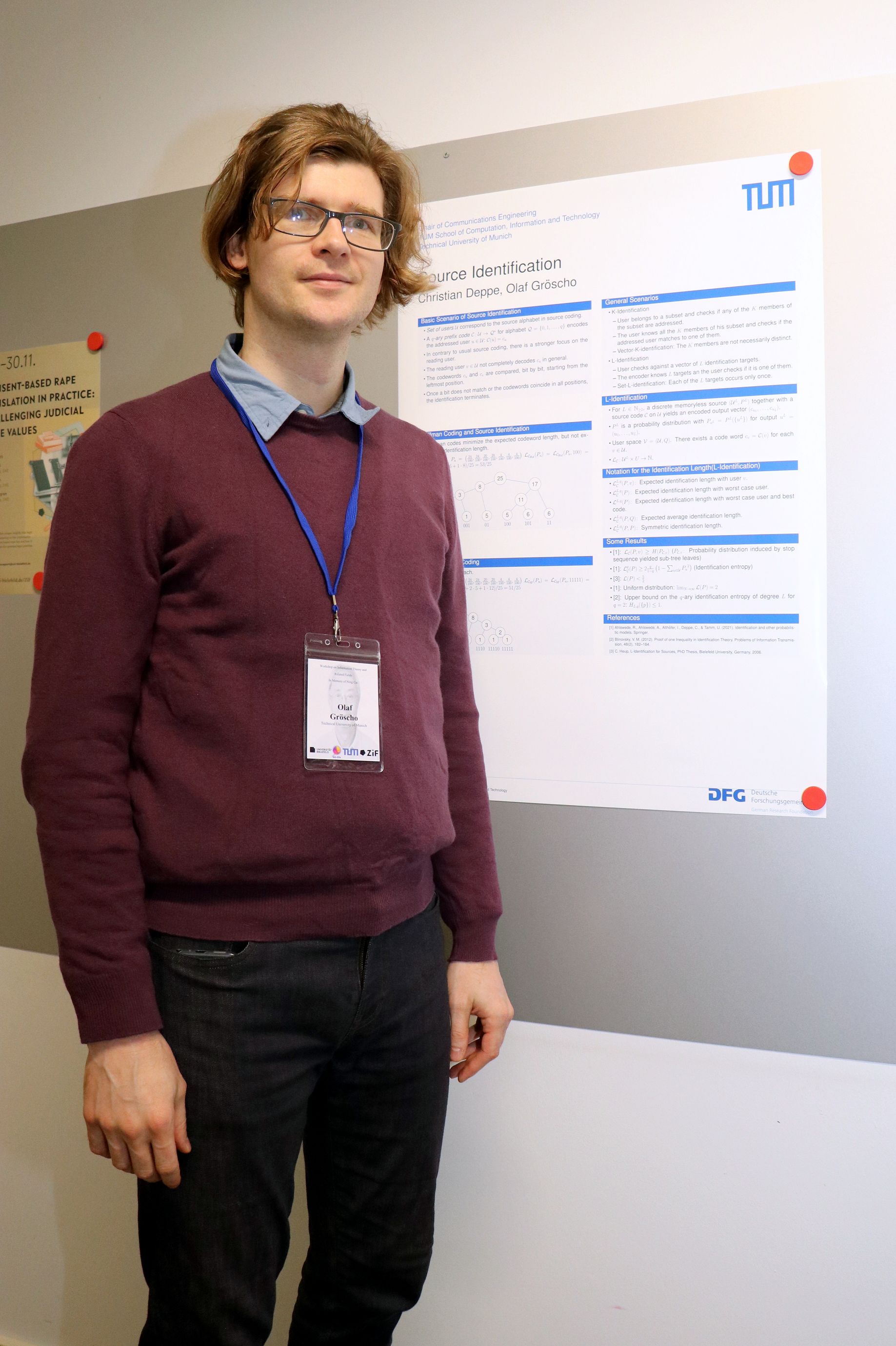}}
\caption{Olaf Gröscho}
\end{floatingfigure}

Olaf Gröscho presented an overview of results in source identification in his poster. In classical information theory, the transmission problem addresses the question of how many distinct messages can be transmitted over a noisy channel. In this context, transmission implies that the receiver can determine the exact message sent. The classical goal is to ensure reliable communication despite the noise, which is addressed by Shannon's Channel Coding Theorem.
In contrast, the identification problem focuses on how many distinct messages the receiver can identify from a noisy channel. Here, identification means that the receiver can determine whether the actual message is a specific candidate, denoted by \( u \), from a set of possible messages. This problem is different from transmission in that the goal is to identify whether the message belongs to a specific subset rather than to transmit the exact message.
It is highlighted that when randomized encoding is allowed, the optimal code size for identification grows double exponentially with the block length. This result is somewhat surprising because, as shown in \cite{ahlswede1989identification}, the second-order capacity for identification matches Shannon’s first-order transmission capacity. This finding indicates a parallel between Shannon's Channel Coding Theorem for transmission and a Channel Coding Theorem for identification.
Furthermore, it is natural to seek a similar parallel for source coding, particularly for noiseless coding (see \cite{ahlswede2006identification}). This idea was suggested by Ahlswede in his work, where he proposed an updated general theory of information transfer. The theory aims to establish parallels between the concepts of transmission and identification in source coding \cite{ahlswede2008general}.

\vspace*{0.5cm}
\newpage

\subsection{Pau Colomer: “Zero-entropy encoders and simultaneous decoders in identification via quantum channels”}
\begin{floatingfigure}[r]{6cm}
\mbox{\includegraphics[width=5.5cm]{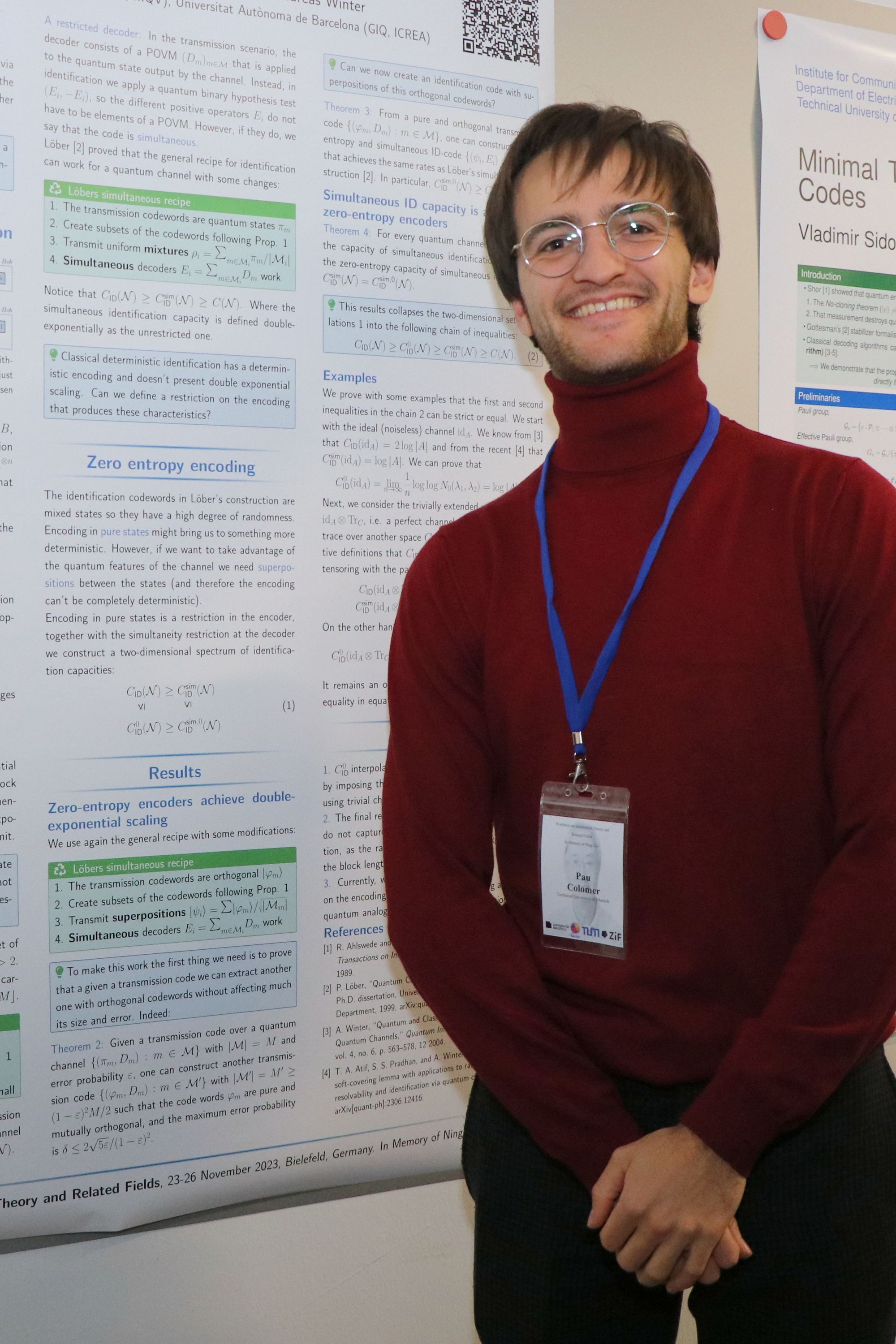}}
\caption{Pau Colomer}
\end{floatingfigure}
In his poster, Pau Colomer presented results on zero-entropy encoders and simultaneous decoders in identification via quantum channels. These results are motivated by deterministic identification via classical channels. In this context, the encoder is restricted from using randomization. This work revisits the problem of identification via quantum channels with the additional constraint that the message encoding must utilize pure quantum states rather than general mixed states. Building on previous distinctions between simultaneous and general decoders, this introduces a two-dimensional spectrum of different identification capacities, which may exhibit significant variability.

The main findings are twofold. Firstly, it is shown that all four combinations of encoder and decoder types (pure/mixed encoder, simultaneous/general decoder) result in a code size that grows double-exponentially. These identification capacities are also lower bounded by the classical transmission capacity for a general quantum channel, as defined by the Holevo-Schumacher-Westmoreland Theorem. Secondly, it is established that the simultaneous identification capacity of a quantum channel is equivalent to the simultaneous identification capacity with pure state encodings. This finding reveals a hierarchy of three linearly ordered identification capacities.

Through analysis of simple examples, it is concluded that these three capacities are distinct: the general identification capacity can exceed the pure-state-encoded identification capacity, which in turn can exceed the pure-state-encoded simultaneous identification capacity. The details of these results are published in \cite{PosterPau}.

\vspace*{0.5cm}
\newpage

\subsection{Rami Ezzine: “Common Randomness Generation over Point-to-Point Channels”}
\begin{floatingfigure}[r]{6cm}
\mbox{\includegraphics[width=5.5cm]{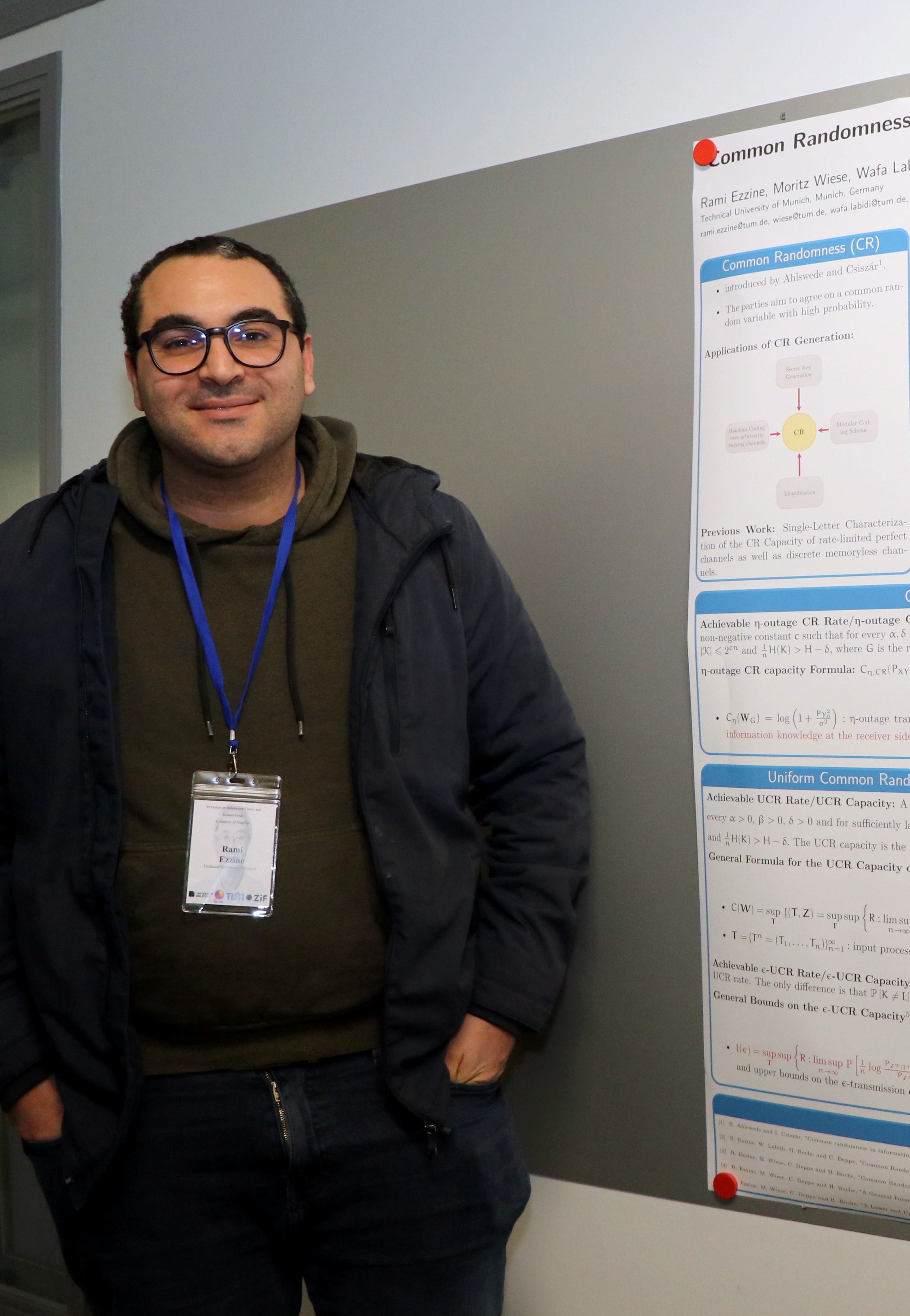}}
\caption{Rami Ezzine}
\end{floatingfigure}
Rami Ezzine's poster examined the problem of generating common randomness (CR) from correlated discrete sources aided by unidirectional communication over point-to-point channels. CR between two terminals refers to a random variable that is observable to both terminals with a low error probability. In many models, one terminal acts as the sender station and the other as the receiver station. The availability of CR enables the implementation of correlated random protocols, potentially leading to faster and more efficient algorithms.
CR generation is crucial in sequential secret key generation, where it is often referred to as information reconciliation. The generated CR can enhance performance gains in identification schemes, known for their efficiency compared to classical transmission schemes in applications requiring ultra-reliable low-latency communication such as 6G communications systems. Moreover, CR is highly relevant in the modular coding scheme for secure communication. A common scenario in seeded modular coding is when legitimate parties have access to CR as an additional resource that can be used as a seed. This poster provides a survey of our research on the problem of CR generation sources from finite sources with one-way communication over Gaussian channels with input constraint \cite{CRgaussian}, slow fading channels with arbitrary state distribution \cite{CRslowfading} as well as arbitrary point-to-point channels \cite{UCR,epsilonUCR}. CR capacity results are presented for each of the proposed models. 
\vspace*{0.5cm}
\newpage

\subsection{Sifat Rezwan: "Functional Compression for goal-oriented communication in control applications"}
\begin{floatingfigure}[r]{6cm}
\mbox{\includegraphics[width=5.5cm]{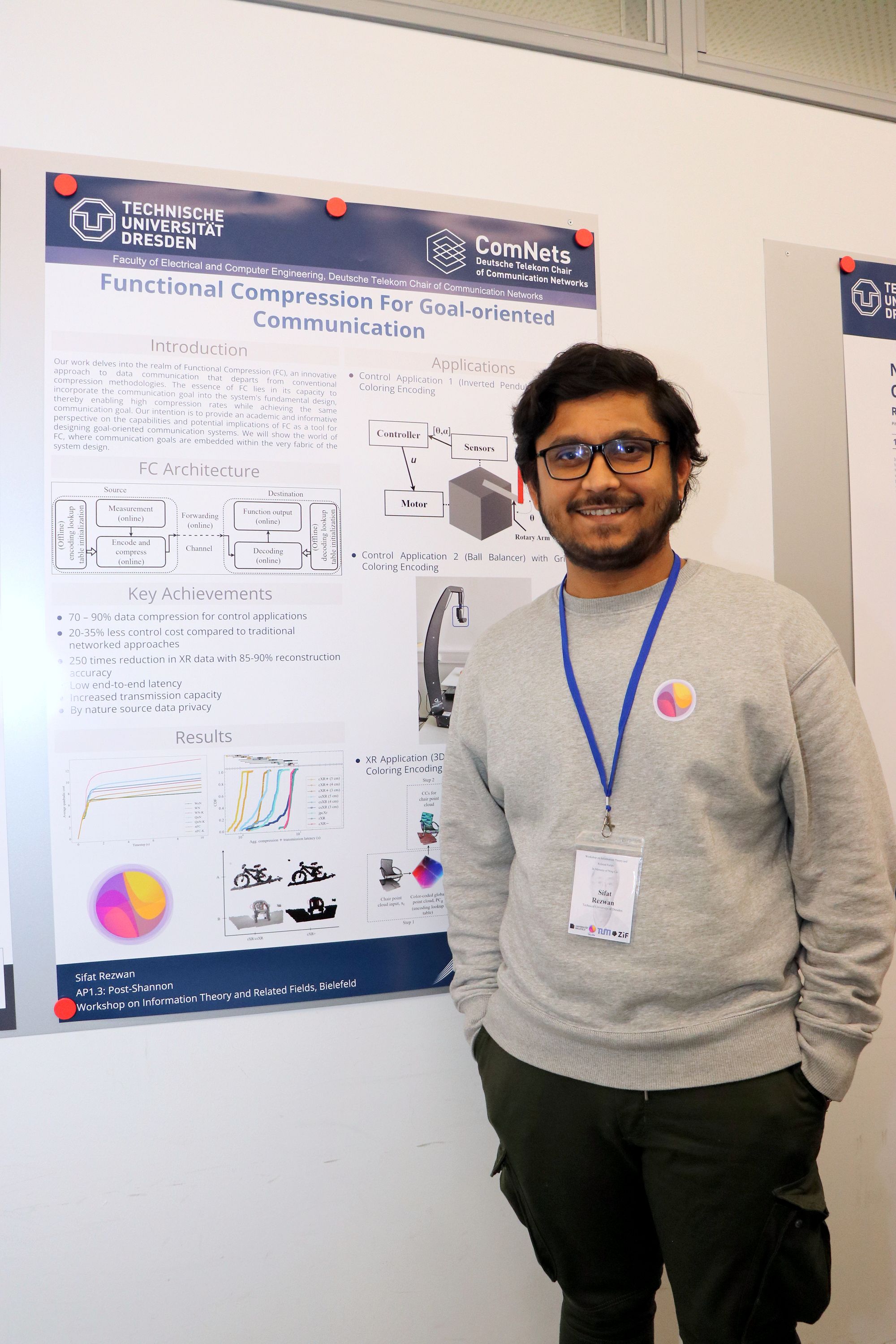}}
\caption{Sifat Rezwan}
\end{floatingfigure}

Sifat Rezwan presented a poster on functional compression for goal-oriented communication in control applications. The focus of future communication systems is increasingly shifting towards managing and controlling cyber-physical systems, particularly in contexts such as Industry 4.0, where systems can experience rapid congestion. In these environments, a large volume of sensor data is transmitted wirelessly to multiple in-network controllers that compute control functions for cyber-physical systems.

The poster introduces an implementation of network Functional Compression (FC) as a proof of concept to significantly reduce data traffic in such settings. FC is a goal-oriented communication strategy designed to transmit only the minimum amount of information necessary for function computation at the receiver end. In this setup, senders transmit an encoded and compressed version of sensor data to an in-network controller that is tasked with computing a target function, specifically a PID controller.

The results demonstrate that FC can achieve compression rates exceeding 50\% in certain scenarios. Additionally, using FC in a distributed cascade manner can further enhance compression rates and reduce computational costs. The findings are detailed in \cite{PosterSifat}.

\vspace*{0.5cm}
\newpage

\subsection{Tobias Oechtering: "Pointwise Maximal Leakage”}
\begin{floatingfigure}[r]{6cm}
\mbox{\includegraphics[width=5.5cm]{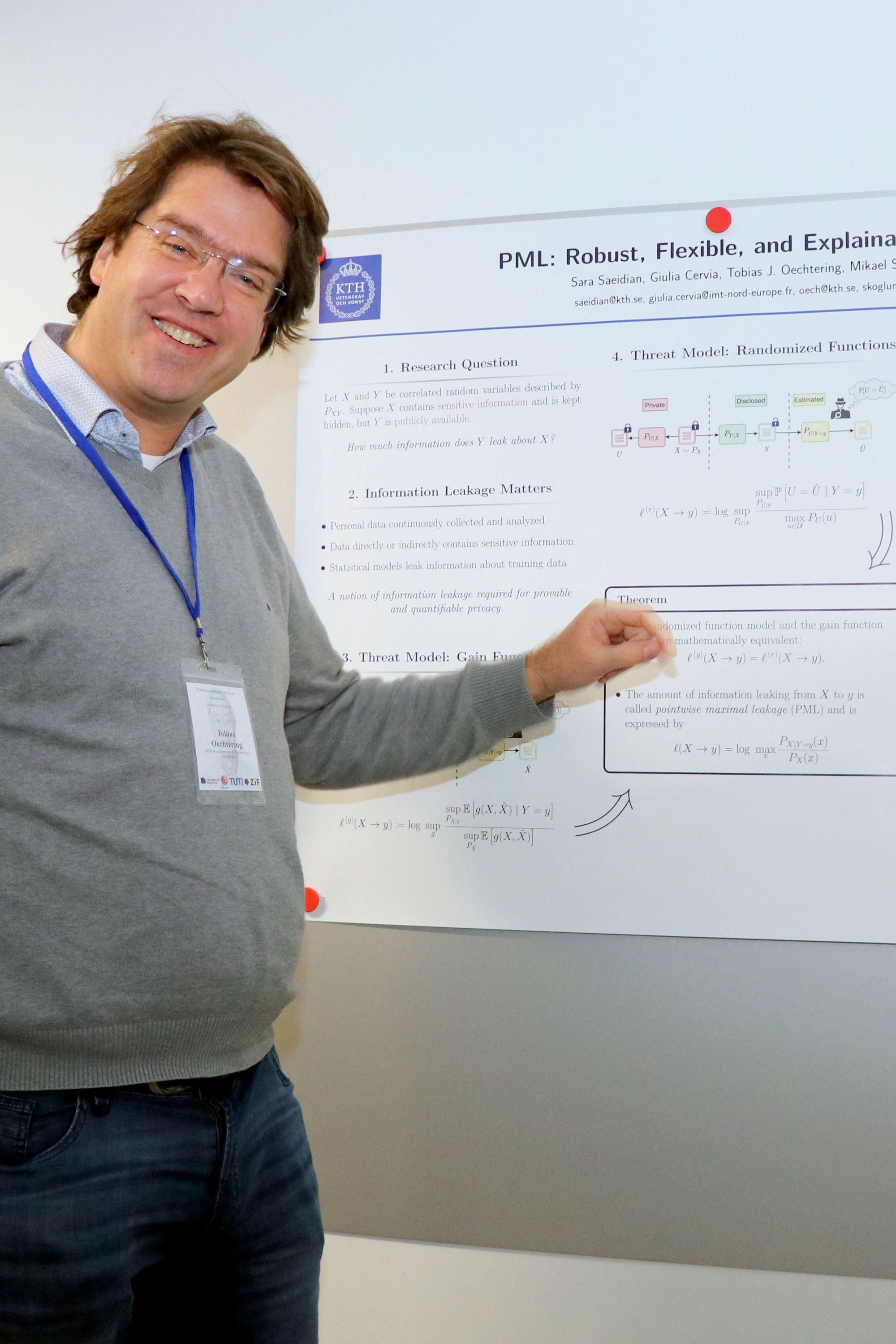}}
\caption{Tobias Oechtering}
\end{floatingfigure}

Tobias Oechtering presented a poster on the topic of pointwise maximal leakage, which is a new privacy measure that builds upon the concept of maximal leakage. This measure quantifies the amount of information disclosed about a secret \( X \) when a single outcome of a (randomized) function computed on \( X \) is revealed. Pointwise maximal leakage is designed to be robust and operationally significant, as it captures the maximum information leakage about \( X \) to adversaries who aim to guess arbitrary (potentially randomized) functions of \( X \), or equivalently, seek to maximize arbitrary gain functions.

The study explores various properties of pointwise maximal leakage, such as its aggregation across multiple outcomes and the effects of pre- and post-processing operations. Additionally, the research proposes conceptualizing information leakage as a random variable, allowing privacy guarantees to be viewed as constraints on different statistical properties of this random variable. Several privacy guarantees are defined and their behavior under pre-processing, post-processing, and composition is examined.

The relationship between pointwise maximal leakage and other privacy concepts, such as local differential privacy, local information privacy, and \( \text{ff} \)-information, is also explored. The work establishes a comprehensive and adaptable framework for assessing privacy risks, centered around a concept with strong operational implications suitable for various applications and practical scenarios. The details of these findings are presented in \cite{PosterTobias}.

\vspace*{0.5cm}

\newpage
\subsection{Vlad Andrei: “Resilient, Federated Large Language Models over Wireless Networks: An OFDM approach”}
\begin{floatingfigure}[r]{6cm}
\mbox{\includegraphics[width=5.5cm]{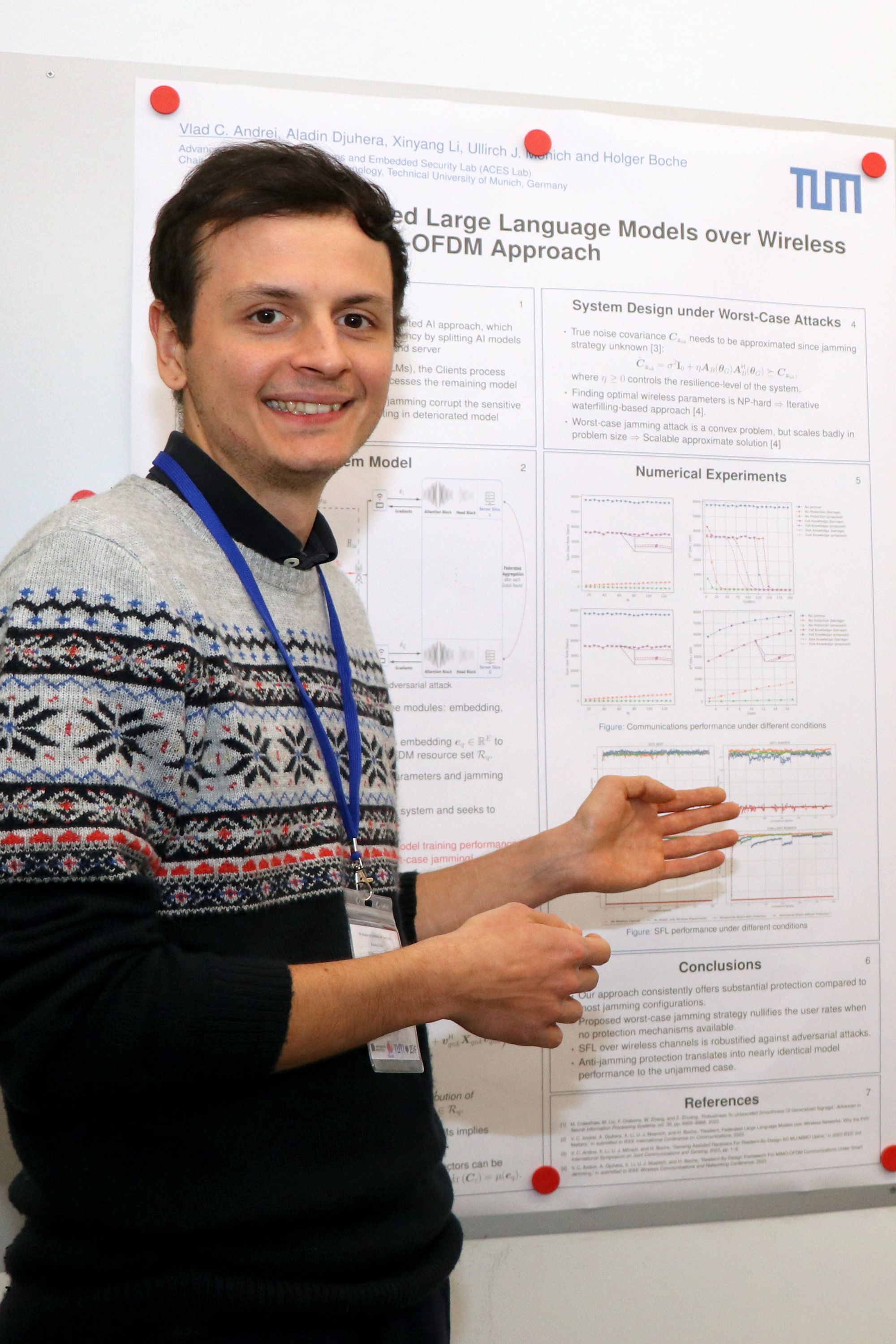}}
\caption{Vlad Andrei}
\end{floatingfigure}

Vlad Andrei's poster addressed the topic of resilient federated large language models over wireless networks, specifically through an OFDM approach. Ensuring security and resilience against native jamming is crucial for future 6G wireless networks. This presents a framework designed to enhance anti-jamming capabilities in MIMO-OFDM wireless communications. This framework introduces a novel method that integrates data from wireless sensing services to develop anti-jamming strategies, without requiring prior information or assumptions about the adversary's setup.

The proposed method involves substituting traditional noise covariance estimation techniques with a surrogate covariance model. This model uses sensing information on the directions-of-arrival (DoAs) of the jamming signals to effectively approximate the true jamming strategy. The study also explores how this sensing-assisted approach can be incorporated into the joint optimization of beamforming, user scheduling, and power allocation for a multi-user MIMO-OFDM uplink scenario. Despite the NP-hard nature of this optimization problem, an iterative water-filling approach is used to solve it efficiently.

To assess the effectiveness of the proposed sensing-assisted jamming mitigation, the worst-case jamming strategy designed to minimize the total user sum-rate is examined. Experimental simulations demonstrate that the approach is robust against both worst-case and barrage jamming scenarios, highlighting its potential to handle a variety of jamming situations. Since this approach integrates sensing-assisted information directly at the physical layer, it inherently incorporates resilience by design. Further details are provided in \cite{PosterVlad}.

\vspace*{0.5cm}
\newpage
\subsection{Wafa Labidi: "Joint Sensing and Identification for Discrete Memoryless Channels"}
\begin{floatingfigure}[r]{6cm}
\mbox{\includegraphics[width=5.5cm]{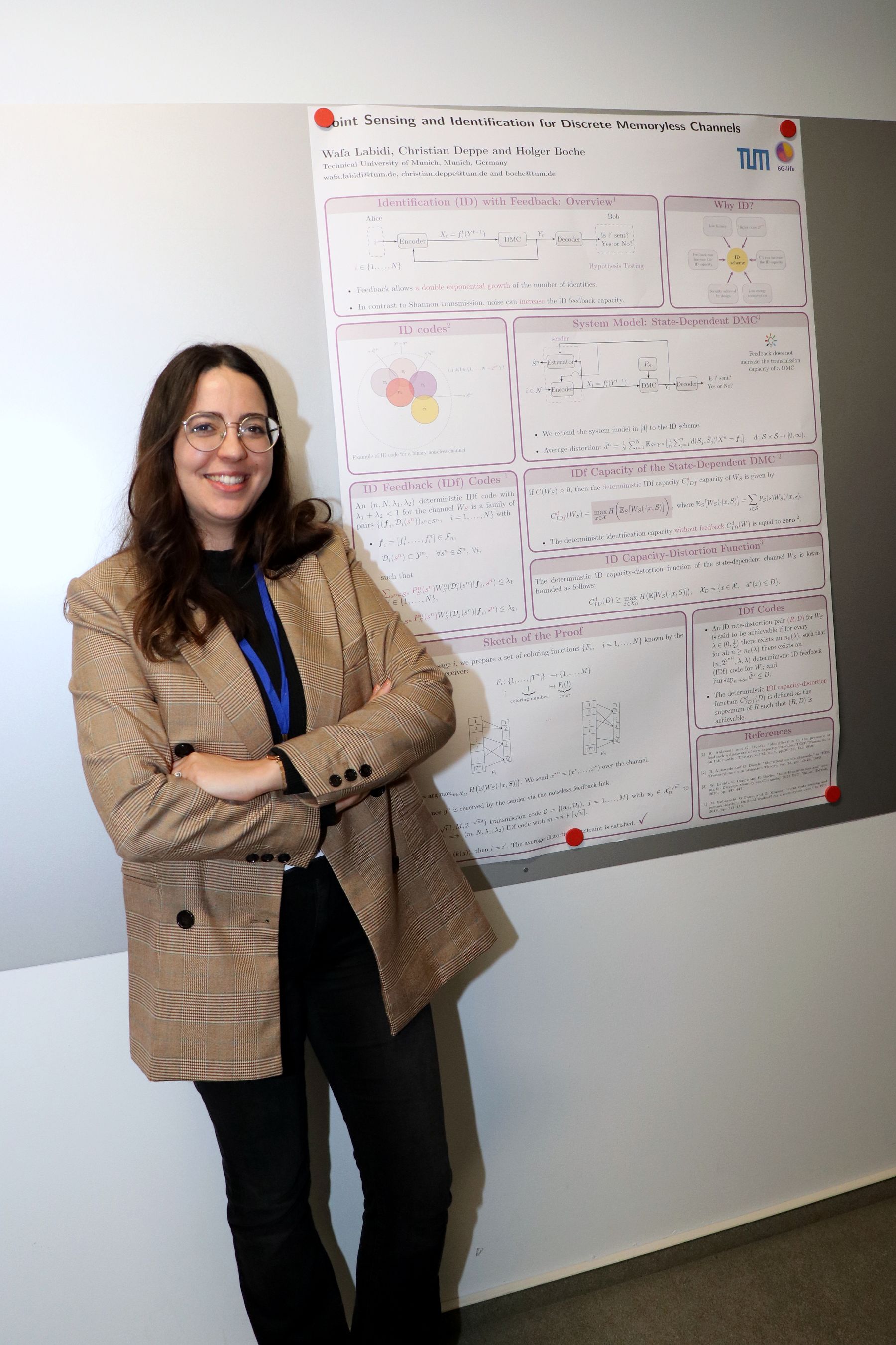}}
\caption{Wafa Labidi}
\end{floatingfigure}

Wafa Labidi's poster addressed the topic of joint sensing and identification for discrete memoryless channels (DMCs). The identification (ID) scheme proposed by Ahlswede and Dueck focuses on the receiver's task of determining whether a specific message of interest has been transmitted. Unlike traditional Shannon transmission codes, the size of ID codes for a DMC grows doubly exponentially with the blocklength when using randomized encoding. This notable result positions the ID paradigm as more efficient than classical Shannon transmission in terms of energy and hardware requirements. Additional advantages may be obtained by incorporating supplementary resources such as feedback.

The research investigates joint ID and channel state estimation over a DMC with independent and identically distributed (i.i.d.) state sequences. In this setup, the sender transmits an ID message alongside a random channel state, while the channel state is estimated strictly causally based on the channel output. Neither the sender nor the receiver has prior knowledge of the random channel state.

For this proposed system model, a lower bound on the ID capacity-distortion function is established. Further details are provided in \cite{PosterWafa}.

\vspace*{0.5cm}

\newpage
\subsection{Xinyang Li: “An Analysis of Capacity-Distortion Trade-Offs in Memoryless Joint Sensing and Communication Systems”}
\begin{floatingfigure}[r]{6cm}
\mbox{\includegraphics[width=5.5cm]{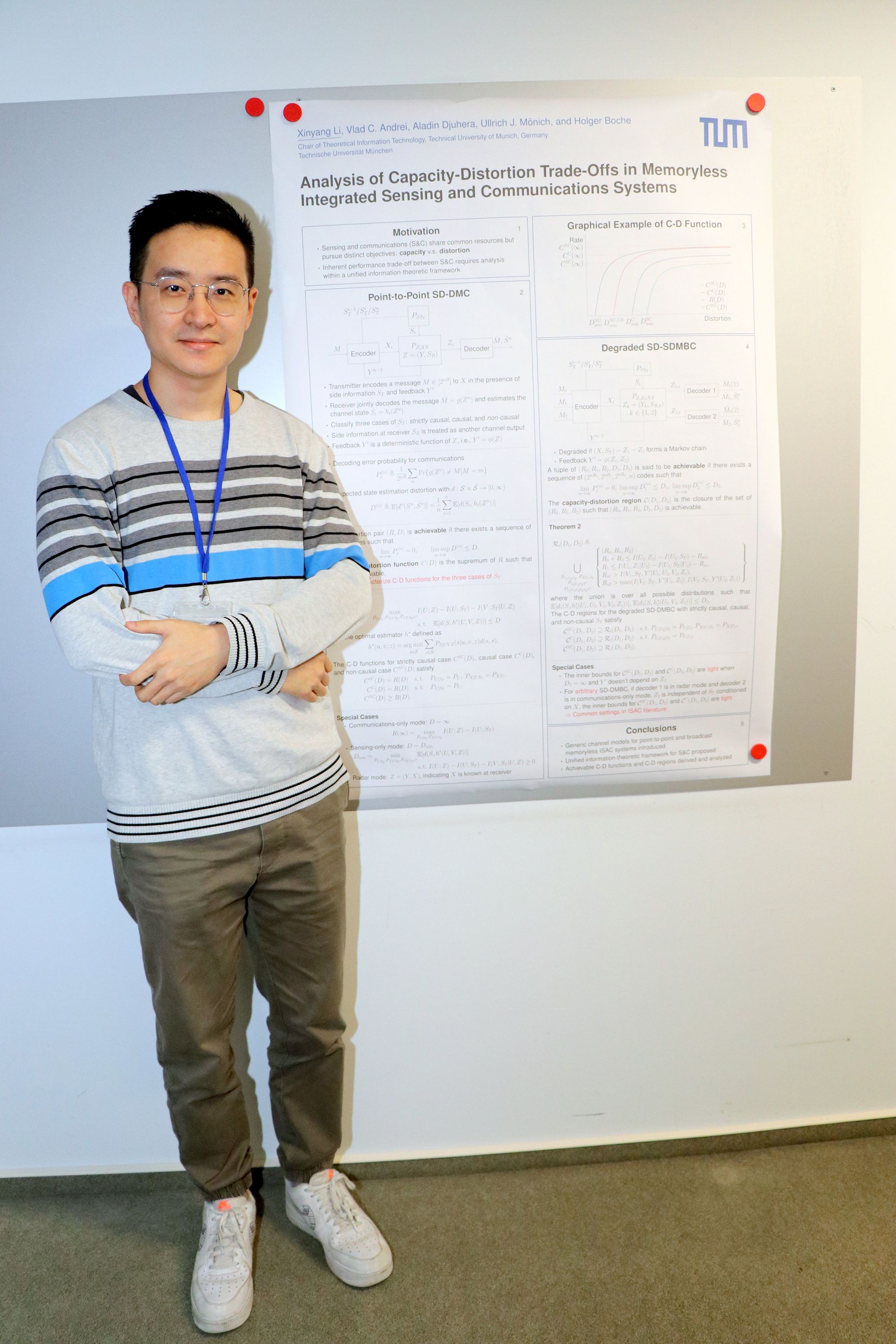}}
\caption{Xinyang Li}
\end{floatingfigure}

Xinyang Li's poster examined the information-theoretic limits of integrated sensing and communications (ISAC), with the goal of achieving both reliable communication and accurate channel state estimation simultaneously. The system is modeled using a state-dependent discrete memoryless channel (SD-DMC), considering scenarios with and without channel feedback and generalized side information available at both the transmitter and receiver. The receiver's role involves joint decoding of messages and estimation of the channel state.

The study investigates the trade-off between achievable communication rates and estimation errors, characterizing the capacity-distortion (C-D) trade-off for various levels of causality in side information. This framework is adaptable to different practical scenarios by interpreting side information in various ways, including applications in monostatic and bistatic radar systems.

The analysis is extended to a two-user degraded broadcast channel, where an achievable C-D region is derived under specific conditions, demonstrating tightness. To address the optimization problem for computing C-D functions and regions, a proximal block coordinate descent (BCD) method is proposed. The method's convergence to a stationary point is proved, and a stopping criterion is established.

Several illustrative examples are provided to highlight the flexibility of the framework and the effectiveness of the proposed algorithm. The results are detailed in \cite{PosterXingyang}.

\vspace*{0.5cm}
\newpage

\subsection{Yannik Böck: “Turing meets Machine Learning: Uncomputability of Zero-Error Classifiers”}
\begin{floatingfigure}[r]{6cm}
\mbox{\includegraphics[width=5.5cm]{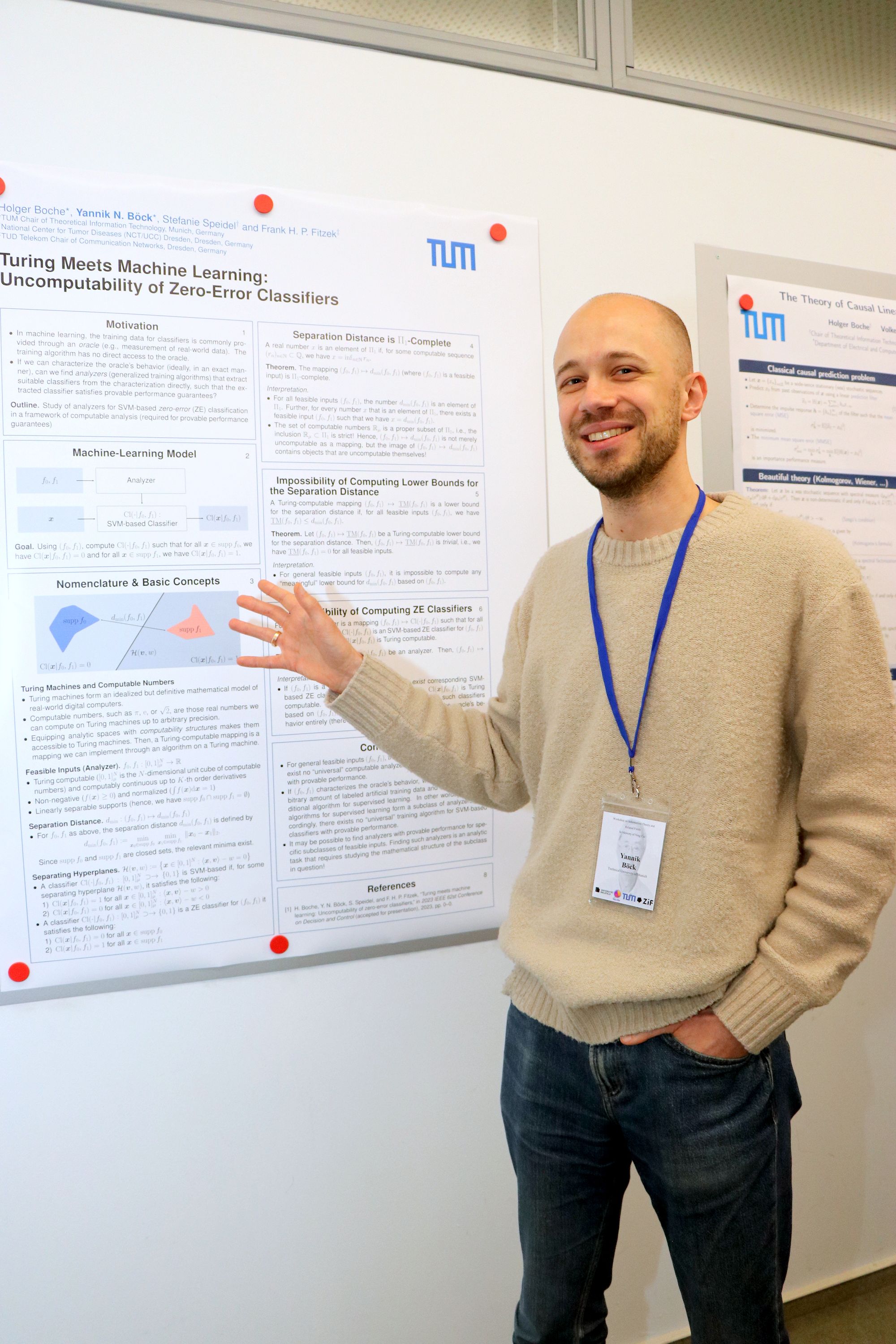}}
\caption{Yannik Böck}
\end{floatingfigure}

Yannik Böck's poster explored the intersection of Turing theory and machine learning, focusing on the uncomputability of zero-error classifiers. As automated decision-making systems powered by intelligent algorithms become increasingly significant across various domains of information technology (especially in applications involving critical infrastructure or sensitive human interests), establishing a robust theoretical foundation for assessing their integrity is crucial. This foundation is important not only for the technical evaluation of these systems but also for ensuring legal accountability of system operators.

The poster aims to deepen the understanding of the integrity of automated decision-making by applying fundamental mathematical models for computing hardware, particularly through the theory of Turing machines. It addresses the challenge of separating the support sets of smooth functions and provides a mathematically rigorous framework for implementing support-vector machines on digital computers. Additionally, it examines the computability properties of key quantities and objects, such as the distance between separated support sets and the separating hyperplanes.

The poster also offers non-technical interpretations of the findings in the context of machine learning and technological trustworthiness. The results are detailed in \cite{PosterYannick}.

\vspace*{0.5cm}

\newpage
\subsection{Zahra Khanian: "Rate Distortion Theory for Mixed States"}
\begin{floatingfigure}[r]{6cm}
\mbox{\includegraphics[width=5.5cm]{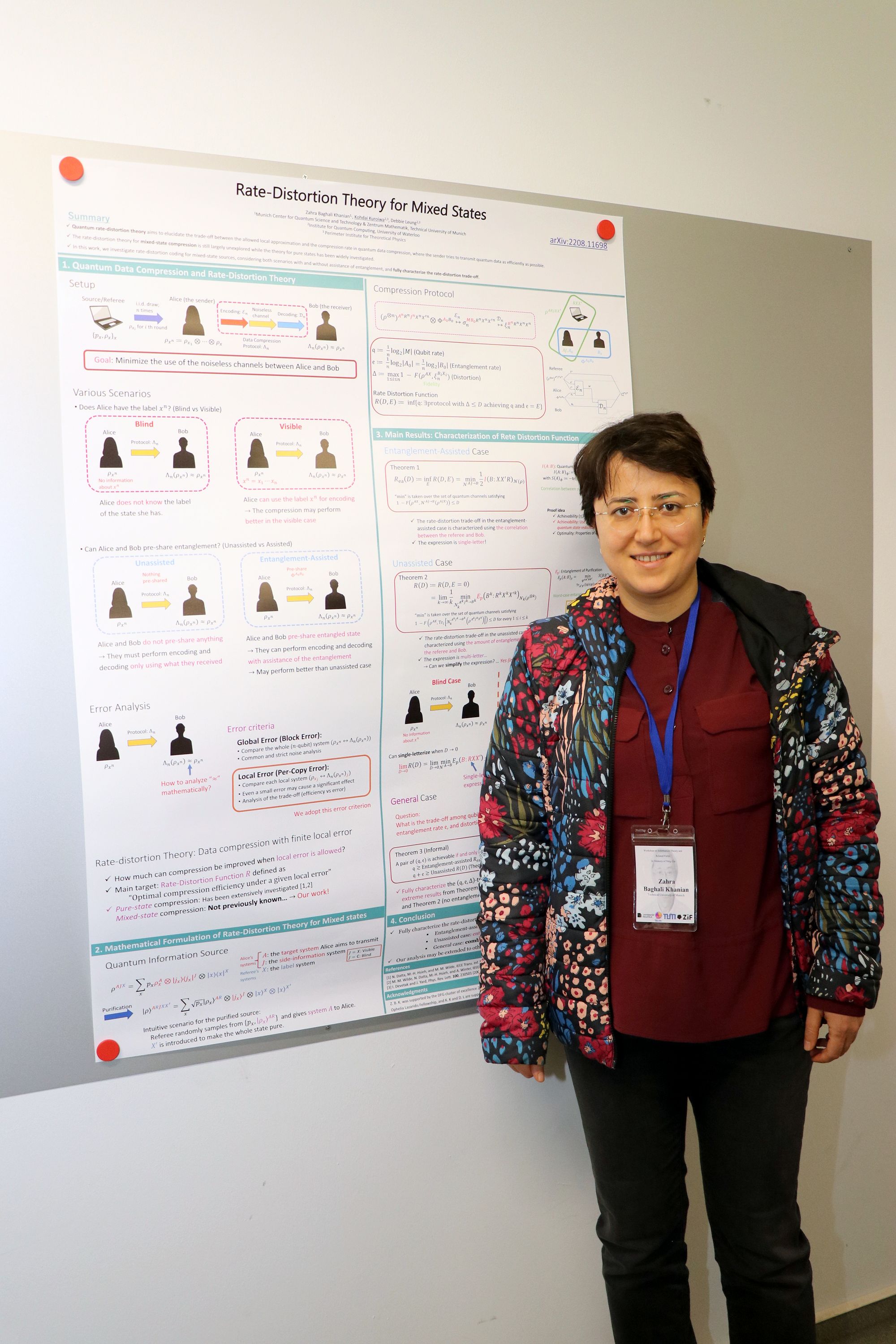}}
\caption{Zahra Khanian}
\end{floatingfigure}

In her poster, Zahra Khanian investigated the compression of asymptotically many i.i.d. copies of ensembles of mixed quantum states, with the encoder having access to a side information system. The performance measure under consideration is the per-copy or local error criterion. Rate-distortion theory is used to explore the trade-off between the compression rate and the per-copy error. The optimal trade-off is defined by the rate-distortion function, which indicates the best achievable rate for a given level of distortion.

The key contribution presented is the derivation of the rate-distortion function for mixed-state compression. Specifically, the rate-distortion functions are expressed for entanglement-assisted and unassisted scenarios in terms of a single-letter mutual information quantity and the regularized entanglement of purification, respectively. For the general scenario involving both communication and entanglement resources, the complete qubit-entanglement rate region is presented.

The proposed compression scheme supports various models, including blind and visible compression, as well as intermediate models, depending on the structure of the side information system. The results are detailed in \cite{PosterZahra}.

\vspace*{0.5cm}
\newpage

\subsection{Zuhra Amiri: “Comparing Latency and Power Consumption: Quantum vs. Classical Preprocessing”}
\begin{floatingfigure}[r]{6cm}
\mbox{\includegraphics[width=5.5cm]{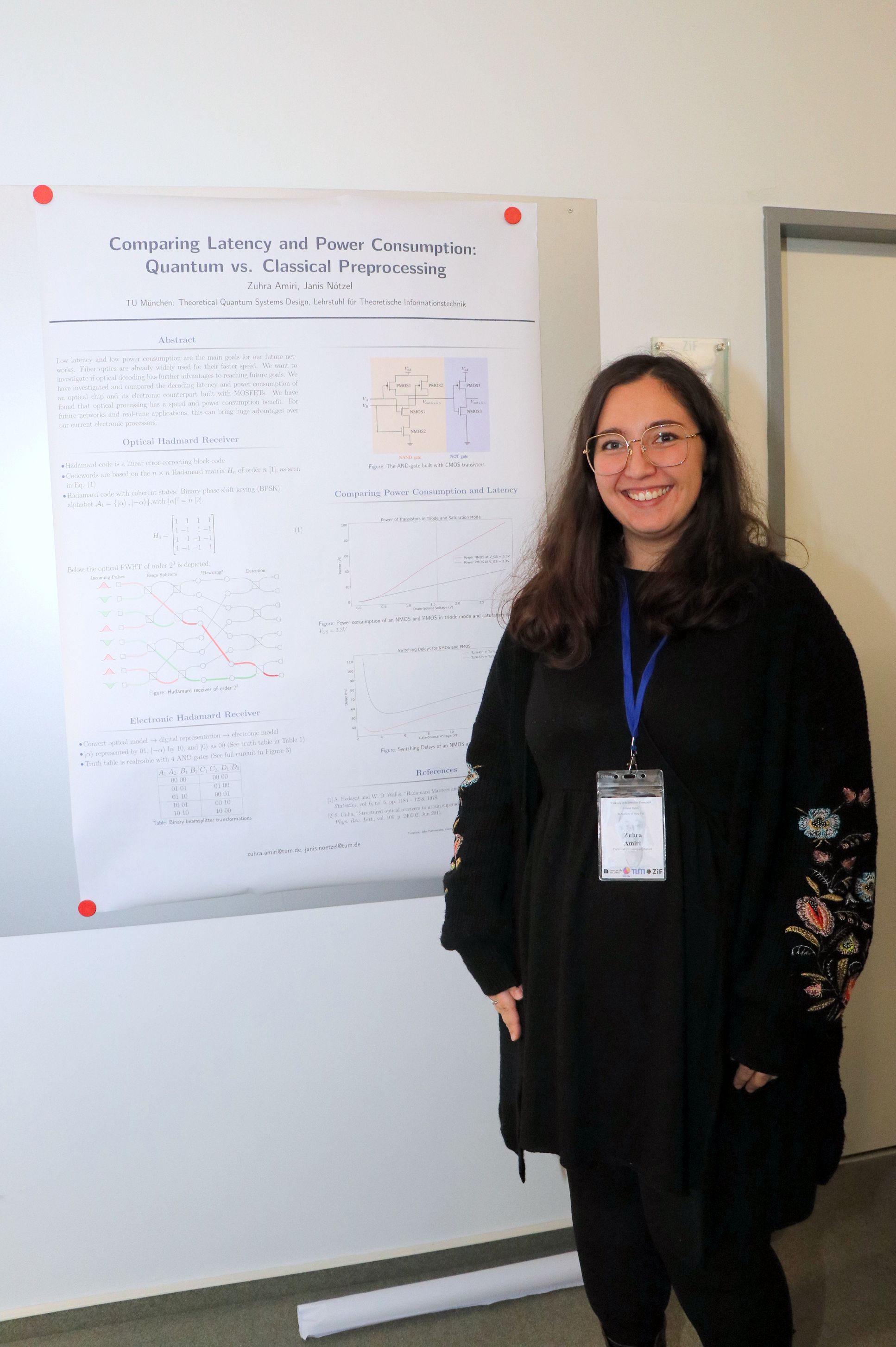}}
\caption{Zuhra Amiri}
\end{floatingfigure}

Zuhra Amiri's poster compared latency and power consumption between quantum and classical preprocessing methods. In optical computing, there has been ongoing interest in using light as the primary resource for information processing, which began with the discovery of lasers in the 1960s. By the 1980s, essential logic functions could be performed with optical devices. Today, advancements in photonics have led to the development of photonic integrated circuits (PICs), such as those produced by Quix Quantum. These circuits integrate optical components and electronics on a single chip, facilitating efficient optical signal processing and computation. Optical processing is particularly beneficial for high-speed communication, including applications in the Internet of Things (IoT), where it can enhance data transmission and computation.

The poster details the objectives of the research, which focus on achieving low latency and low power consumption for future networks. While fiber optics are already recognized for their high-speed transmission capabilities, the research investigates whether optical decoding can provide additional benefits in achieving these goals.

The study compares the decoding latency and power consumption of an optical chip with those of an electronic chip based on MOSFETs. The results show that optical processing offers significant advantages in terms of speed and power consumption, potentially leading to substantial improvements for future networks and real-time applications compared to current electronic processors. Further details are available in \cite{PosterZuhra}.

\newpage

\section{Workshop Talks}
Now we present the talks that were held during the conference.
\subsection{Jens Stoye - Welcome}
\begin{floatingfigure}[r]{6cm}
\mbox{\includegraphics[width=5.5cm]{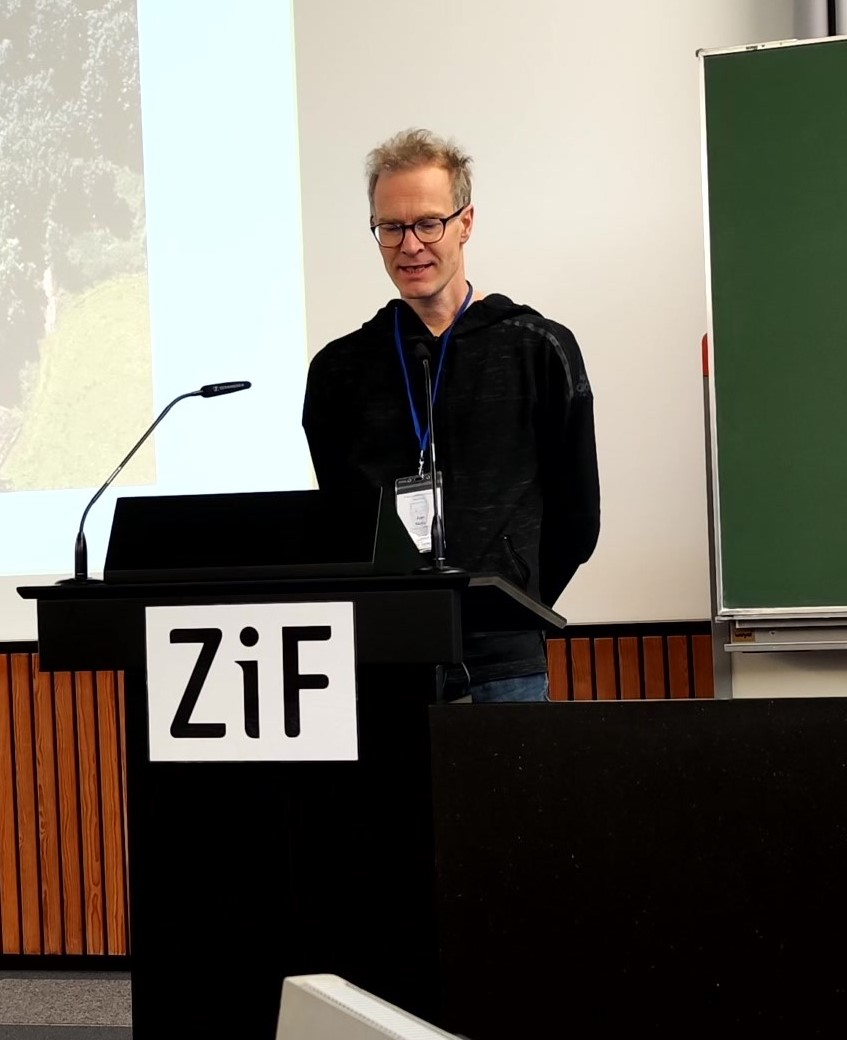}}
\caption{Jens Stoye}
\end{floatingfigure}
Jens Stoye (Director of ZiF) was the first to welcome the participants. He and Ning worked together on Rudolf Ahlswede's ZiF project "General Theory of Information Transfer" from 2002 to 2006. As they worked on different topics, they did not have much contact but Jens did a lot to ensure that the workshop could take place at ZiF. He also gave a brief introduction to ZiF and the research opportunities available there. Ning really appreciated working at ZiF and maintaining contacts with international scientists from all over the world. 

The Center for Interdisciplinary Research (ZiF) is an institute of Bielefeld University. It was founded in 1968 at Rheda Castle as the first institute of its kind in Germany, and as the nucleus of the new university. 
Details on the founding and orientation of the ZiF can be found in \cite{zif1994, padberg2014center}. Since then, it has been the inspiration for numerous similar establishments throughout Europe. It is located directly south of the main university building on the northern slope of the Teutoburg Forest.
The ZiF promotes and hosts international and interdisciplinary research groups. Scientists from various disciplines can carry out joint research projects at the ZiF, from large one-year research groups to conferences and workshops. Well-known researchers include Norbert Elias and the Nobel Prize winners Reinhard Selten, John Charles Harsanyi, Roger B. Myerson and Elinor Ostrom. 

\newpage
\subsection{Masahito Hayashi - Secure network coding with adaptive attack}
\begin{floatingfigure}[r]{6cm}
\mbox{\includegraphics[width=5.5cm]{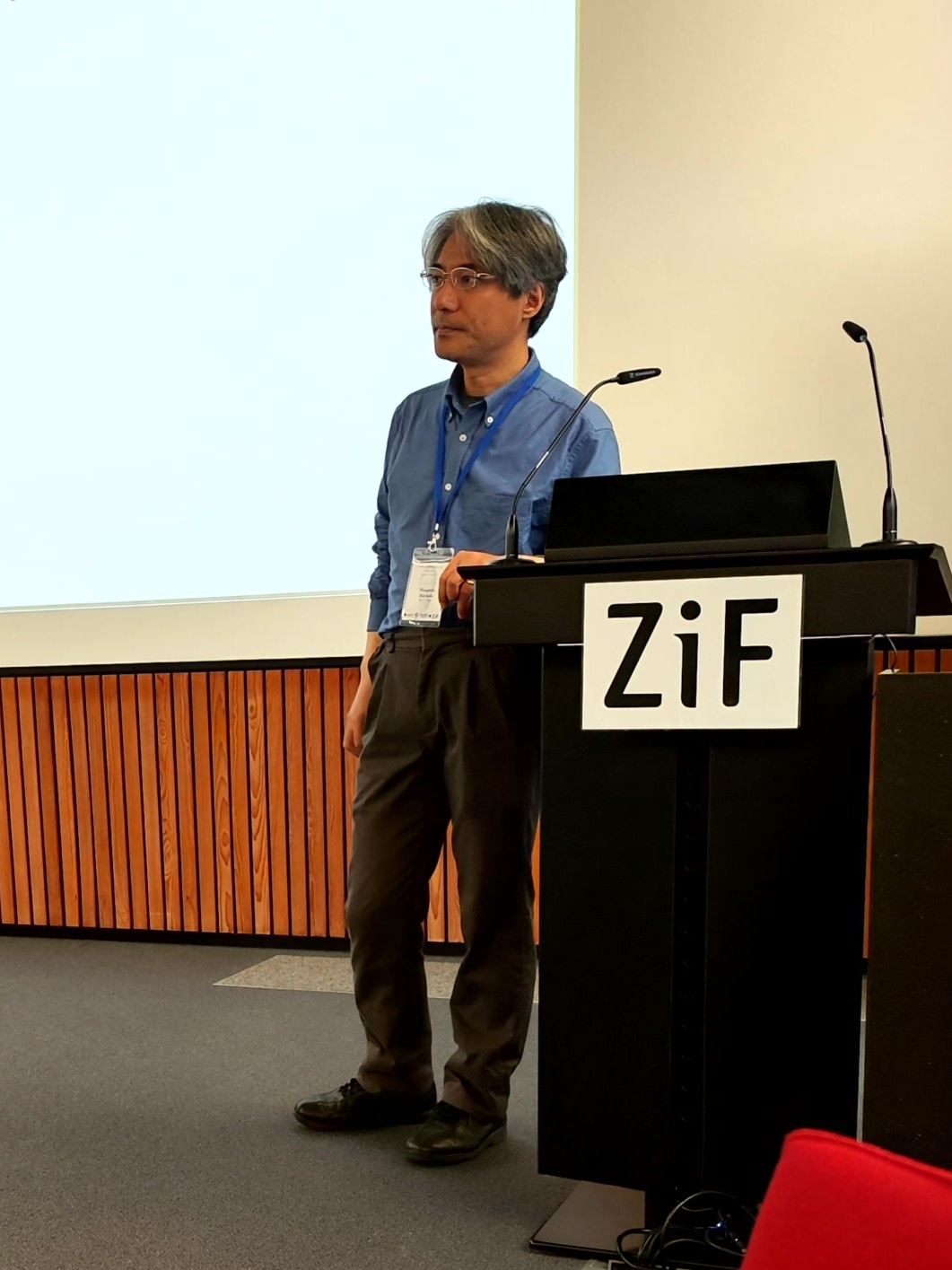}}
\caption{Masahito Hayashi}
\end{floatingfigure}
The first scientific lecture was delivered by Masahito Hayashi, focusing on the topic of "Secure Network Coding with Adaptive Attack." Masahito Hayashi, in collaboration with Ning, presented their latest research findings. Network coding is a crucial technology for enabling efficient communication over a network, ensuring that data is transmitted securely and reliably. Secure network coding specifically aims to protect our communication from various forms of attacks. However, most existing studies have primarily addressed deterministic and passive attacks, neglecting the implications of adaptive changes to the attacked edges and active modifications to the information on the attacked edge. Hayashi and Ning's research fills this gap by exploring how secure network coding can be fortified against these adaptive and active threats, thereby enhancing the robustness and security of network communication. The results are available in \cite{HayashiTalk}.

\bigskip

\begin{figure}
    \centering
    \includegraphics[width=8cm]{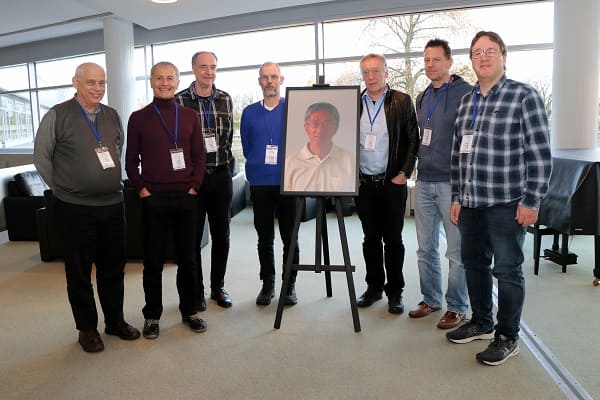}
    \caption{A meeting of the people who were involved in the start of the Ahlswede Group's ZiF Project together with Ning in 2003.}
    \label{fig:enter-label}
\end{figure}

\newpage
\subsection{Gyula Katona - Finding one valuable element in case of unreliable tests}
\begin{floatingfigure}[r]{6cm}
\mbox{\includegraphics[width=5.5cm]{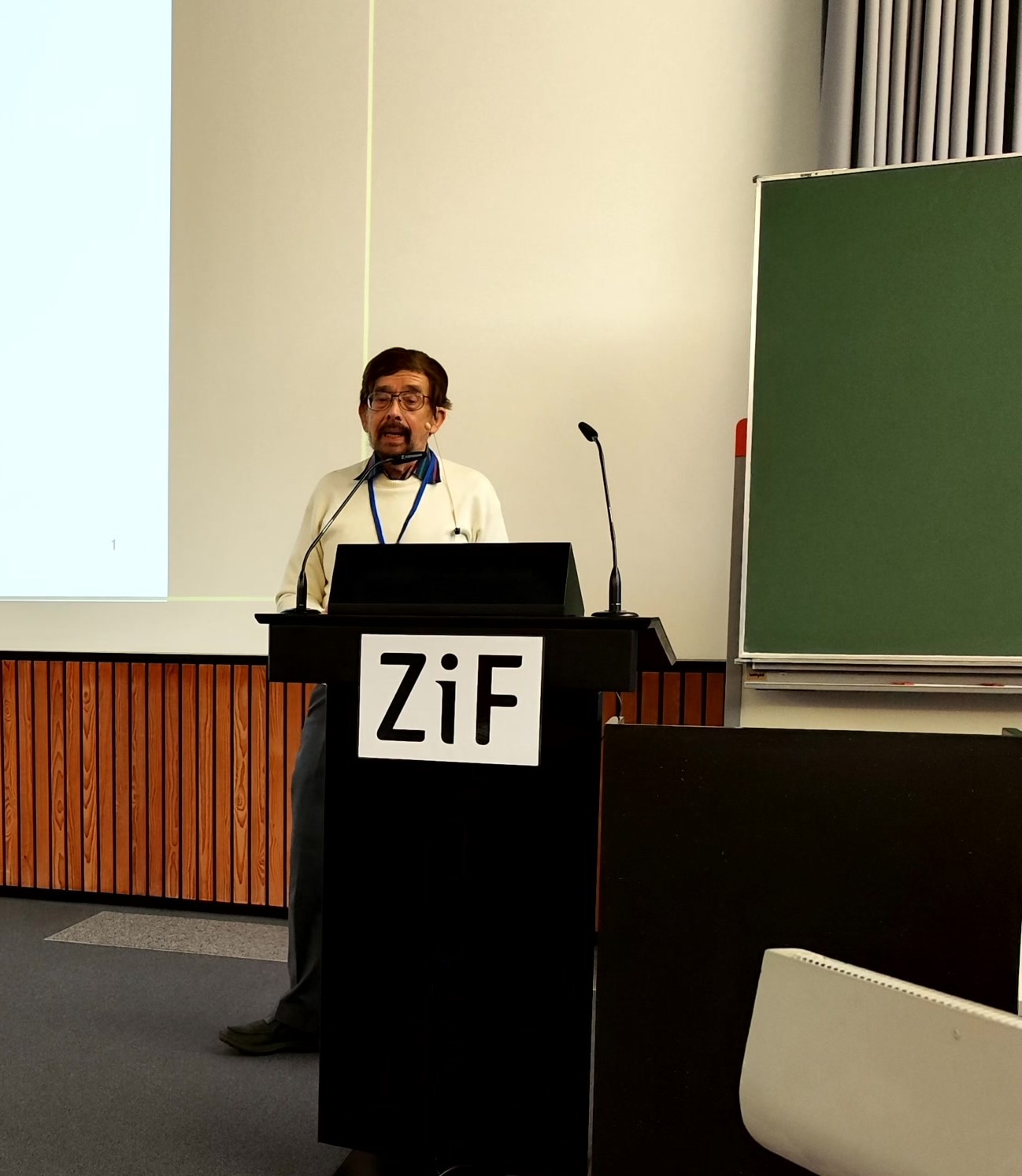}}
\caption{Gyula Katona}
\end{floatingfigure}
Gyula Katona began his lecture with a warm Chinese greeting and a heartfelt commemoration of Ning Cai. His presentation focused on the topic of combinatorial search for defective elements in a finite set. He commenced by providing a comprehensive overview of both adaptive and non-adaptive search methods. Katona then delved into the complexities that arise when some of the questions posed during the search process are answered incorrectly, exploring the impact of these inaccuracies on the overall search strategy.
Following this, he ask how many inquiries are necessary to identify at least one defective element within the set. This part of his lecture was particularly detailed, as he analyzed various scenarios and their respective search requirements.
Concluding his presentation, Katona shared a significant new finding related to non-adaptive search methods. Katona wrapped up his lecture by expressing his gratitude to the audience and bidding farewell in Chinese, mirroring the respectful tone with which he had begun.
\bigskip

\begin{figure}
    \centering
    \includegraphics[width=8cm]{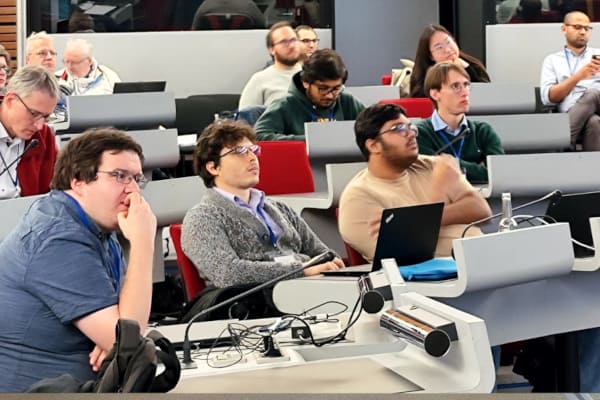}
    \caption{Interested listeners during the presentation.}
    \label{fig:enter-label}
\end{figure}

\newpage

\subsection{Ulrich Tamm - Memories of Ning}
\begin{floatingfigure}[r]{6cm}
\mbox{\includegraphics[width=5.5cm]{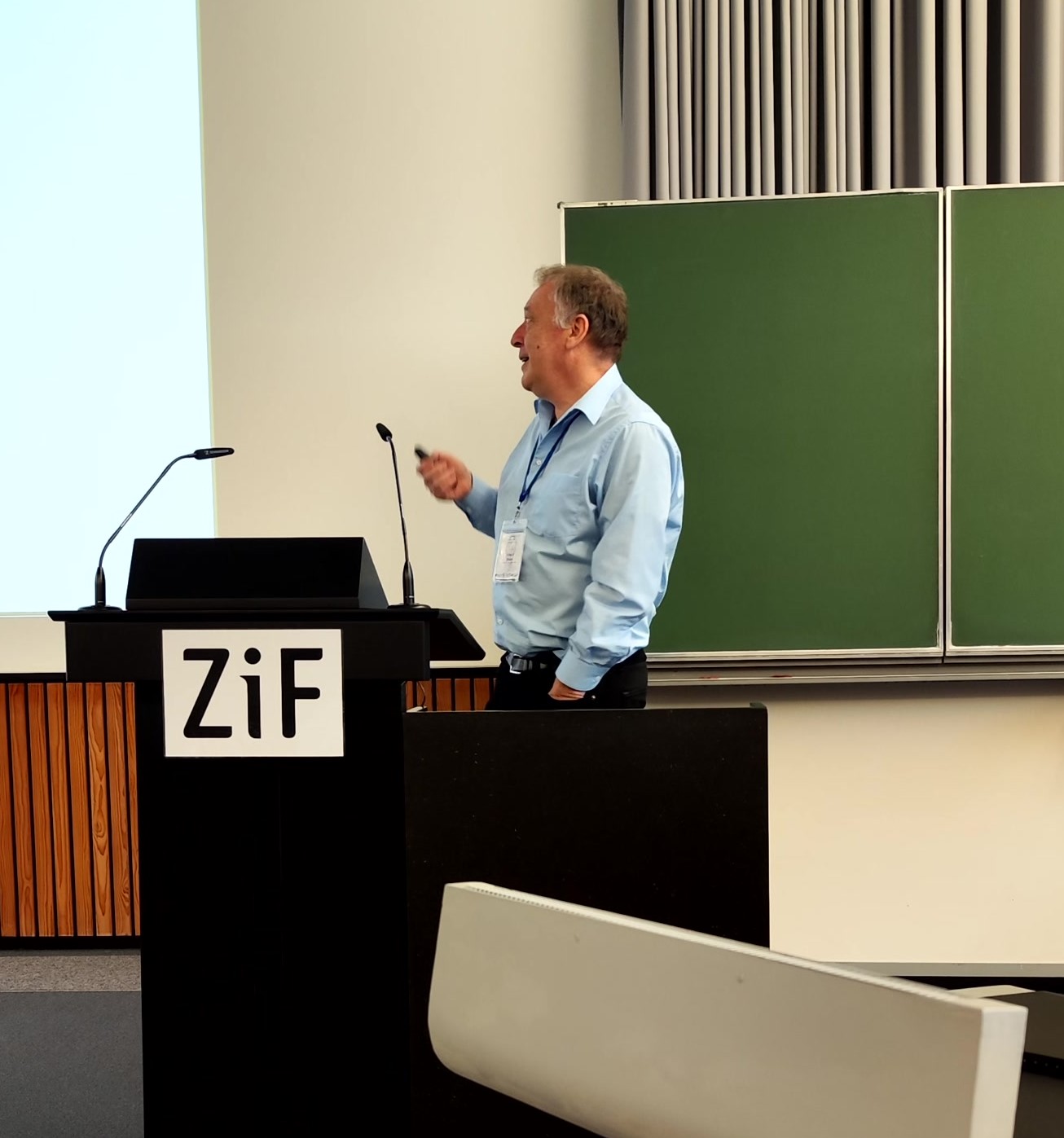}}
\caption{Ulrich Tamm}
\end{floatingfigure}
Following Gyula Katona's lecture, Ulrich Tamm shared his personal memories of Ning Cai. He recounted their time together as part of Rudolf Ahlswede's group, reflecting on several significant events. Tamm mentioned the winter schools they attended in Eindhoven in 1991 and 1994, the ISIT conference in Ulm, and numerous other conferences they participated in over the years. He fondly recalled their visit to Japan and highlighted the initiation of the renowned Coding Paper network during Raymond Yeung's visit to Bielefeld in 1997. Tamm concluded his reminiscence by talking about the annual Christmas gatherings with old colleagues in Bielefeld, cherishing the camaraderie and lasting connections formed during those times.

\bigskip

\subsection{Moritz Wiese - Universal Hash Functions and Mosaics of Designs}
\begin{floatingfigure}[r]{6cm}
\mbox{\includegraphics[width=5.5cm]{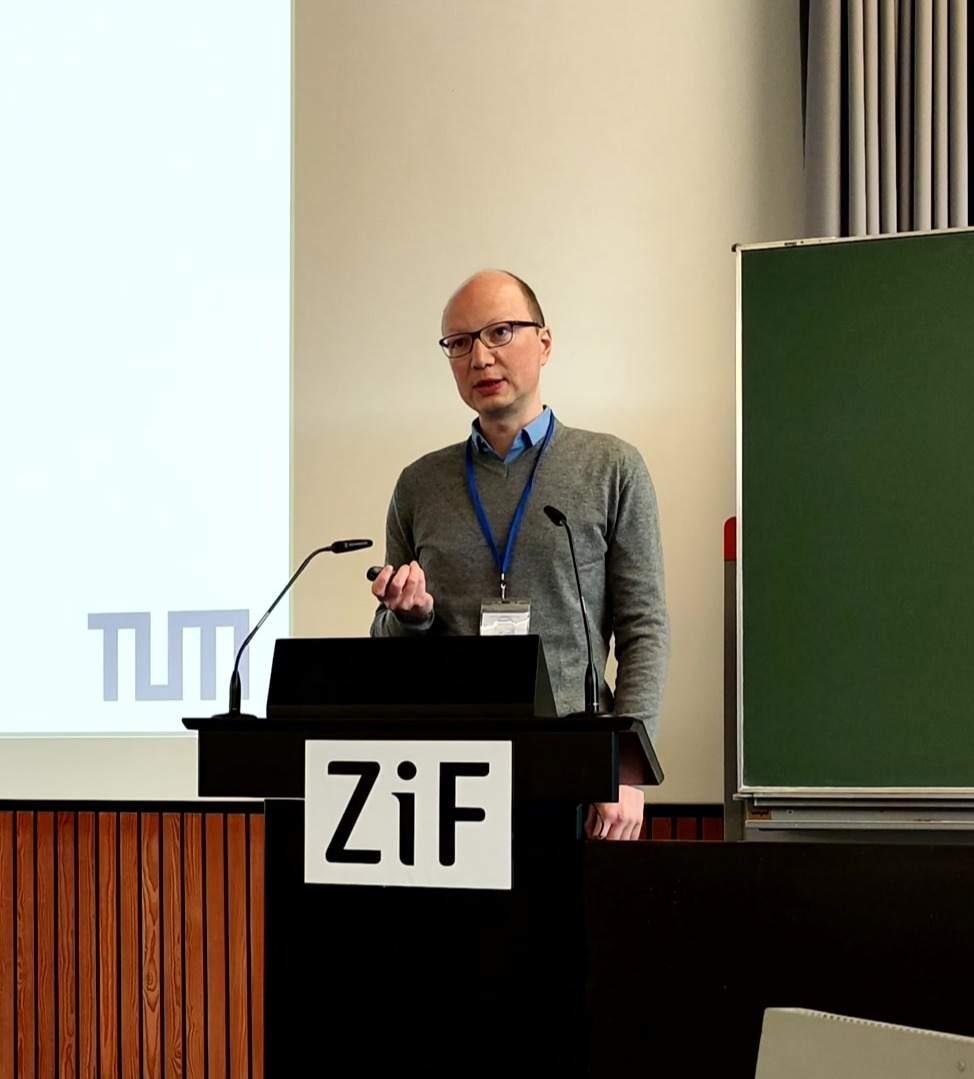}}
\caption{Moritz Wiese}
\end{floatingfigure}
In his scientific talk, Moritz Wiese introduced $\varepsilon$-almost collision-flat universal (ACFU) hash functions $f:\mc X\times\mc S\to\mc A$. Their main property is that the number of collisions in any given value is bounded. Each $\varepsilon$-ACFU hash function is an $\varepsilon$-almost universal (AU) hash function, and every $\varepsilon$-almost strongly universal (ASU) hash function is an $\varepsilon$-ACFU hash function. Moritz showed how the size of the seed set $\mc S$ depends on $\varepsilon,\abs{\mc X}$ and $\abs{\mc A}$. Depending on how these parameters are interrelated, seed-minimizing ACFU hash functions are equivalent to mosaics of balanced incomplete block designs (BIBDs), or to duals of mosaics of quasi-symmetric block designs. In a third case, mosaics of transversal designs and nets yield seed-optimal ACFU hash functions, but a full characterization is missing. By either extending $\mc S$ or $\mc X$, it is possible to obtain an $\varepsilon$-ACFU hash function from an $\varepsilon$-AU hash function or an $\varepsilon$-ASU hash function, generalizing the construction of mosaics of designs from a given resolvable design (Gnilke, Greferath, Pav\v cevi\'c, Des.\ Codes Cryptogr.~86(1)). The concatenation of an ASU and an ACFU hash function again yields an ACFU hash function. Finally, Moritz highlighted the importance of ACFU hash functions by their applicability in privacy amplification. The results are available in \cite{PosterMoritz}.
\bigskip
\subsection{Christian Deppe - Research with Ning}
\begin{floatingfigure}[r]{6cm}
\mbox{\includegraphics[width=5.5cm]{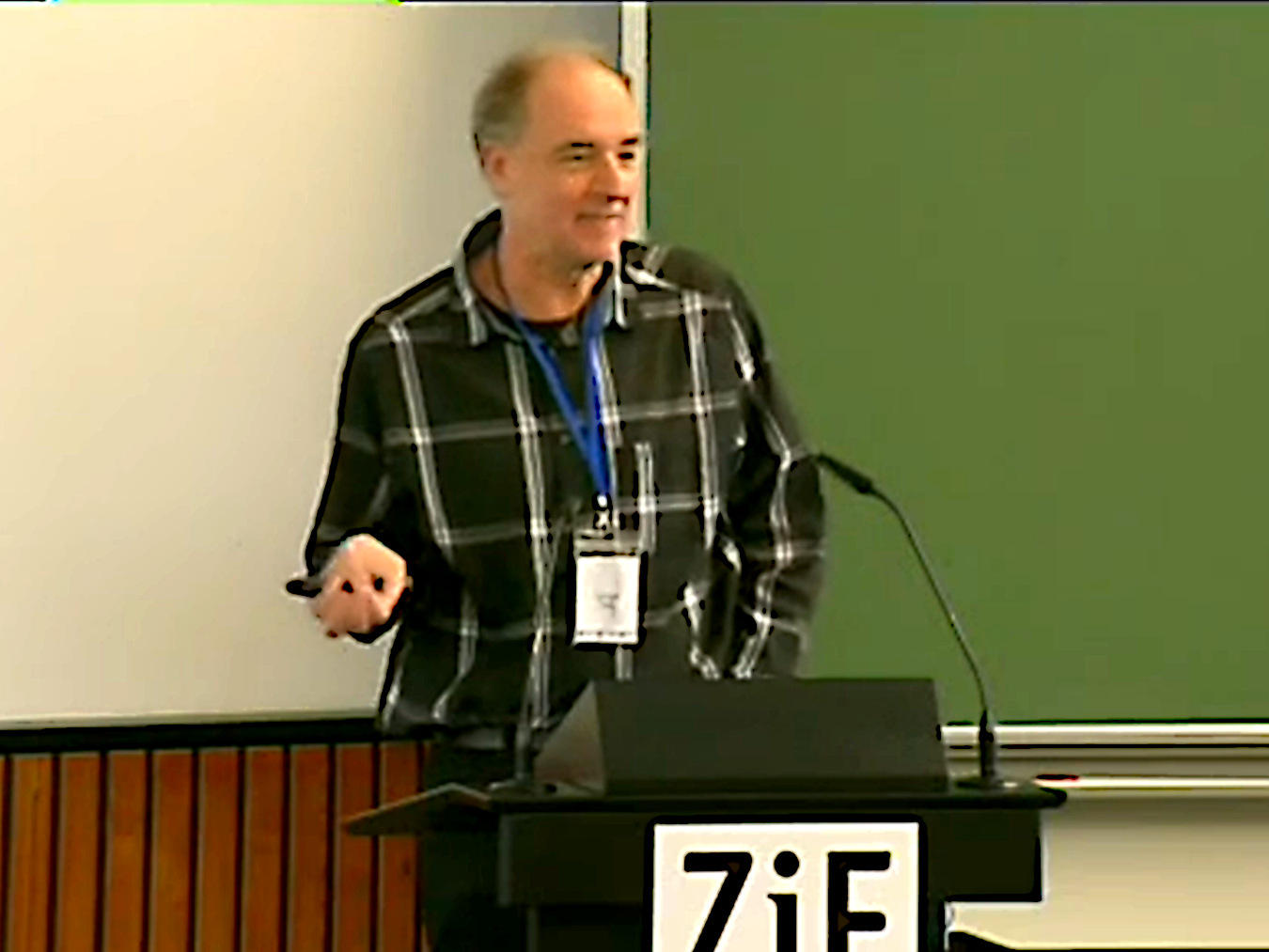}}
\caption{Christian Deppe}
\end{floatingfigure}
Following Moritz Wiese's scientific presentation, Christian Deppe shared insights into his collaborative work with Ning Cai. As a doctoral student, Deppe shared an office with Ning, who provided invaluable support and guidance. Their collaboration led to Deppe's first scientific post-doctorate results, focusing on error-correcting codes \cite{Ning_BC}. After Ning Cai left Bielefeld, his son, Minglai Cai, joined Rudolf Ahlswede's group as a doctoral student. Upon Ahlswede's passing, Deppe assumed leadership of the group, which led to two joint publications by Christian Deppe, Minglai Cai, and Ning Cai \cite{Cai_Compound, Ning_Minglai}. Deppe remembered Ning Cai as a highly supportive colleague and mentor, renowned for his patience. He emphasized that Ning Cai's passing was a significant loss to all who knew him.

\begin{figure}
    \centering
    \includegraphics[width=8cm]{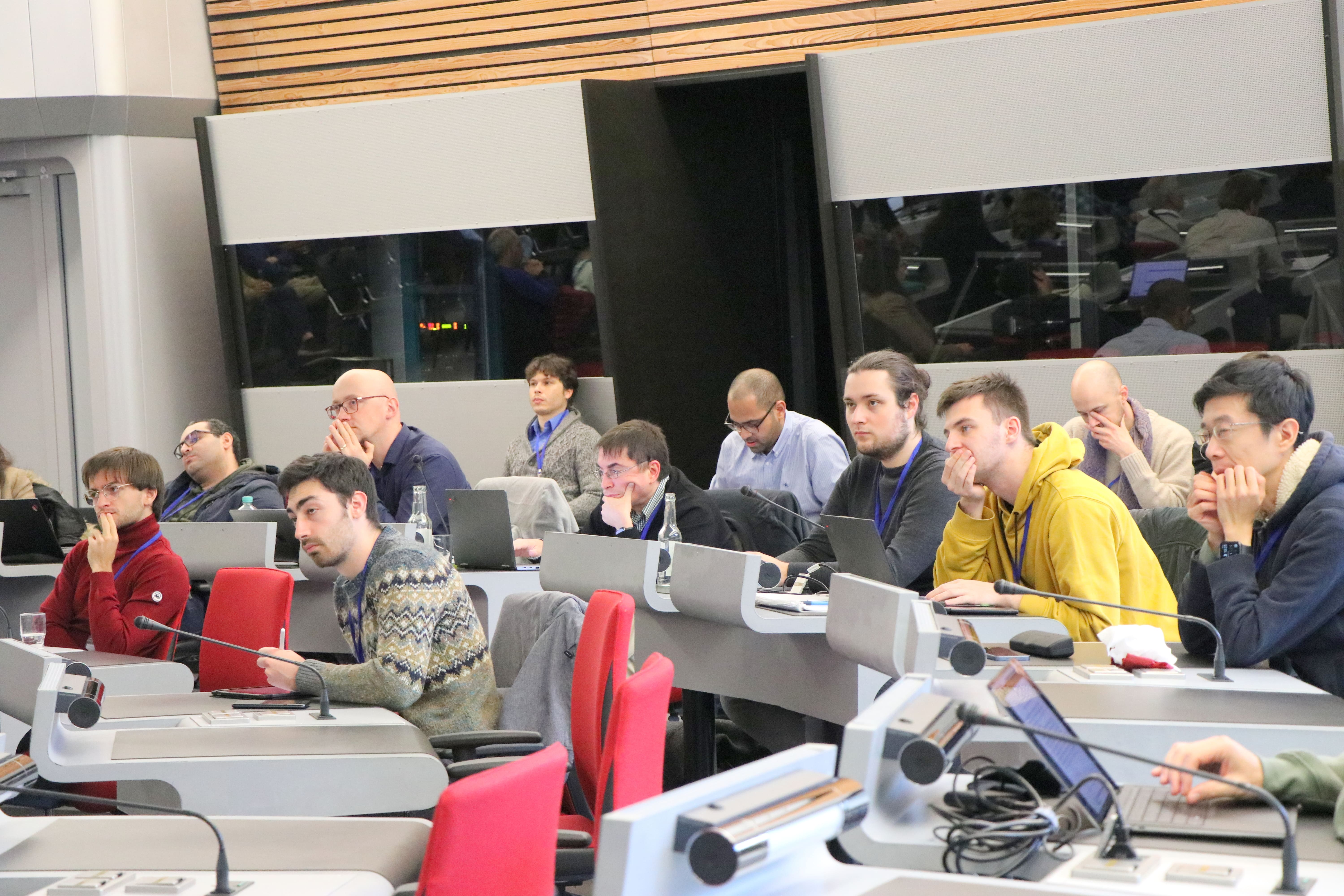}
    \caption{Interested listeners during the presentation.}
    \label{fig:enter-label}
\end{figure}

\newpage

\subsection{Andreas Winter - Ephemeral to fundamental: Ning Cai and the quantum}
\begin{floatingfigure}[r]{6cm}
\mbox{\includegraphics[width=5.5cm]{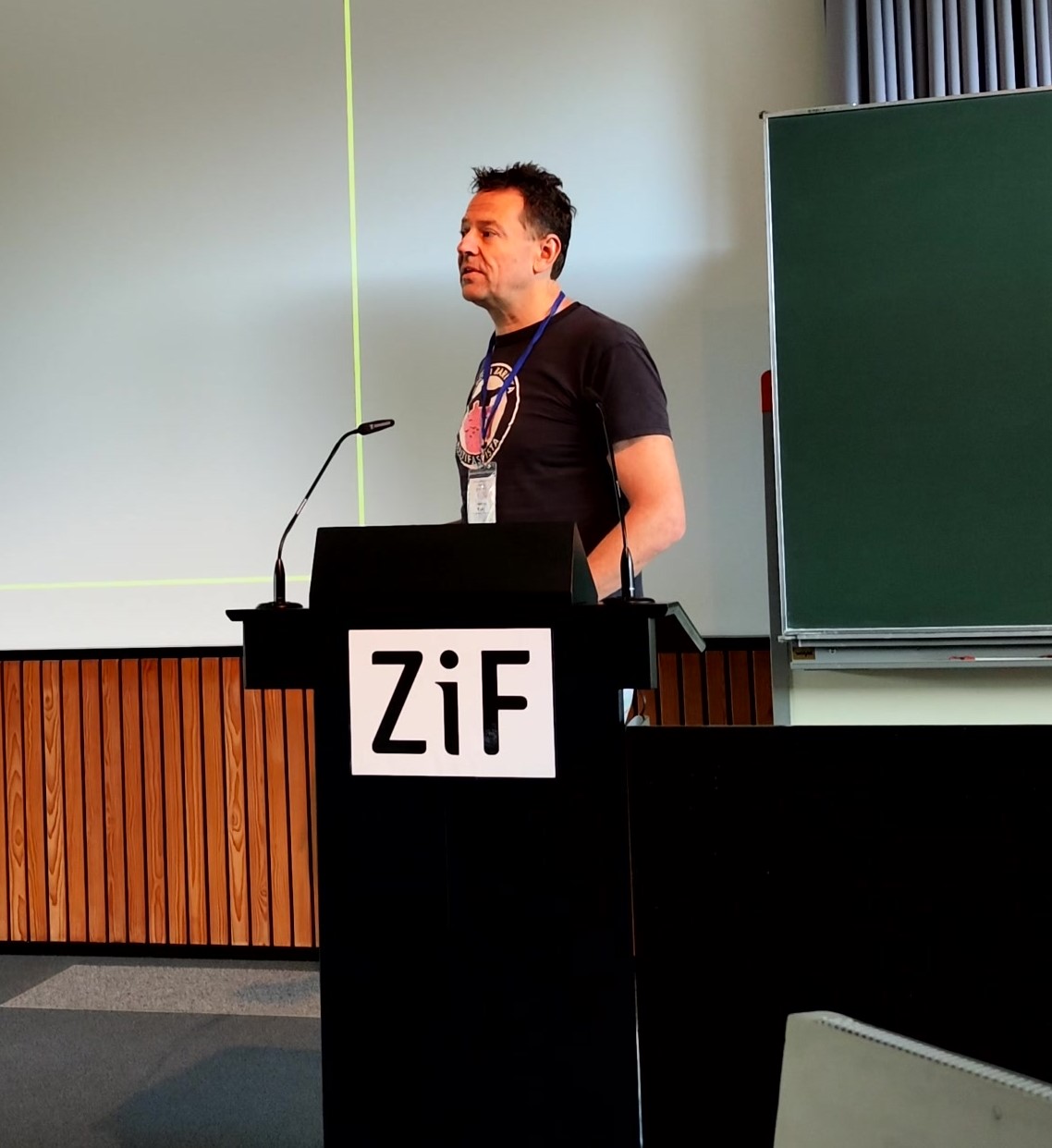}}
\caption{Andreas Winter}
\end{floatingfigure}

Christian Deppe's memories were followed by those of \mbox{Andreas Winter}. When
Andreas Winter arrived in Bielefeld in 1997, his research interests did not overlap significantly with those of Ning Cai. Winter was focused on quantum information theory for his doctoral thesis, a field that Ning found intriguing but one that he did not pursue. In 2004, Ning surprised \mbox{Winter} with an idea related to the quantum wiretap channel. Winter was impressed with the result and provided feedback, leading to his collaboration as a co-author with Raymond Yeung on this fundamental work. Although Winter did not publish further papers with Ning, he noted that Ning began to explore quantum information theory, producing papers with Rudolf Ahlswede, Holger Boche, Christian Deppe, and Janis Nötzel. From 2017 onward, Ning authored seven papers with Masahito Hayashi. Despite never managing to visit Ning in China, Winter and Ning met occasionally at conferences or in Bielefeld.

\bigskip

\begin{figure}
    \centering
    \includegraphics[width=8cm]{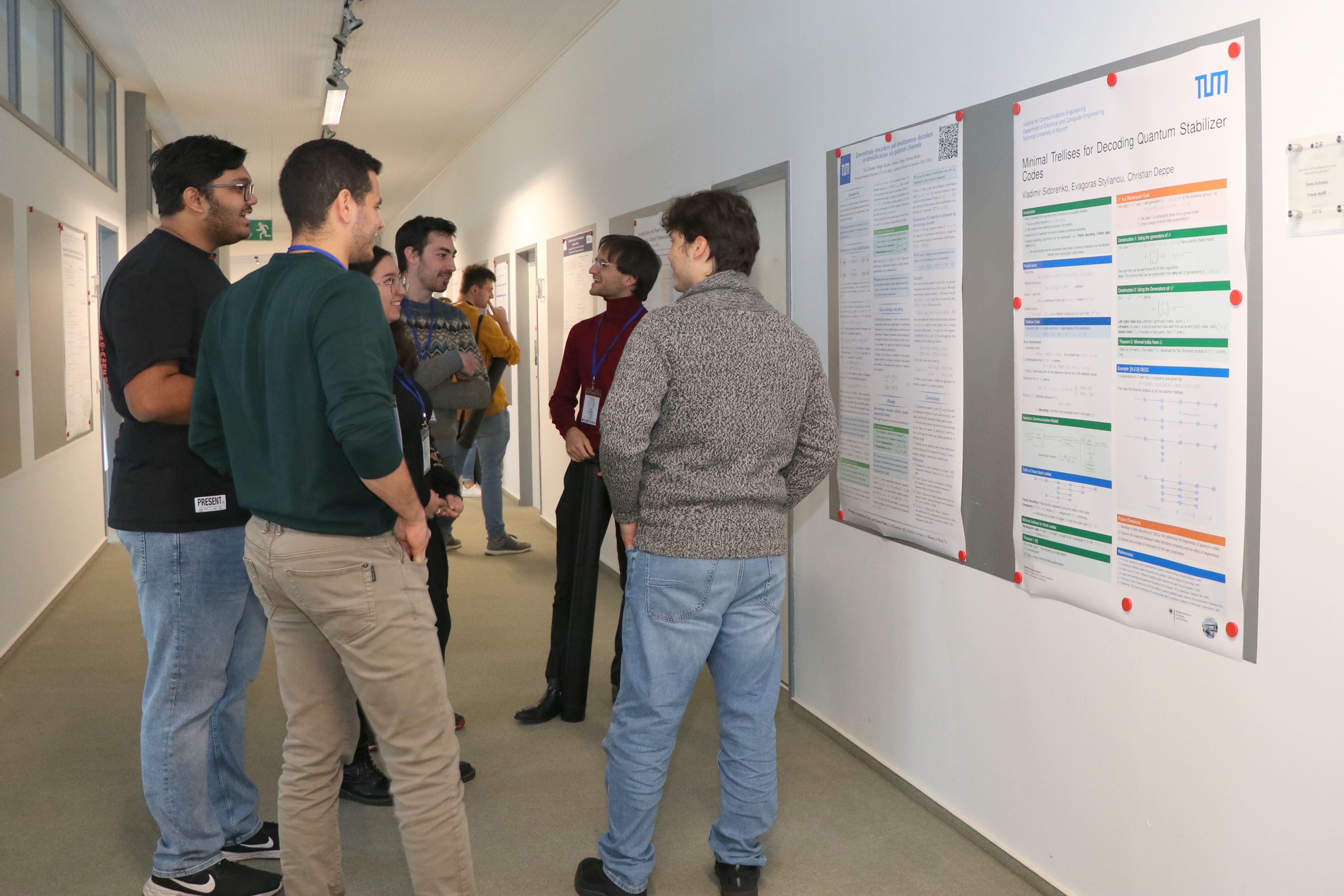}
    \caption{There were also interesting discussions at the posters during the breaks.}
    \label{fig:enter-label}
\end{figure}

\newpage
\subsection{Yanling Chen - Wiretap Channel with Correlated Sources}
\begin{floatingfigure}[r]{6cm}
\mbox{\includegraphics[width=5.5cm]{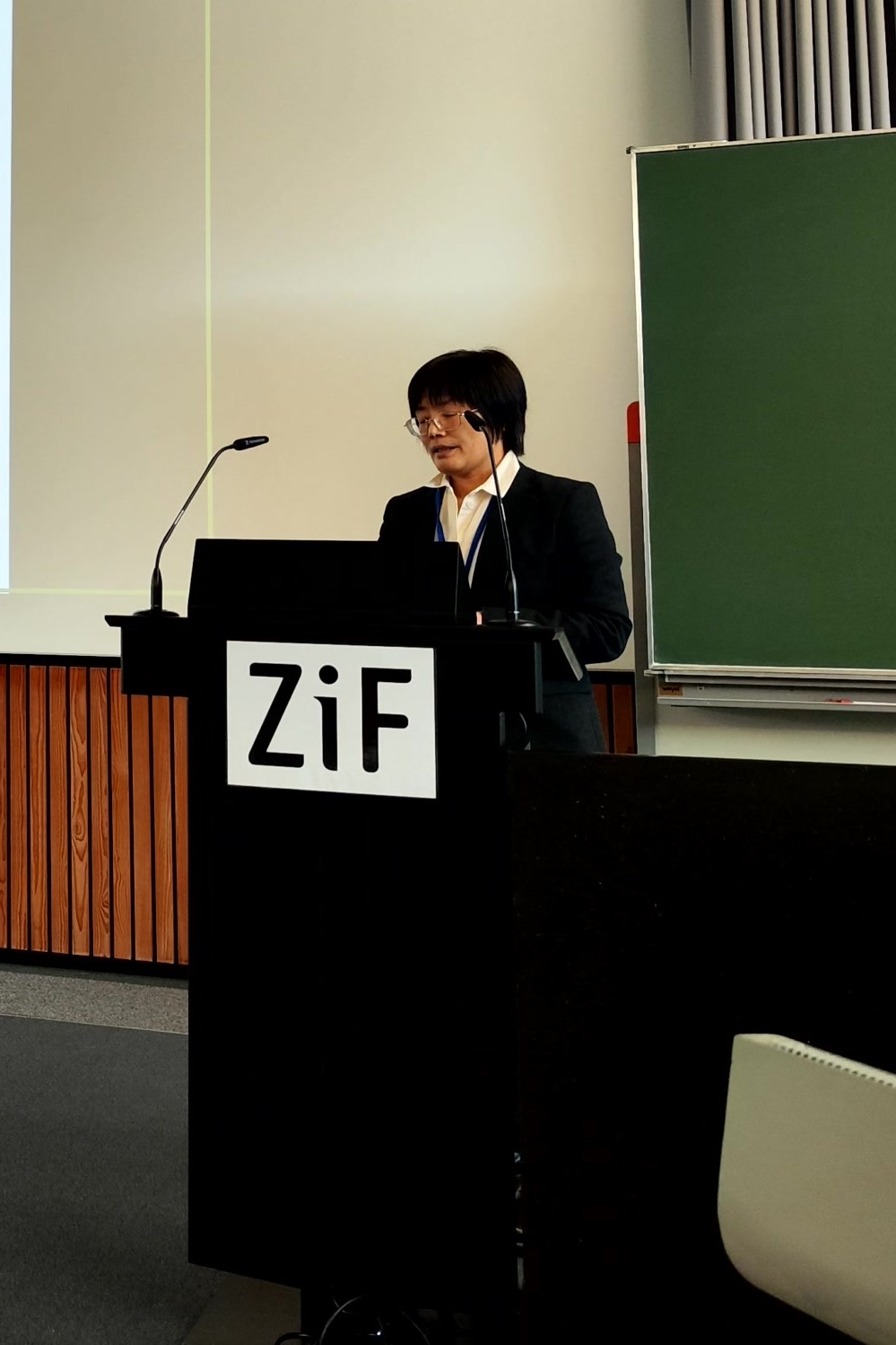}}
\caption{Yanling Chen}
\end{floatingfigure}

In her talk, Yanling Chen addressed the issue of transmitting secret messages over a wiretap channel with correlated sources in the presence of an eavesdropper who lacks access to the source observation. 
The talk was related to joint work with Ning Cai.
The proposed coding scheme meticulously combines three key techniques: 1) Wyner-Ziv source coding, which generates a secret key from correlated sources considering a specific channel cost; 2) a one-time pad, which secures messages without incurring additional cost; and 3) Wyner's secrecy coding, which ensures secrecy by leveraging the legitimate receiver's channel advantage over the eavesdropper's. Yanling gave insights into optimal strategies for designing practical codes for secure communication and storage systems. The results are detailed in \cite{ChenTalk}.

\begin{figure}
    \centering
    \includegraphics[width=8cm]{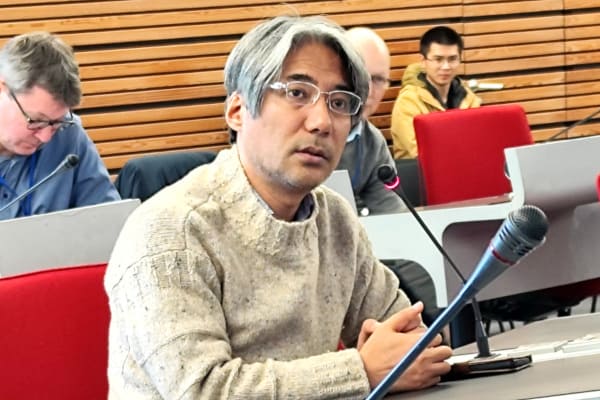}
    \caption{Masahito Hayashi posing a question.}
    \label{fig:enter-label}
\end{figure}

\newpage

\subsection{P\'eter L. Erd\"os - On the Small Quasi-kernel conjecture}
\begin{floatingfigure}[r]{6cm}
\mbox{\includegraphics[width=5.5cm]{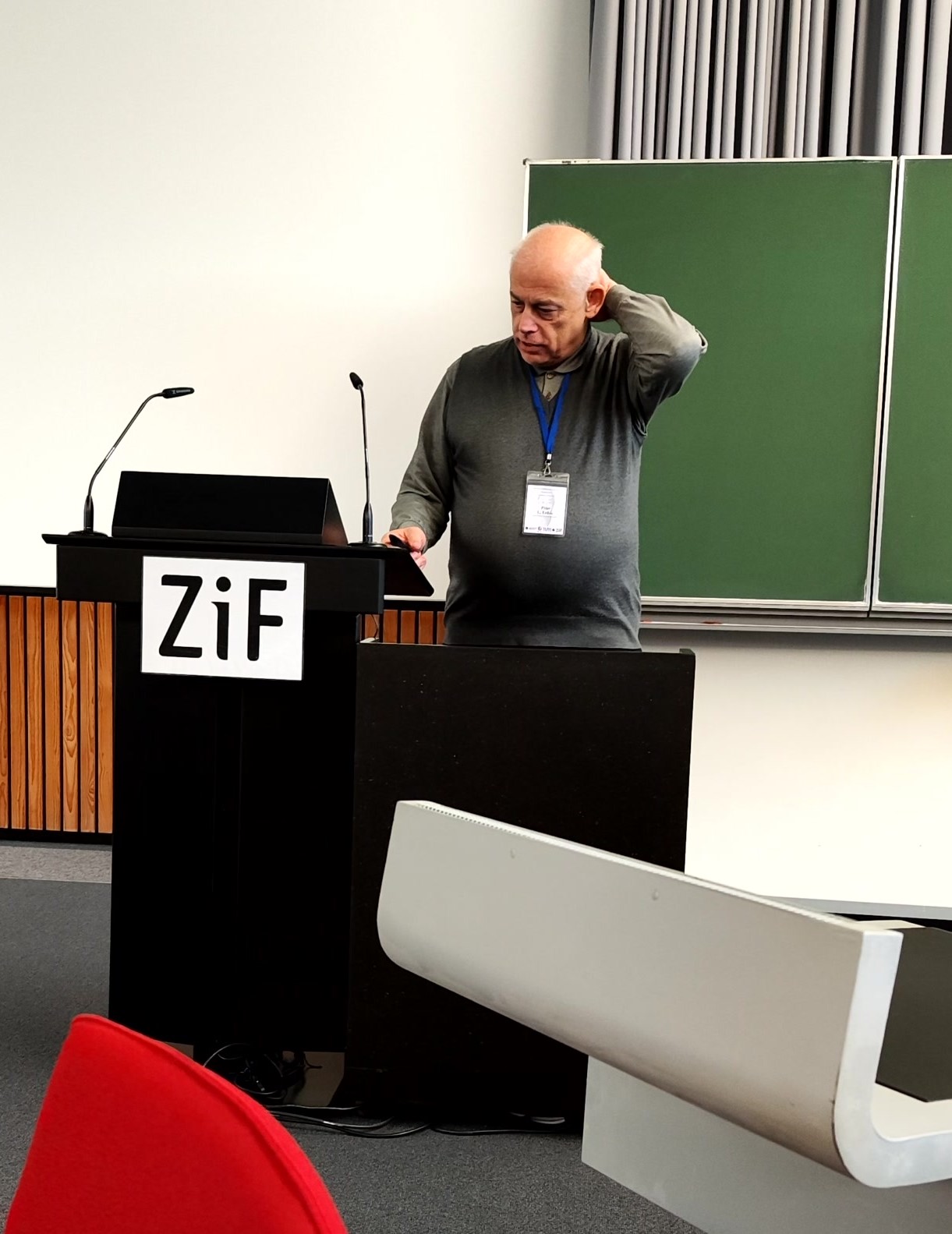}}
\caption{P\'eter L. Erd\"os}
\end{floatingfigure}

P\'eter L. Erd\"os, together with Ning Cai, was a visitor of the ZiF, doing research. He gave a survey talk on the Small Quasi-kernel conjecture.
An independent vertex subset $S$ of the directed graph $G$ is a \emph{kernel} if the set of out-neighbors of $S$ is $V(G)\setminus S$. An independent vertex subset $Q$ of $G$ is a \emph{quasi-kernel} if the union of the first and second out-neighbors contains $V(G)\setminus S$ as a subset. Deciding whether a directed graph has a kernel is an NP-hard problem. In stark contrast, each directed graph has quasi-kernel(s) and one can be found in linear time. He surveyed the results on quasi-kernels and their connection with kernels. The focus is on the \emph{small quasi-kernel} conjecture which states that if the graph has no vertex of zero in-degree, then there exists a quasi-kernel of size not larger than half of the order of the graph. At the end of his talk P\'eter L. Erd\"os also presented some new results. These results can be found in \cite{PeterMemorialNing}.

\bigskip

\begin{figure}
    \centering
    \includegraphics[width=8cm]{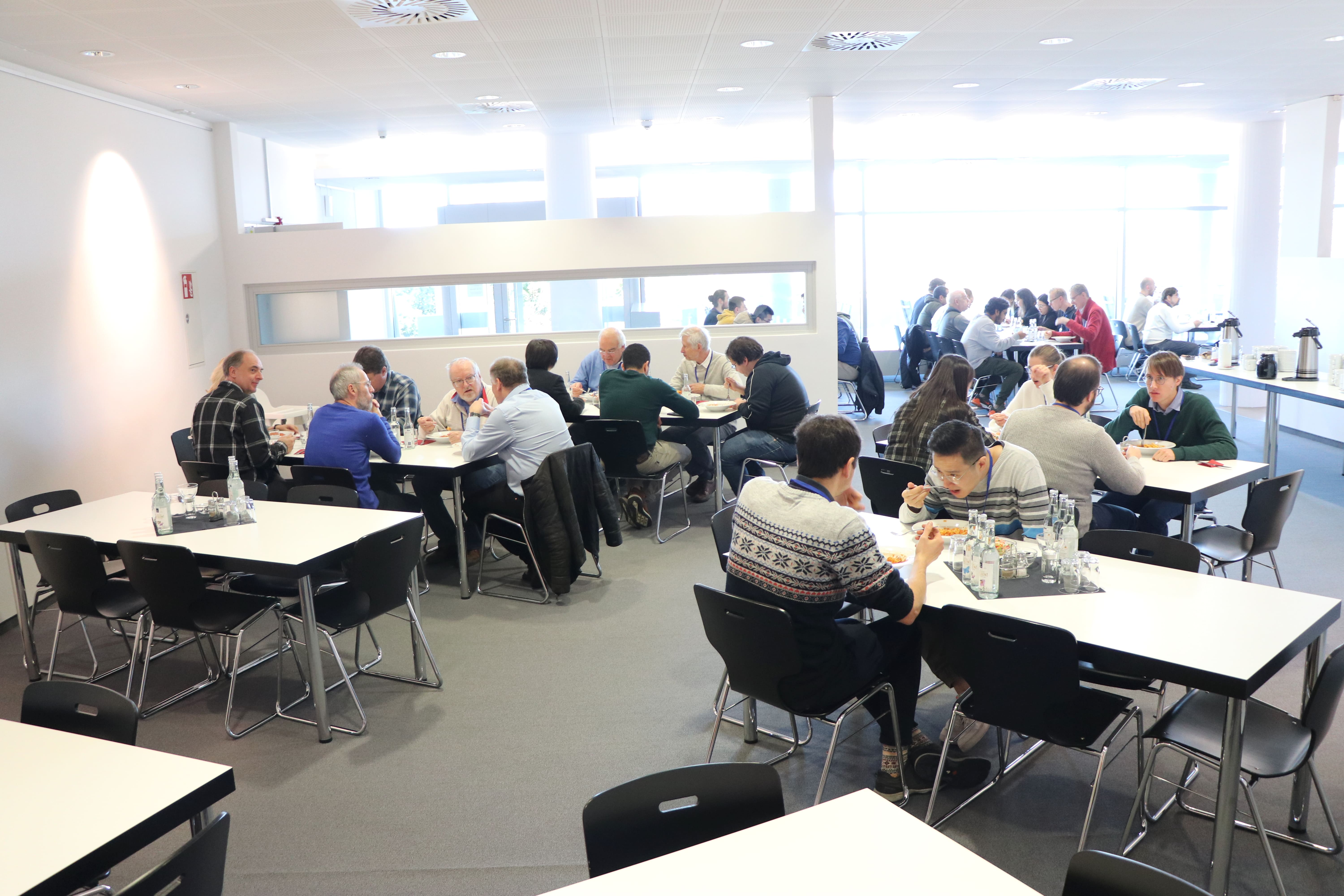}
    \caption{Lunch break in the ZiF cafeteria.}
    \label{fig:enter-label}
\end{figure}

\newpage
\subsection{Boulat Bash - Covert Communication of Classical-Quantum Channels}
\begin{floatingfigure}[r]{6cm}
\mbox{\includegraphics[width=5.5cm]{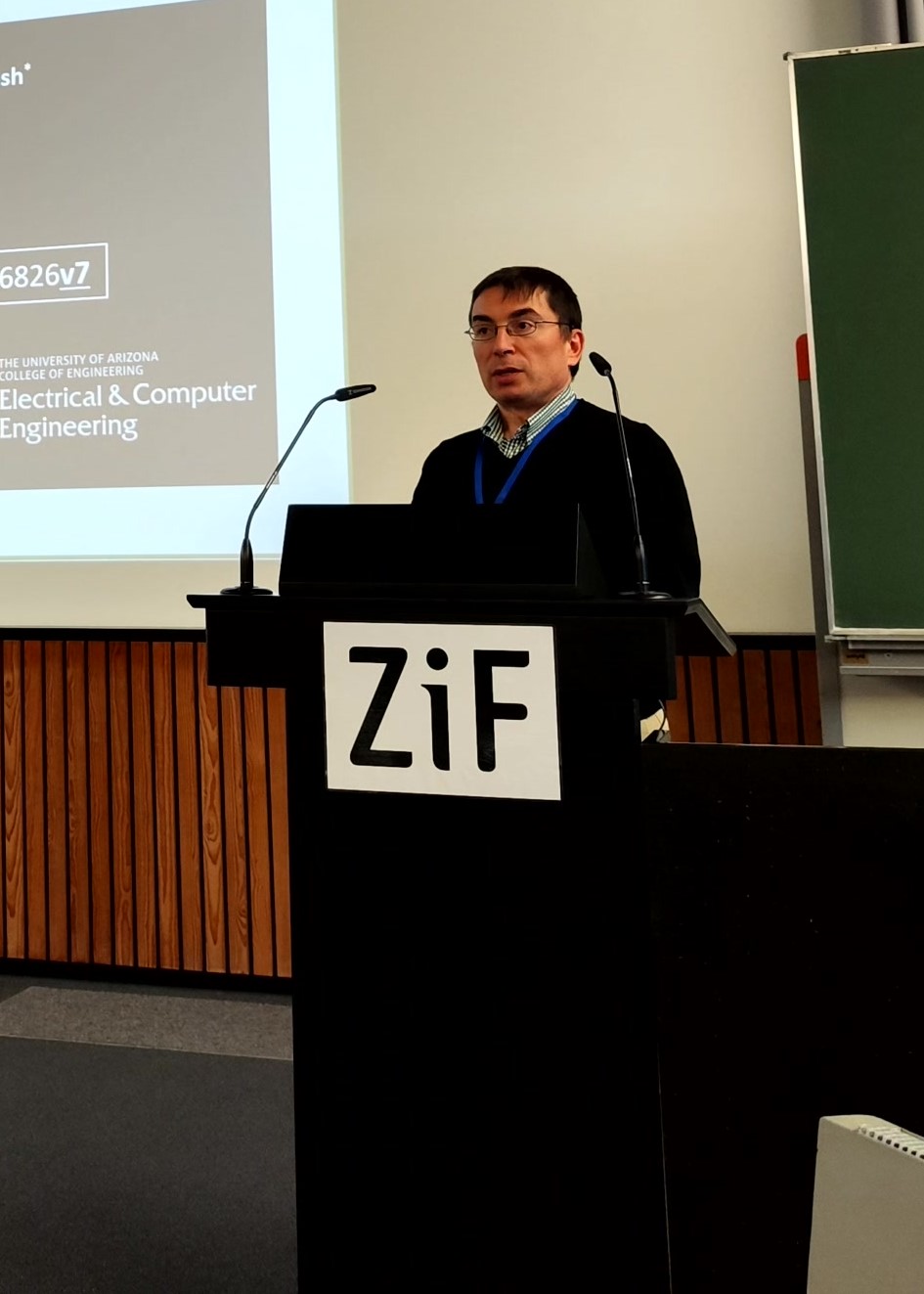}}
\caption{Boulat Bash}
\end{floatingfigure}

Boulat Bash continued with a talk about communication security.
He discussed covert communication over general memoryless classical-quantum channels with fixed finite-size input alphabets. 
He showed that the square root law (SRL) governs covert communication in this setting when the product of $n$ input states is used: $L_{\rm SRL}\sqrt{n}+o(\sqrt{n})$ covert bits (but no more) can be reliably transmitted in $n$ uses of a classical-quantum channel, where $L_{\rm SRL}>0$ is a channel-dependent constant that is called a \emph{covert capacity}.
Furthermore, he showed that ensuring covertness requires $J_{\rm SRL}\sqrt{n}+o(\sqrt{n})$ bits, a secret shared by the communicating parties prior to transmission, where $J_{\rm SRL}\geq 0$ is a channel-dependent constant.
It is assumed that a quantum-powerful adversary can perform an arbitrary joint (entangling) measurement on all $n$ channel uses.
Finally, Boulat determined the single-letter expressions for $L_{\rm SRL}$ and $J_{\rm SRL}$, and established conditions when $J_{\rm SRL}=0$ (i.e., no pre-shared secret is needed). We refer the reader to \cite{BoulatTalk} for more details.

\begin{figure}
    \centering
    \includegraphics[width=8cm]{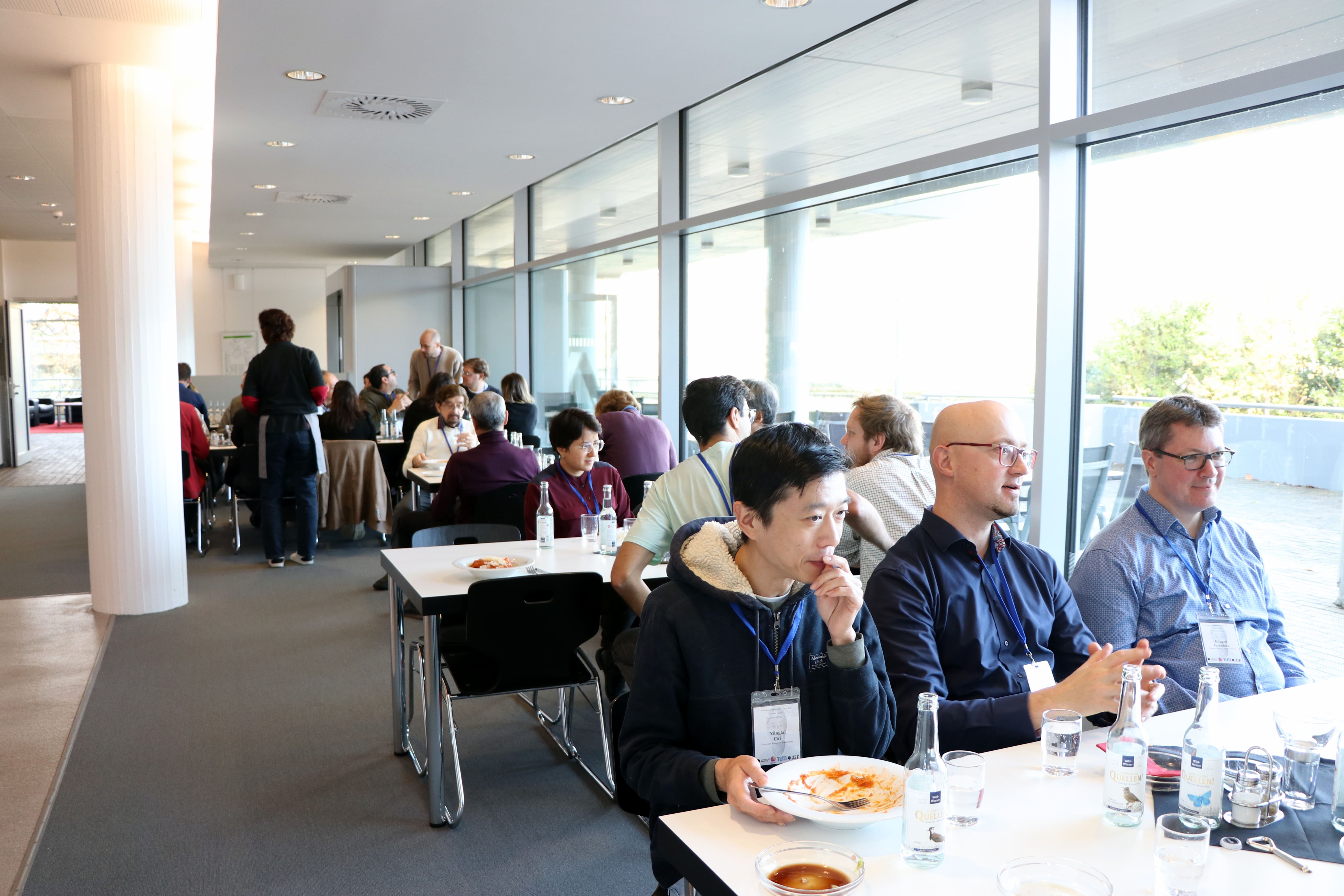}
    \caption{Lunch break in the ZiF cafeteria.}
    \label{fig:enter-label}
\end{figure}

\newpage

\subsection{Ilya Vorobyev - Correcting one error in channels with feedback}
\begin{floatingfigure}[r]{6cm}
\mbox{\includegraphics[width=5.5cm]{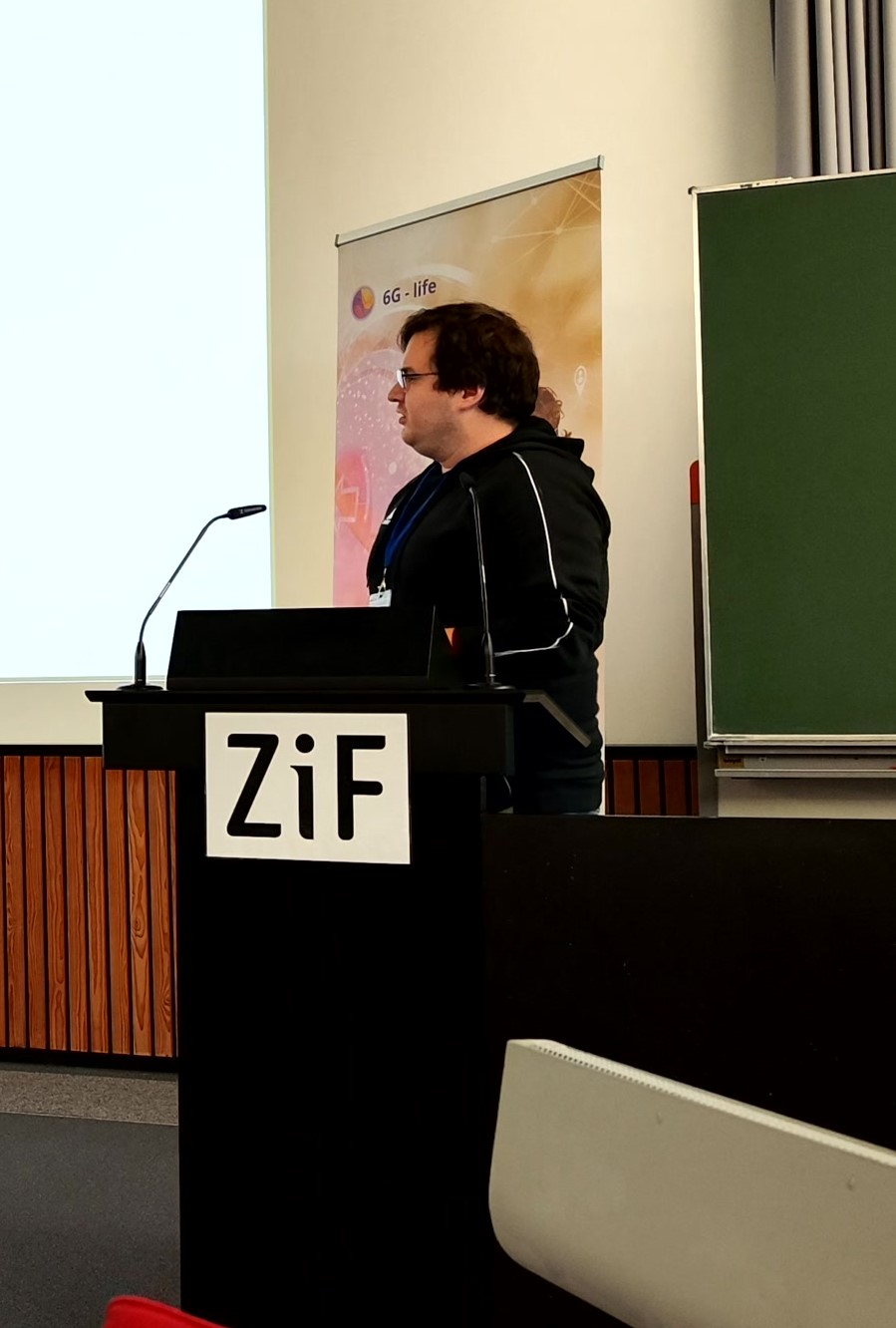}}
\caption{Ilya Vorobyev}
\end{floatingfigure}
In his talk, Ilya Vorobyev addressed the challenge of correcting a single error in an arbitrary discrete memoryless channel with error-free instantaneous feedback. The focus was primarily on binary symmetric and asymmetric channels.
In a binary symmetric channel, any symbol can be transmitted erroneously; for instance, a 0 may be received instead of a 1, or vice versa. The term "symmetric" is often omitted, and such a channel is typically referred to simply as a binary channel. In contrast, a binary asymmetric channel is one where a 0 can be received instead of a transmitted 1, but a 0 is always transmitted correctly. Vorobyev considered a combinatorial model of such a channel with feedback and a single transmission error.
It is known that the problem of correcting \( t \) errors in a binary channel with complete feedback is equivalent to a combinatorial search problem. In this problem, the goal is to find an element \( x \in M \) using \( n \) questions of the form: “Does an element \( x \) belong to a subset \( A \) of a set \( M \)?” Questions are asked sequentially, meaning each subsequent question may depend on the answers to previous ones. The opponent, who knows the element \( x \), is permitted to lie at most \( t \) times.
Vorobyev proposed a method for constructing optimal transmission strategies in scenarios involving one-time feedback. His results demonstrate that for a binary channel, two feedbacks are sufficient to transmit the same number of messages as with complete feedback. Additionally, Vorobyev applied these techniques to a binary asymmetric channel to develop transmission strategies for short message lengths. The result is provided in \cite{IlyaTalk}.

\bigskip
\newpage
\subsection{Abdalla Ibrahim - Identification with effective secrecy}
\begin{floatingfigure}[r]{6cm}
\mbox{\includegraphics[width=5.5cm]{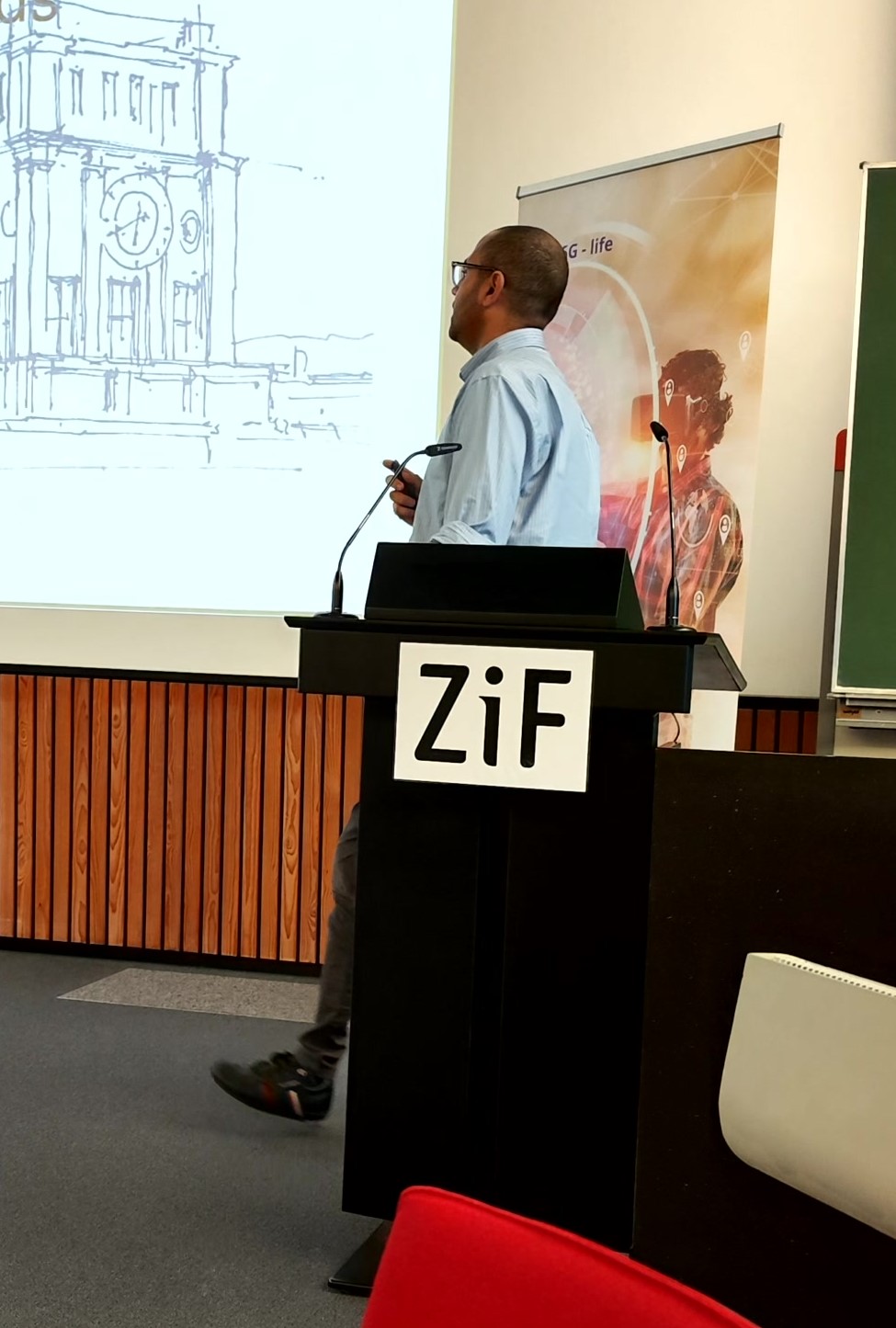}}
\caption{Abdalla Ibrahim}
\end{floatingfigure}
In his talk, Abdalla Ibrahim discussed identification via discrete memoryless wiretap channels under the requirement of semantic effective secrecy, which combines semantic secrecy and stealth constraints. He improved a previously established lower bound by applying it to a prefix channel, created by concatenating an auxiliary channel with the actual channel. These bounds are tight when the legitimate channel is more capable than the eavesdropper's channel. To illustrate this, he provided an example involving a wiretap channel composed of a point-to-point channel and a parallel, reversely degraded wiretap channel. He also compared these results with those for message transmission and for identification with only a secrecy constraint \cite{Abdallatalk}. This talk concluded the first day of the conference.

\newpage

\subsection{Gerhard Kramer - Network Coding and Edge Cut Bounds}
\begin{floatingfigure}[r]{6cm}
\mbox{\includegraphics[width=5.5cm]{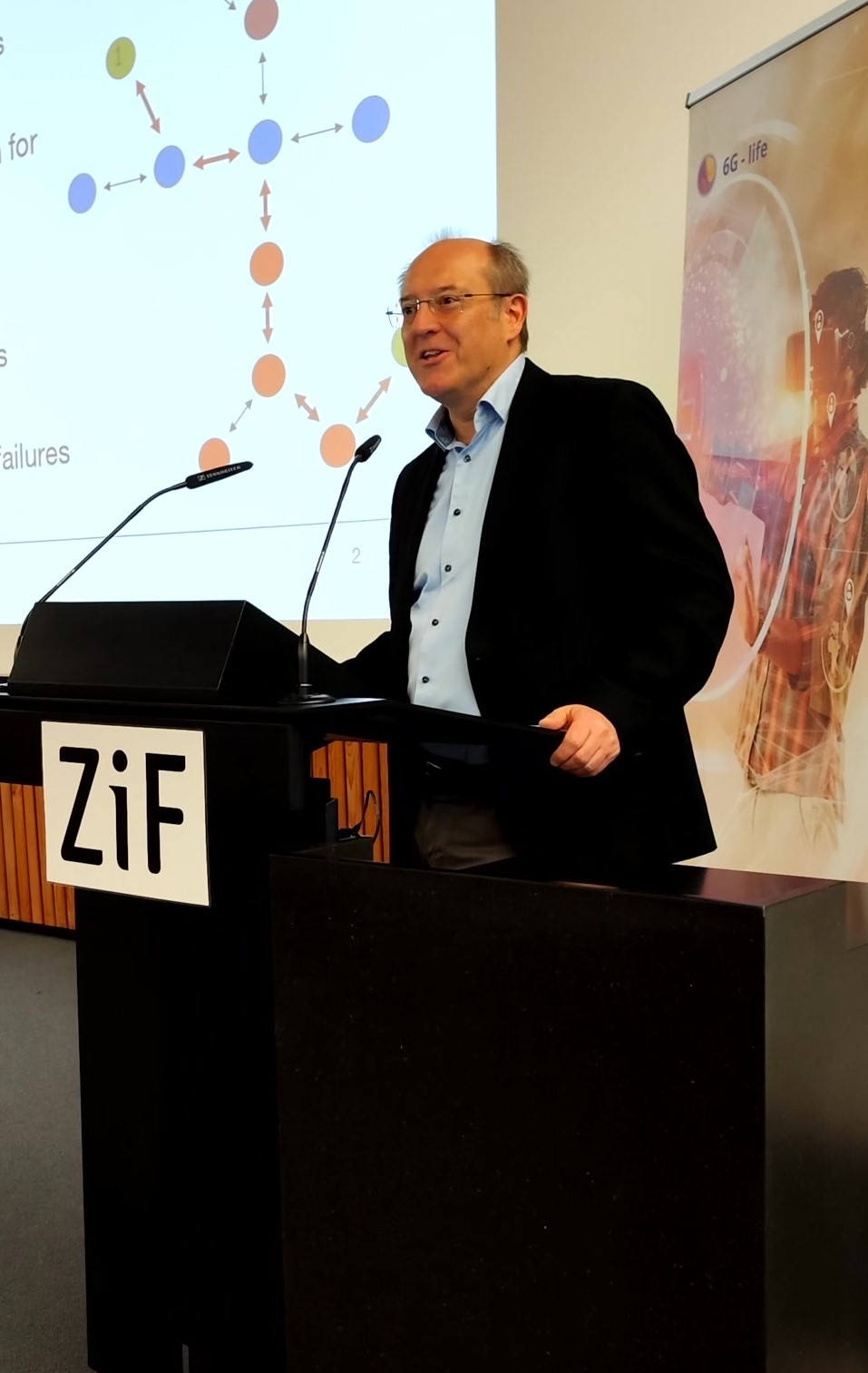}}
\caption{Gerhard Kramer}
\end{floatingfigure}
Gerhard Kramer gave his talk at the beginning of the second day of the workshop. He discussed network coding and edge cut bounds \cite{Kramertalk}. Active networks, characterized by architectures with processors capable of executing code carried by passing packets, present a critical network management concern focused on optimization. Tight performance bounds serve as useful design benchmarks for these networks. Kramer introduced a new bound on communication rates relevant to network coding, an active network application where processors transmit packets that are general functions (such as bit-wise XOR) of selected received packets. This bound generalizes an edge-cut bound on routing rates by progressively removing edges from the network graph and verifying whether certain strengthened d-separation conditions are met. It improves upon the cut-set bound, demonstrating its effectiveness by showing that routing achieves rate-optimality for several well-known examples in the networking literature.

\newpage

\subsection{Frank Fitzek - Pitfalls and Joys of Network Coding Implementation}
\begin{floatingfigure}[r]{6cm}
\mbox{\includegraphics[width=5.5cm]{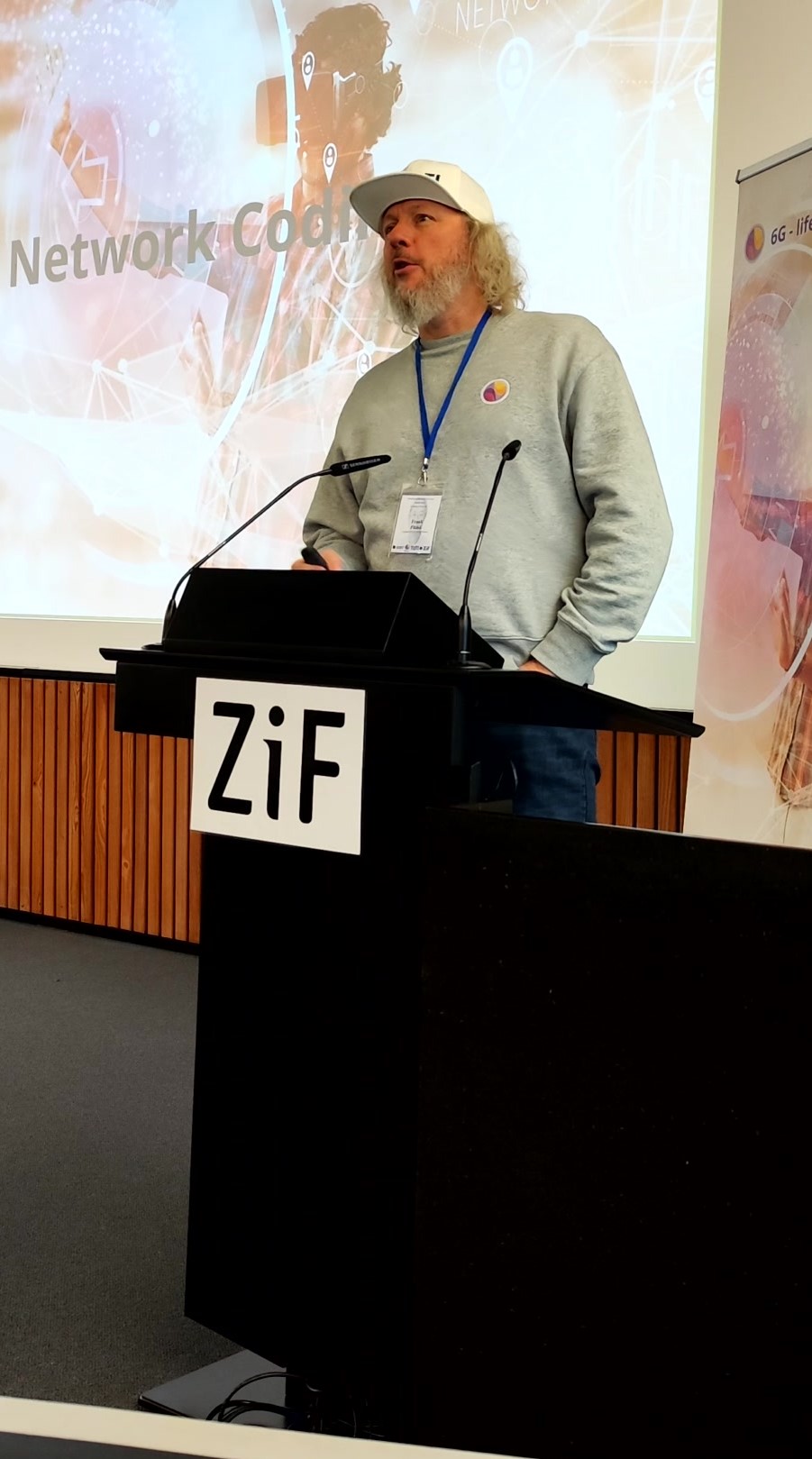}}
\caption{Frank Fitzek}
\end{floatingfigure}
The second talk, delivered by Frank Fitzek, delved deeply into Ning's most influential paper on network coding, which established the theoretical underpinnings of the field. Fitzek discussed his extensive practical attempts to implement network coding, a concept that has had a profound impact across numerous research areas. Despite its conceptual divergence from traditional forms like source and channel coding, the practical realization of network coding continues to pose significant challenges.

In his presentation, Fitzek detailed the ongoing research activities in network coding, focusing particularly on its applications within commercial platforms. He shared his personal experiences, shedding light on both the difficulties and the successes encountered during the implementation process. Fitzek emphasized that network coding demands a paradigm shift in how we perceive communication networks. Unlike conventional networks, which operate according to Kirchhoff's laws and maintain a straightforward relationship between inputs and outputs, network coding requires a more nuanced understanding.
Fitzek argued that to fully comprehend and leverage network coding, engineers need to rethink their traditional views on networks. He used a simple meshed network example to illustrate this point, demonstrating how network coding can significantly reduce the number of time slots needed for data exchange. This example not only highlighted the efficiency gains offered by network coding but also provided deeper insights into its potential applications and benefits.
In summary, Fitzek's talk underscored the transformative potential of network coding in the realm of communication networks. By sharing his practical insights and experiences, he aimed to inspire a shift in perspective that could lead to more innovative and effective implementations of network coding in the future \cite{Franktalk}.
\bigskip
\newpage

\subsection{Han Vinck - On the Binary Symmetric Channel with a Transition Probability Determined by a Poisson Distribution}
\begin{floatingfigure}[r]{6cm}
\mbox{\includegraphics[width=5.5cm]{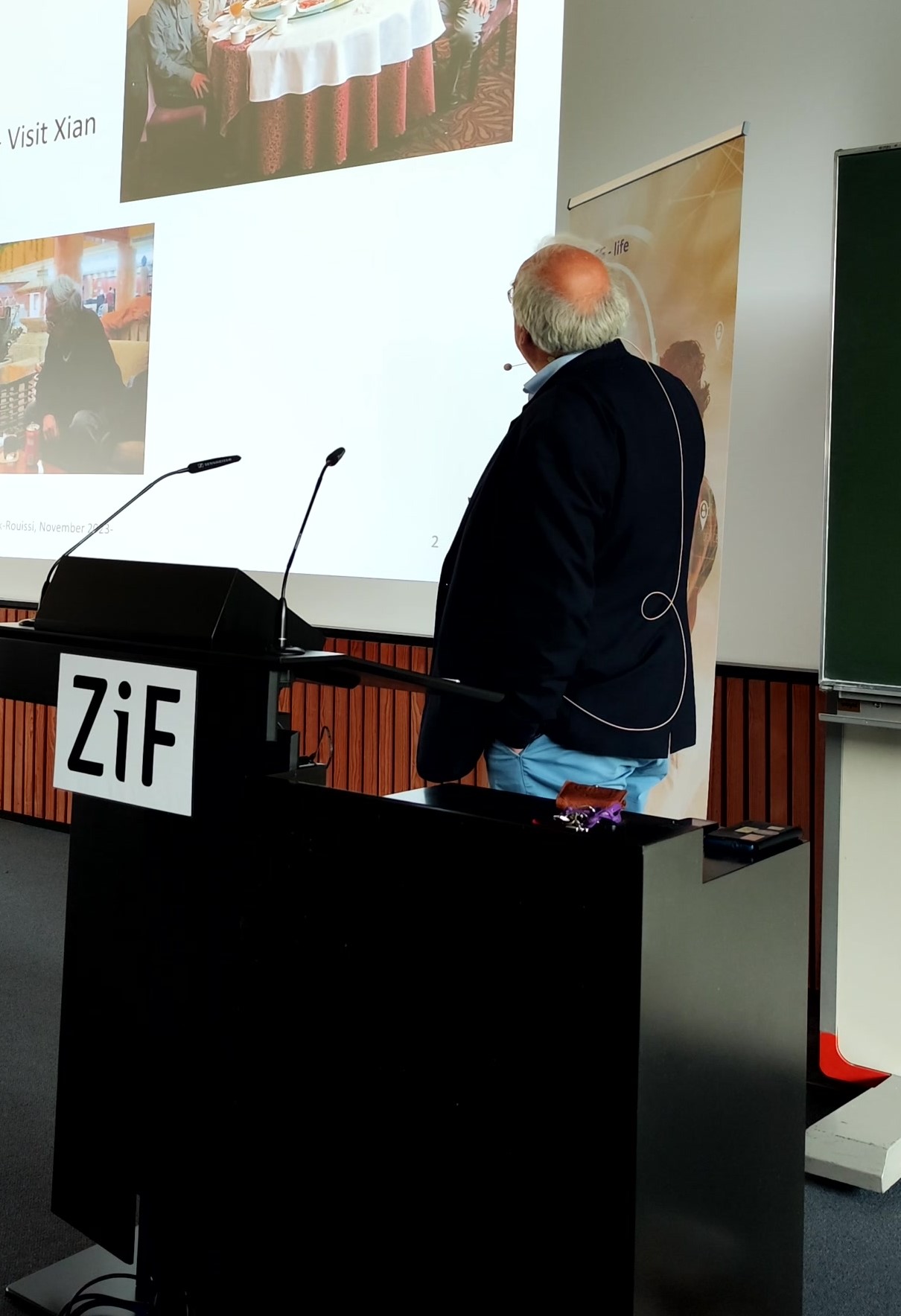}}
\caption{Han Vinck}
\end{floatingfigure}

The third talk on the second day was delivered by Han Vinck. Ning Cai had frequently visited Han and his group, and as previously mentioned by Ulrich Tamm, they often collaborated at conferences. Han Vinck's talk focused on the binary symmetric channel where the transition probability is determined by a Poisson distribution.
The classical Binary Symmetric Channel (BSC) features a fixed transition probability. In this discussion, we explore a BSC with a variable transition probability governed by a Poisson distribution. We determine the error rate for this channel and provide bounds for its capacity. The motivation for this model stems from Middleton's Class-A impulse noise model. Additionally, this channel model can be extended to the Additive White Gaussian Noise (AWGN) channel, where the noise variance is also influenced by a Poisson distribution \cite{Vincktalk}.


\begin{figure}
    \centering
    \includegraphics[width=8cm]{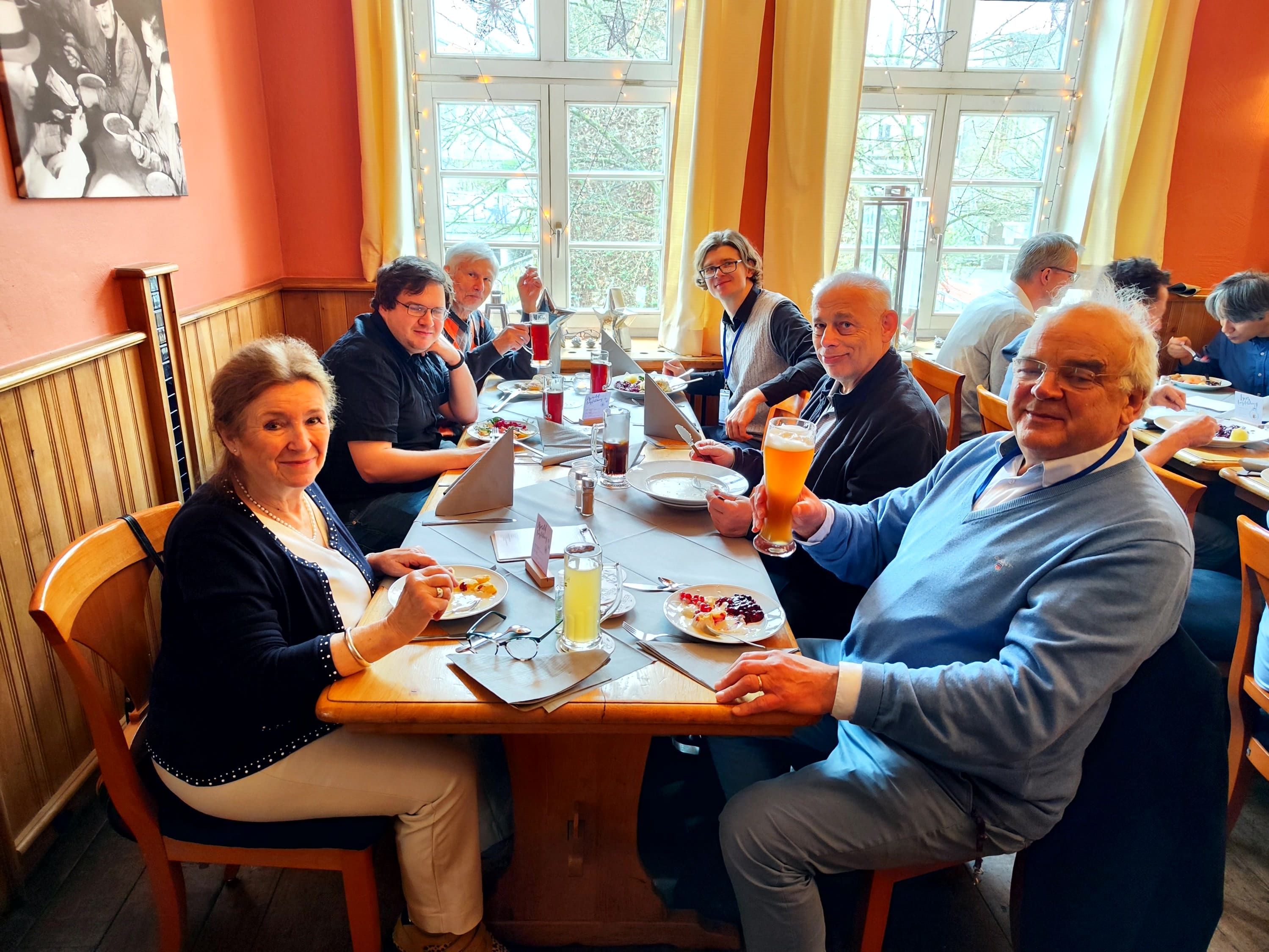}
    \caption{Han Vinck (right) at the memorial lunch.}
    \label{fig:enter-label}
\end{figure}

\newpage

\subsection{Raymond Yeung - Network Information Flow}
\begin{floatingfigure}[r]{6cm}
\mbox{\includegraphics[width=5.5cm]{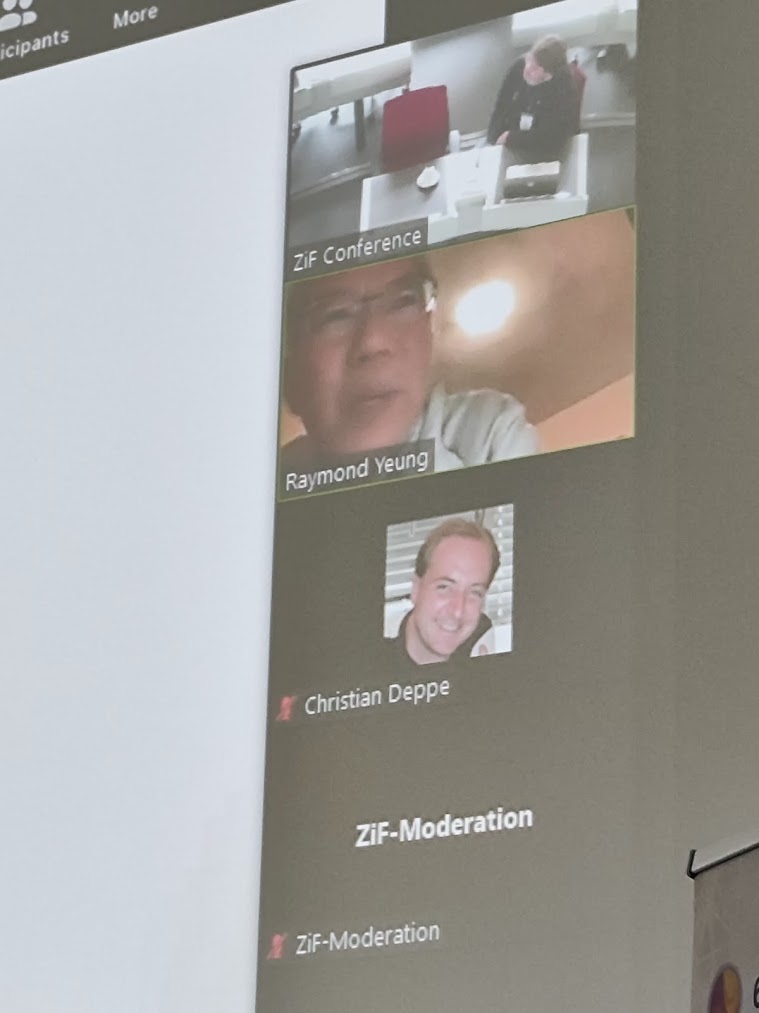}}
\caption{Raymond Yeung}
\end{floatingfigure}

Ning Cai achieved his most significant scientific success through his collaboration with Robert Li, Raymond Yeung and Rudolf Ahlswede. Their joint paper \cite{DBLP:journals/tit/AhlswedeCLY00}, titled \textit{"Network Information Flow,"} was groundbreaking and has been cited over 10,000 times. Ning Cai was the primary author of this influential work. However, as is customary in mathematics, the authors' names are listed alphabetically, resulting in Ning Cai being listed as the second author. Despite this, it is widely recognized that this paper represents Ning Cai's most impactful contribution to the field of information theory.

Unfortunately, Raymond Yeung was unable to attend the workshop in person and instead delivered the only online talk. In his presentation, Yeung covered the fundamental concepts outlined in their seminal paper, provided historical context regarding its development, and discussed the ongoing advancements stemming from their research findings. This presentation highlighted the profound and lasting influence of Ning Cai's work in the field of information theory.

\begin{figure}
    \centering
    \includegraphics[width=8cm]{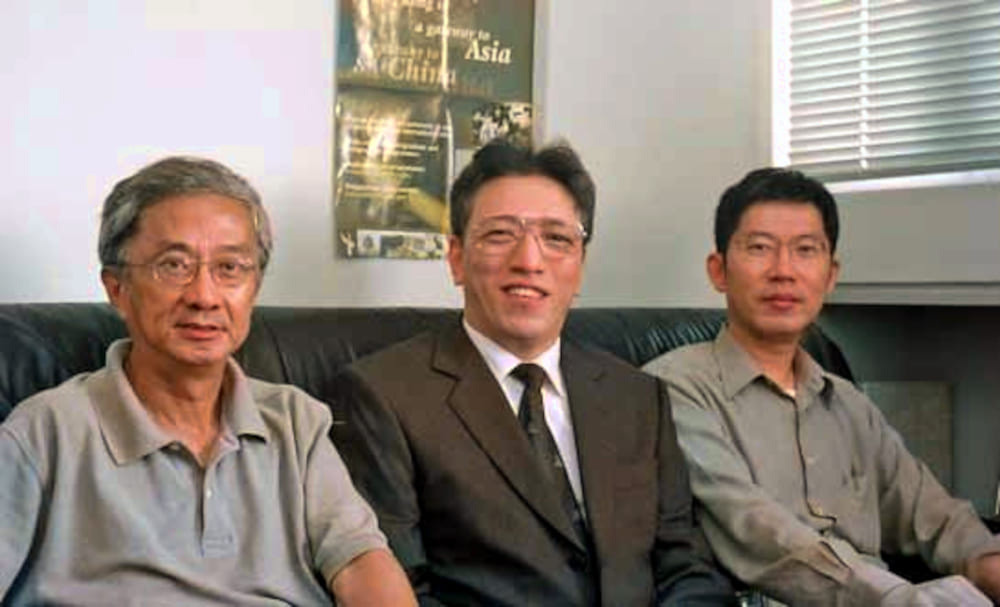}
    \caption{A photo from years gone by with Ning Cai, Robert Li and Raymond Yeung.}
    \label{fig:enter-label}
\end{figure}

\bigskip
\newpage
\subsection{Janis Nötzel - Entanglement-Assisted Classical Communication}
\begin{floatingfigure}[r]{6cm}
\mbox{\includegraphics[width=5.5cm]{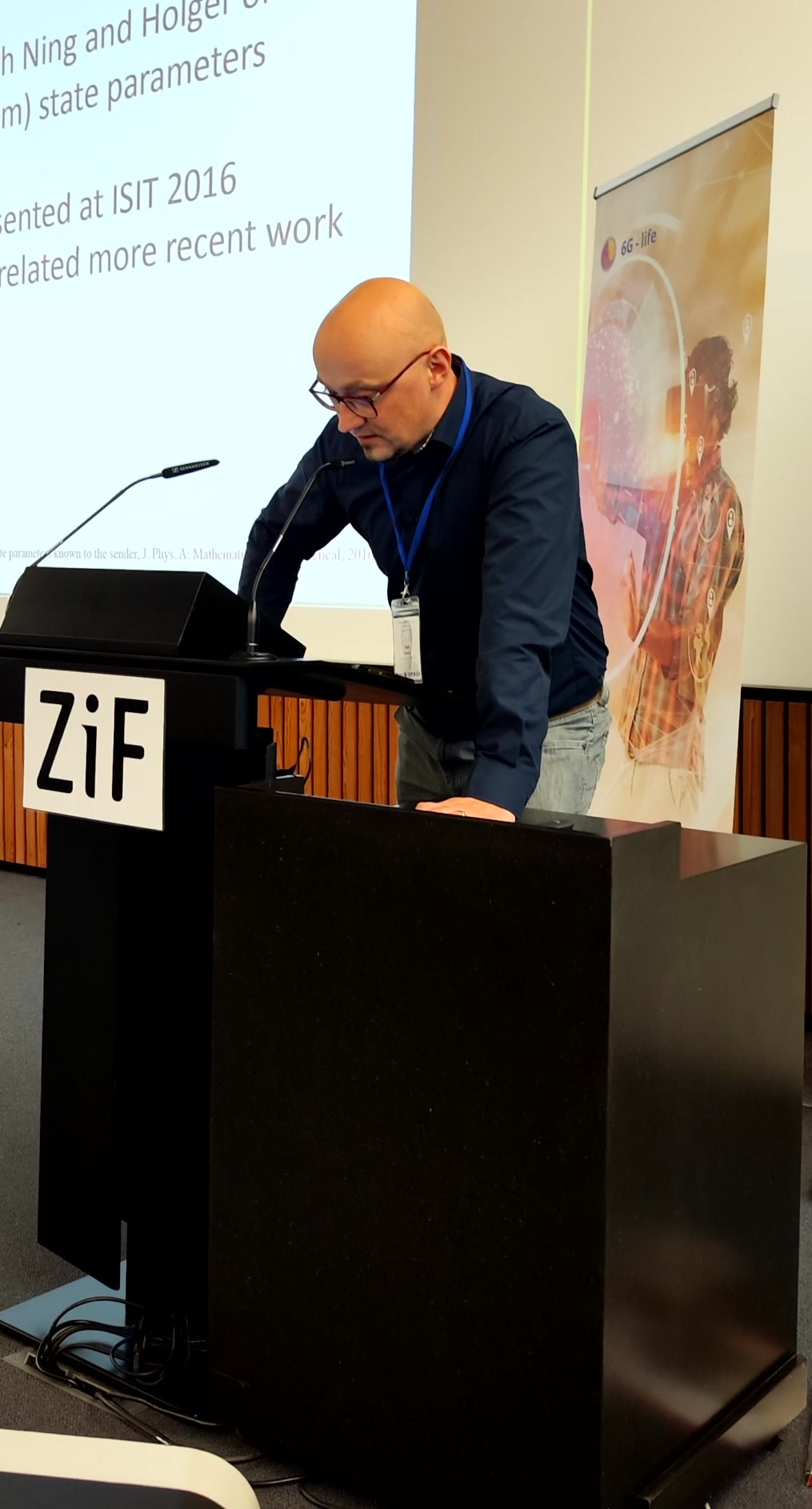}}
\caption{Janis Nötzel}
\end{floatingfigure}
The next talk was given by Janis Nötzel, a collaborator of Ning Cai in quantum communication. He explained how quantum technology can enhance classical communication. As the contributions of quantum technologies to future networks become a focal point of engineering research, efforts are underway to overcome limitations in current network infrastructures. One proposal for using quantum methods in future networks is entanglement-assisted data transmission. However, simulating the functionalities of proposed quantum links under realistic traffic patterns has become a research bottleneck. In his talk, Janis proposed a simulation tool called the Discrete Traffic Simulator (GO-DTS) for evaluating the performance of communication links using the Generate Entanglement while Idle (GEWI) method. This tool aims to accelerate data transmission by generating and later utilizing entanglement. GO-DTS is designed to analyze generic real-world traffic data. Janis demonstrated its performance under self-similar traffic and heavy-tailed burst traffic, highlighting the parameter regions where the GEWI link outperforms a hypothetical purely classical communication link that does not utilize shared entanglement \cite{Janistalk}.



\newpage

\subsection{Eduard Jorswieck - Arbitrarily Varying Wiretap Channels with Jammer}
\begin{floatingfigure}[r]{6cm}
\mbox{\includegraphics[width=5.5cm]{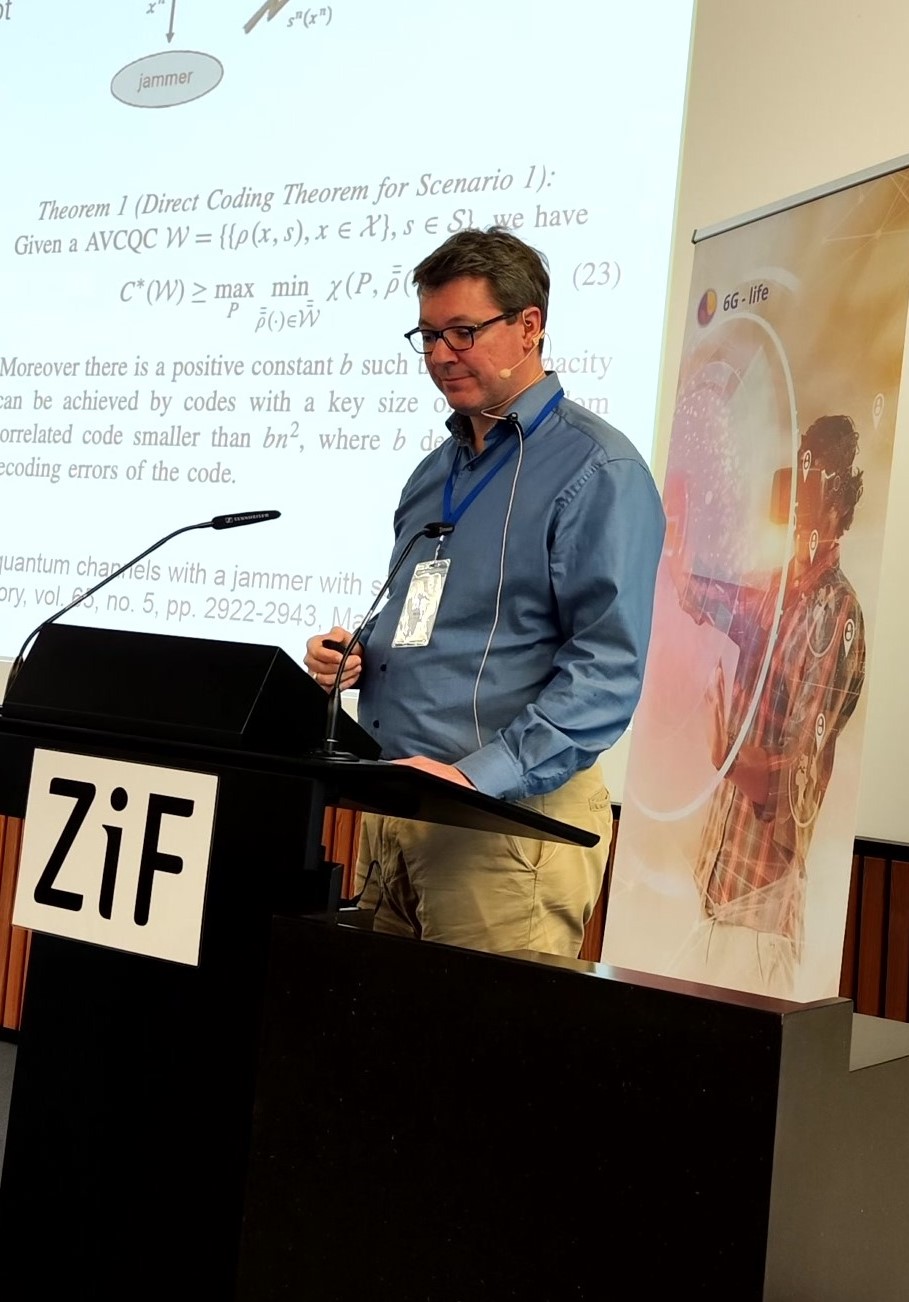}}
\caption{Eduard Jorswieck}
\end{floatingfigure}

Eduard Jorswieck's talk focused on a topic Ning Cai frequently explored: the arbitrarily varying wiretap channel, likely introduced to him by Ahlswede. Ning Cai also examined this topic in the context of quantum communication. The arbitrarily varying wiretap channel (AVWC) provides information-theoretical bounds on the amount of information that can be exchanged despite the presence of an active attacker. If the attacker has non-causal side information, it can model situations where a legitimate communication system has been compromised. Jorswieck investigated the AVWC with non-causal side information at the jammer, assuming the existence of an optimal channel to the eavesdropper. Non-causal side information implies that the transmitted codeword is known to an active adversary before transmission. By considering the maximum error criterion, the analysis also includes scenarios where messages are known to the jammer before the corresponding codeword is transmitted.
Jorswieck derived a single-letter formula for the common randomness secrecy capacity. He also provided this formula for cases where the channel to the eavesdropper is strongly degraded, strongly noisier, or significantly less capable compared to the main channel. Furthermore, he compared these results to the random code secrecy capacity under different conditions: with the maximum error criterion but without non-causal side information at the jammer, with the maximum error criterion and non-causal side information of the messages at the jammer, and with the average error criterion without non-causal side information at the jammer \cite{Jorswiecktalk}. 
\bigskip
\newpage
\subsection{Christian Meise - Music und Mathematics - a Piano Concert for Ning}
\begin{floatingfigure}[r]{6cm}
\mbox{\includegraphics[width=5.5cm]{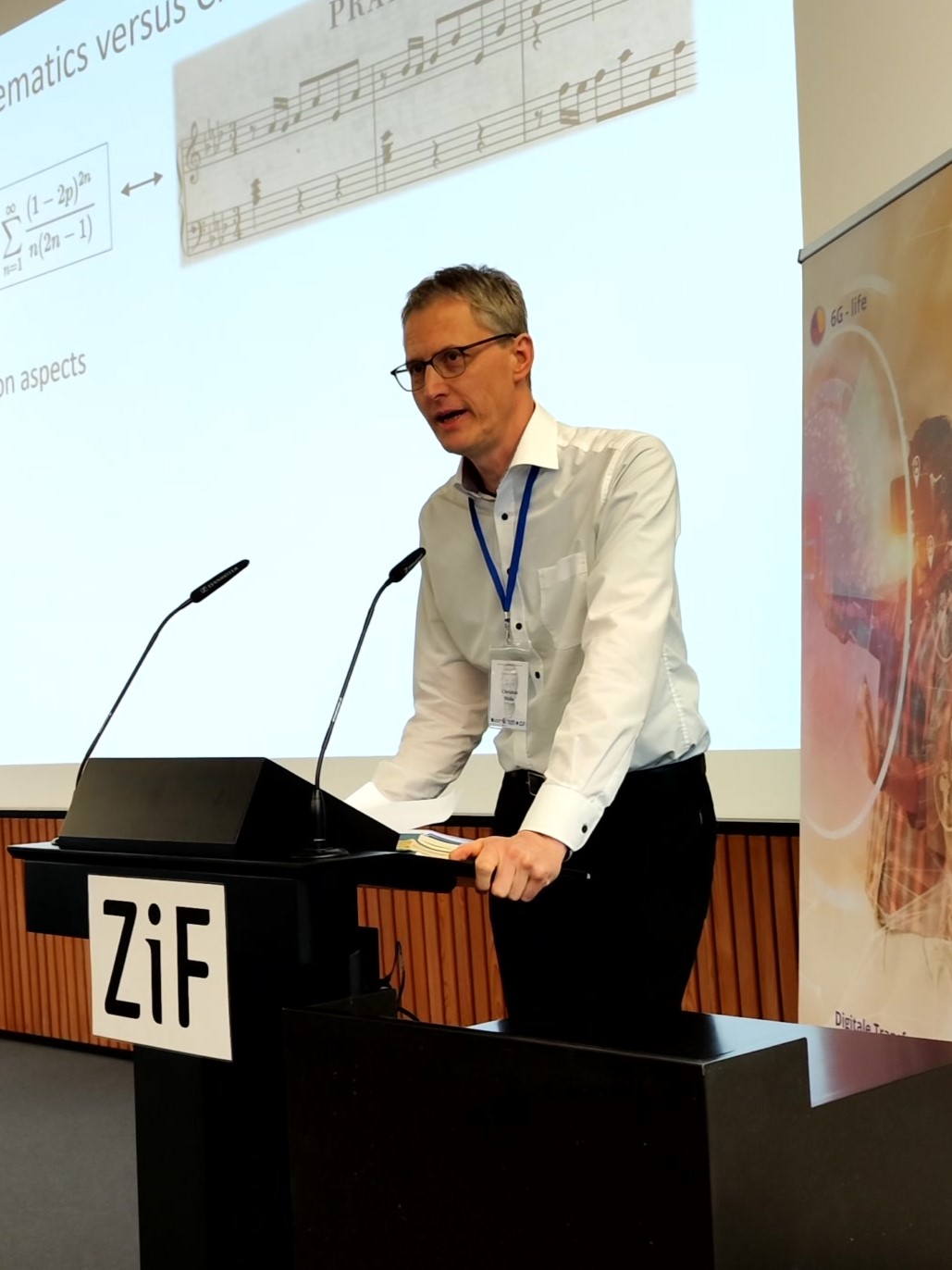}}
\caption{Christian Meise}
\end{floatingfigure}

Ning Cai has always had a deep appreciation for music, with a particular fondness for piano pieces. During the time when Ning Cai was part of the Ahlswede Group in Bielefeld, Christian Meise was a doctoral student under the supervision of Prof. Götze at Bielefeld University. In addition to his mathematical pursuits, Christian has a passion for playing the piano.

We were able to persuade Christian to share his musical talents by giving a concert. Before his performance, Christian delivered a brief talk discussing the intriguing connections between mathematics and music, providing insights into how these two disciplines intersect and influence each other.

Following his talk, Christian sat down at the piano and performed a selection of pieces that showcased his skill and the beauty of classical music. His repertoire included:
\begin{itemize}
    \item J.S. Bach's \textit{Prelude and Fugue in A-flat major} from \textit{The Well-Tempered Clavier I}
    \item Bach's chorale arrangement \textit{"Jesu, bleibet meine Freude"}
    \item Two of Mendelssohn's \textit{Songs Without Words}
    \item Schubert's \textit{Impromptu in A-flat major}
    \item The first movement of Bach's \textit{English Suite in G minor}
\end{itemize}

The audience was captivated by the performance, enjoying the harmonious blend of mathematics and music. The concert not only provided a delightful musical experience but also served as a commemoration of Ning Cai, celebrating his contributions and interests. The participants expressed their appreciation for Christian's performance and the thoughtful integration of mathematical concepts with musical expression.

\newpage

\subsection{Christoph Hirche - Quantum Rényi and $f$-divergences from integral representations}
\begin{floatingfigure}[r]{6cm}
\mbox{\includegraphics[width=5.5cm]{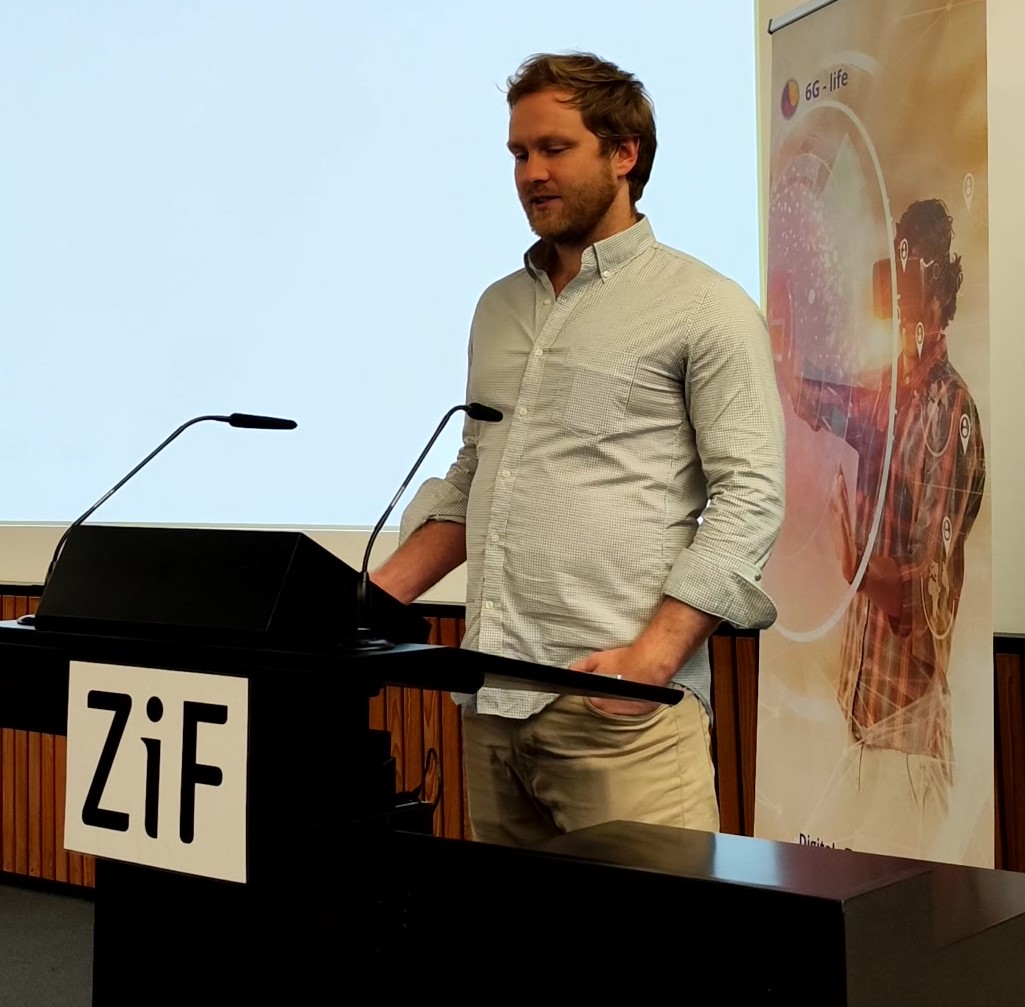}}
\caption{Christoph Hirche}
\end{floatingfigure}
After the piano concert, the event continued with a scientific presentation. Christoph Hirche gave a talk on quantum Rényi and $f$-divergences from integral representations \cite{Hirchetalk}.
Smooth Csiszár $f$-divergences can be expressed as integrals over so-called hockey stick divergences. This provides a natural motivation for a quantum generalization in terms of quantum hockey stick divergences, which we investigate here. Using this approach, the Kullback-Leibler divergence generalizes to the Umegaki relative entropy, in the integral form recently discovered by Frenkel. Our findings reveal that the Rényi divergences, defined via our new quantum $f$-divergences, are not generally additive. However, their regularizations surprisingly yield the Petz Rényi divergence for $\alpha < 1$ and the sandwiched Rényi divergence for $\alpha > 1$, thereby unifying these two significant families of quantum Rényi divergences.
Furthermore, we find that the contraction coefficients for the new quantum $f$-divergences collapse for all $f$ that are operator convex, mirroring classical behavior and resolving some long-standing conjectures by Lesniewski and Ruskai. We derive various inequalities, including new reverse Pinsker inequalities with applications in differential privacy, and explore several other applications of the new divergences.
\bigskip
\newpage
\subsection{Ugo Vaccaro - New Results on Alphabetic Codes}
\begin{floatingfigure}[r]{6cm}
\mbox{\includegraphics[width=5.5cm]{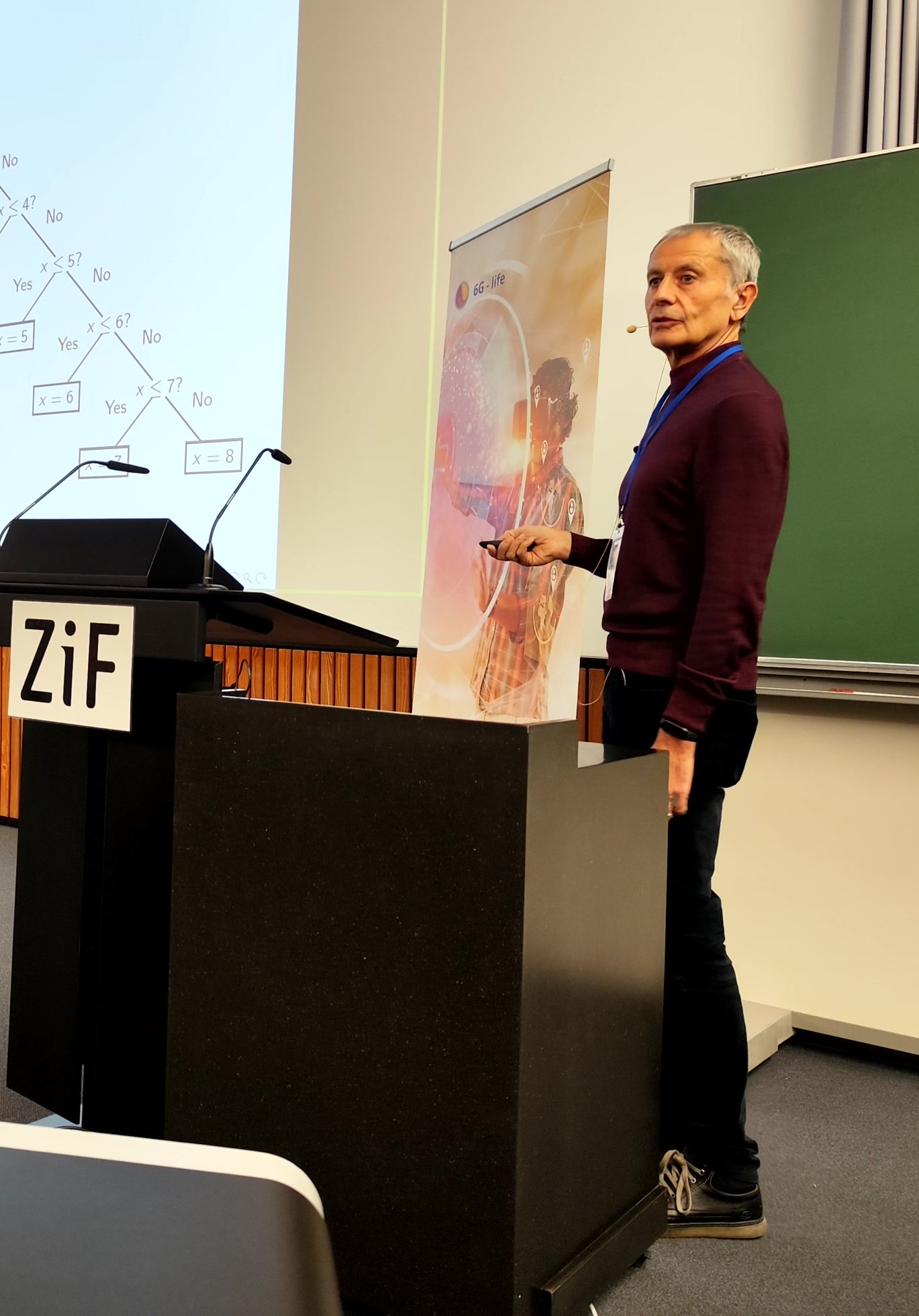}}
\caption{Ugo Vaccaro}
\end{floatingfigure}    
Ugo Vaccaro visited Bielefeld and the ZiF multiple times, particularly during the period when Ning Cai was actively working in Bielefeld. Vaccaro provided an extensive survey of the research field related to alphabetic codes. He meticulously traced the main results from the early inception of alphabetic codes to their current state-of-the-art developments.
Vaccaro's survey included an in-depth analysis of classical alphabetic codes and their numerous variants. He explored the fundamental properties of these codes, as well as the mathematical and algorithmic principles that underpin their design. His analysis shed light on the complexities and innovations that have shaped the evolution of alphabetic codes over the years.
Furthermore, Vaccaro illustrated the wide range of applications for alphabetic codes. 
For those interested in a more detailed account, Vaccaro's comprehensive survey on alphabetic codes is documented in \cite{UgoMemorialNing}.

\bigskip

\begin{figure}[h]
\begin{center}
    \includegraphics[width=8cm]{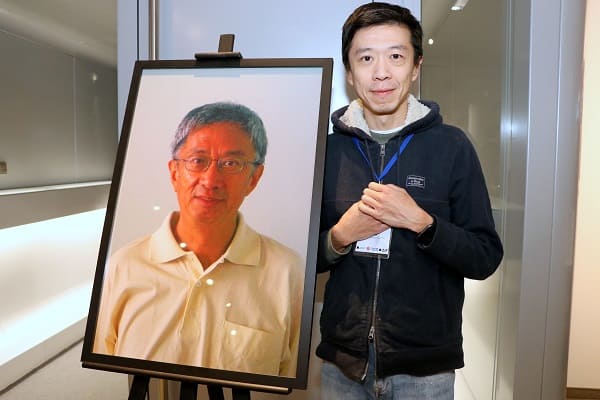}
\end{center}
\caption{Minglai, the son of Ning Cai, next to the picture of his father}
\end{figure}

\newpage

\subsection{Ingo Althöfer - Rudolf Ahlswede in China in 1986 and Combinatorial Space Trajectories}
\begin{floatingfigure}[r]{6cm}
\mbox{\includegraphics[width=5.5cm]{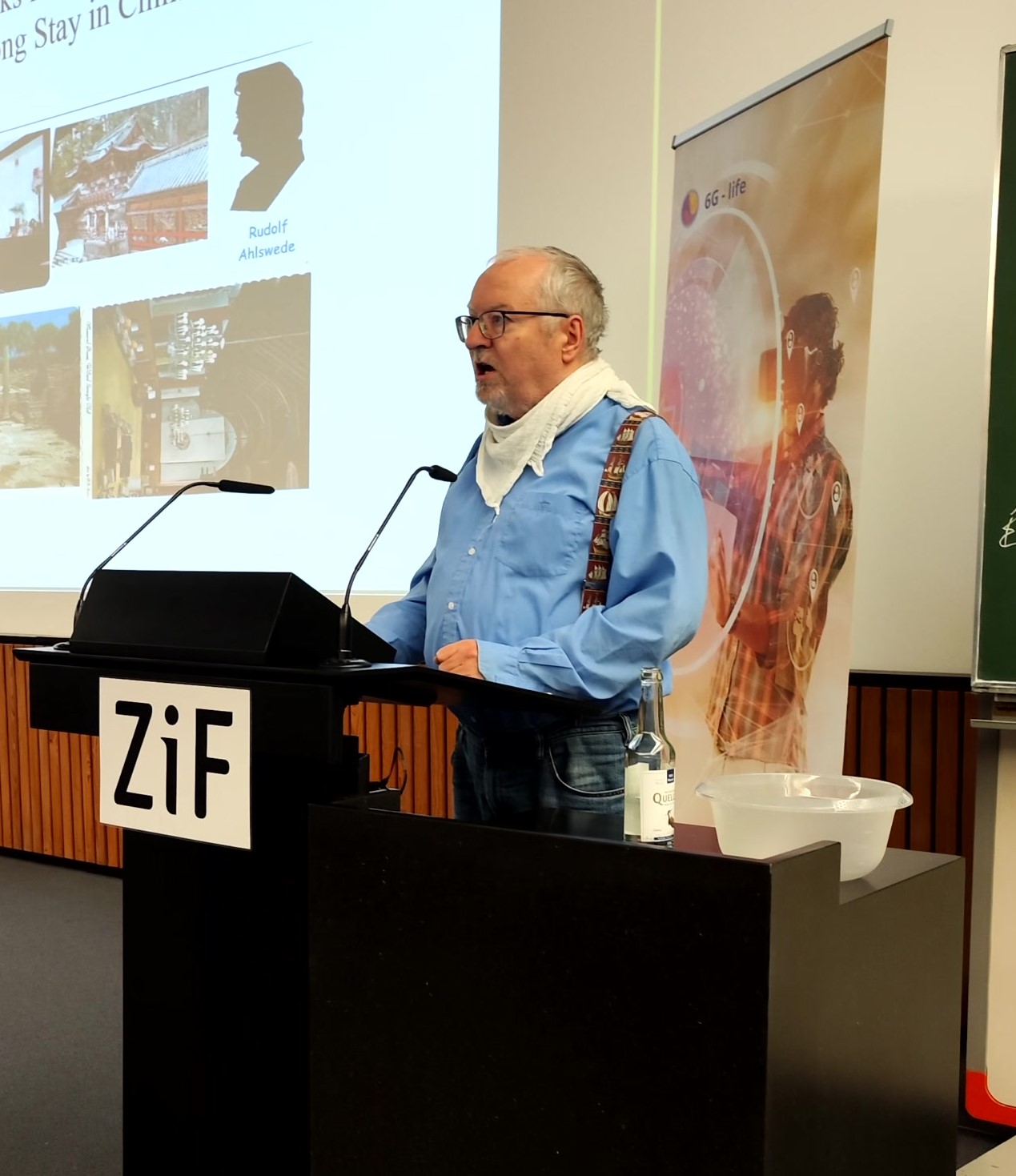}}
\caption{Ingo Althöfer}
\end{floatingfigure}    

At the end of the workshop, Ning Cai's good friend Ingo Althöfer gave a lecture. Ingo Althöfer was one of the colleagues who regularly met with Ning Cai during Christmas in Bielefeld. He began his talk by recounting how Ning Cai and Rudolf Ahlswede met. Rudolf Ahlswede had traveled to China specifically to recruit young talent for his research group. More details about this can be found in the memorial text \cite{MemorialNing}.
Ingo then discussed his research on combinatorial space trajectories. While the design of optimal space trajectories is fundamentally a continuous optimization problem, it also involves aspects of combinatorial optimization when multiple objects (planets, asteroids, comets, space debris) are considered \cite{Althoefertalk}. In 2009, upon learning about the "Global Trajectory Optimization Competitions," Ingo's old and nearly forgotten interest in space missions was rekindled.
The workshop concluded with Ingo Althöfer playing a piece of music on his accordion in memory of Ning Cai.

\bigskip

\begin{figure}
    \centering
    \includegraphics[width=8cm]{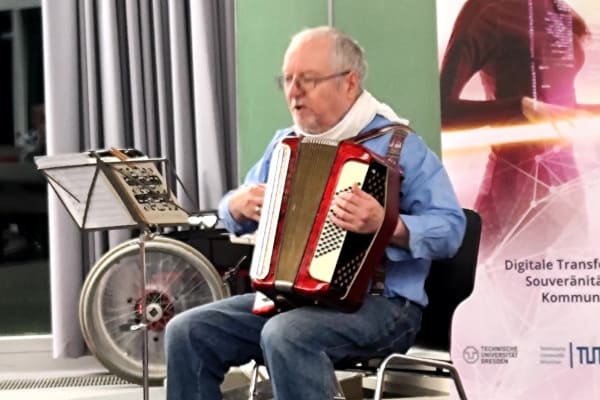}
    \caption{Ingo Althöfer playing a piece of music on his accordion in memory of Ning Cai.}
    \label{fig:enter-label}
\end{figure}

\newpage

\section{Social Events}

The mix of participants at the workshop was very successful, bringing together a diverse group of individuals. Many of Ning Cai's old colleagues and friends were present, alongside numerous young researchers working on topics that Ning had explored or even initiated. 
\begin{figure}[h]
\begin{center}
    \includegraphics[width=12cm]{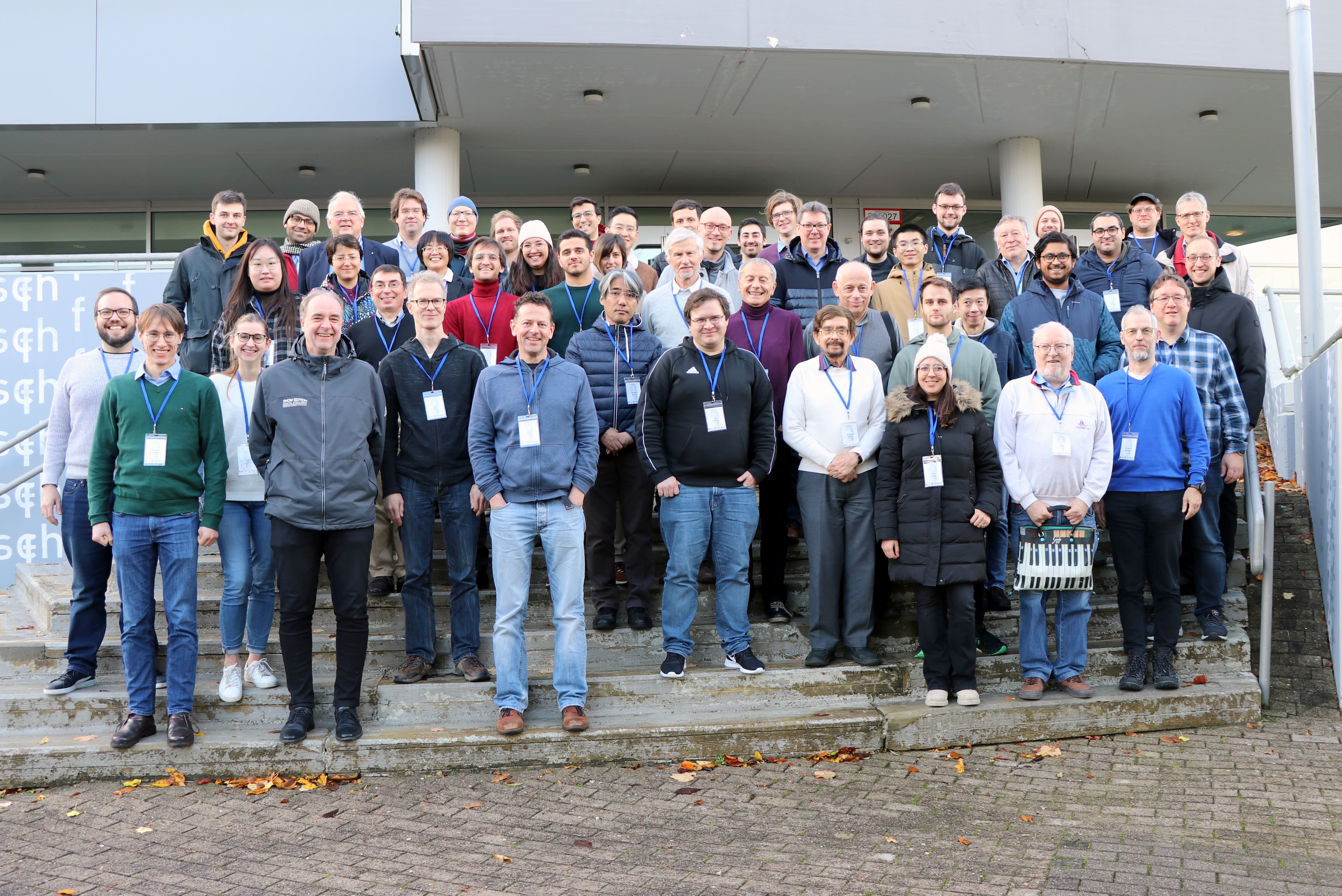}
\end{center}
\caption{Group Photo of the participants at the workshop}
\end{figure}
This blend of generations facilitated a lively scientific exchange and provided a wonderful opportunity for long-standing colleagues to reunite.

The main workshop was preceded by a satellite workshop on post-Shannon theory, an area of research that Ning had also contributed to significantly. To further promote interaction and exchange among researchers, various social activities were organized.

On Friday evening, the participants gathered at the restaurant "Wok \& Roll" for a communal dinner. This restaurant held special significance as it was where Ning had regularly celebrated Christmas with his old colleagues. The dinner allowed the attendees to bond over shared memories and experiences. The evening continued with the younger colleagues heading out for a fun bowling session, fostering camaraderie and team spirit.
\begin{figure}[h]
\begin{center}
    \includegraphics[width=8cm]{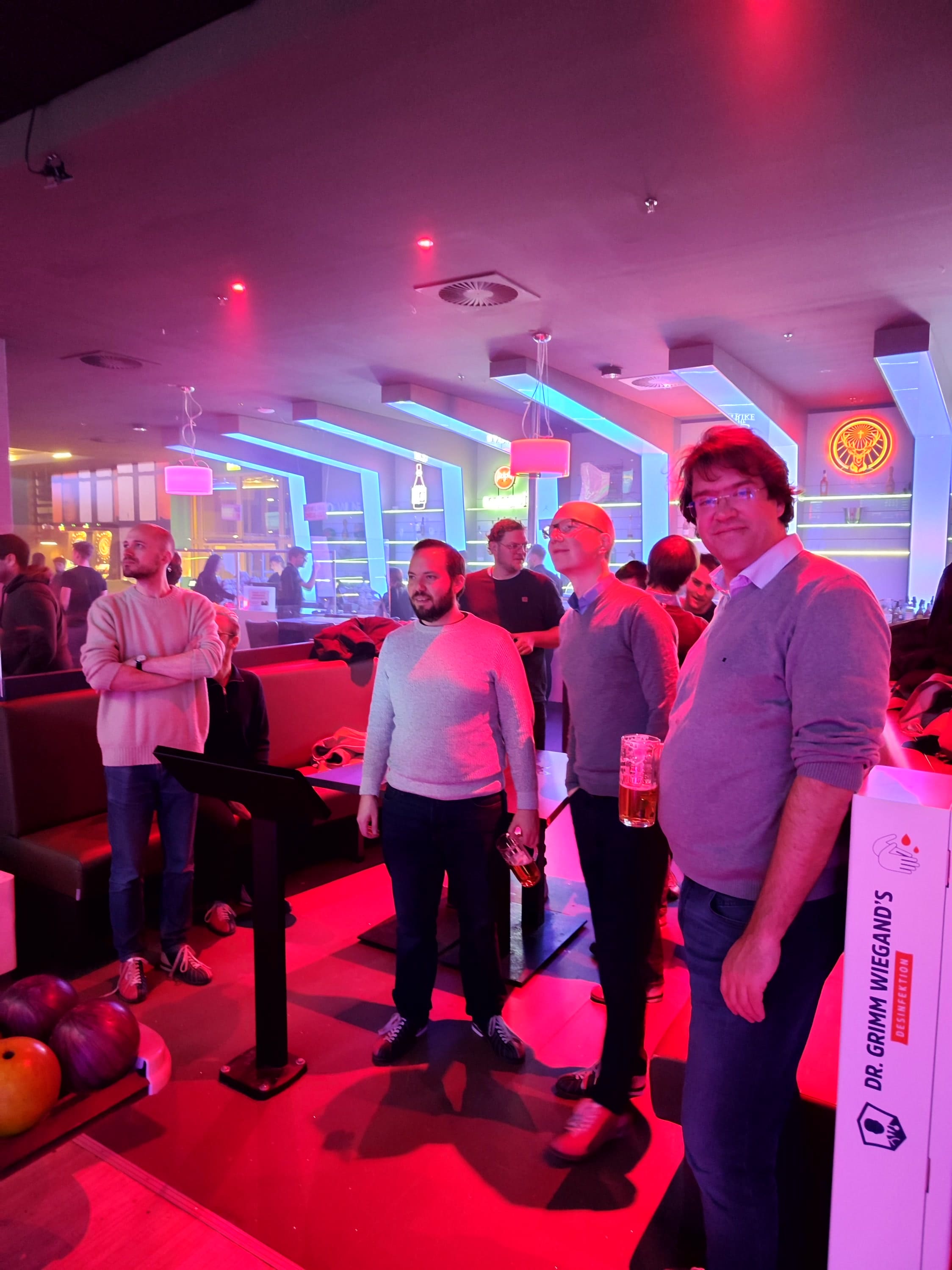}
\end{center}
\caption{Colleagues bowling together.}
\end{figure}

Ning had a deep love for Christian celebrations and was actively involved in the Christian community in Bielefeld. Given that the workshop took place during the pre-Christmas period, Saturday was dedicated to visiting the Christmas market in Bielefeld. This outing was a blend of festive cheer and collegial interaction. The younger colleagues enjoyed trying their hand at curling, adding a playful element to the day.
\begin{figure}[h]
\begin{center}
    \includegraphics[width=8cm]{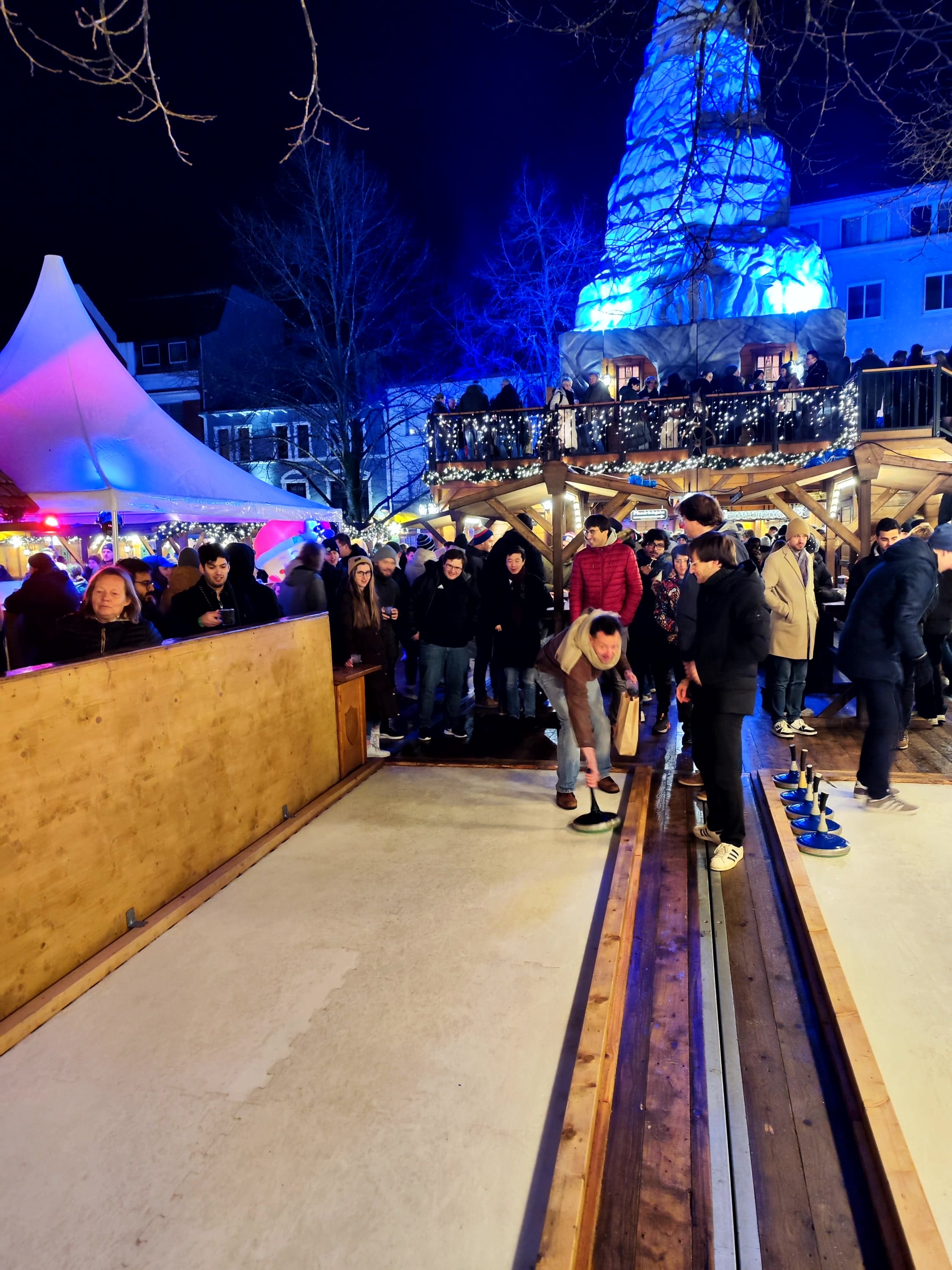}
\end{center}
\caption{Curling event during the workshop on Friday afternoon}
\end{figure}

Throughout the workshop, participants frequently reminisced about Ning, sharing stories and reflecting on the impact he had on their lives and work. This constant thread of remembrance highlighted the deep respect and affection that everyone had for Ning.

The workshop concluded on Sunday with a shared meal at the Brauhaus in Bielefeld. Ning's wife was also invited to this gathering. During the meal, the organizers, who were friends and colleagues of Ning, presented her with a bouquet of flowers as a token of appreciation for Ning's support and in memory of his contributions. They expressed their gratitude for the wonderful collaborations and the positive influence Ning had on their professional and personal lives.

\begin{figure}[h]
\begin{center}
    \includegraphics[width=8cm]{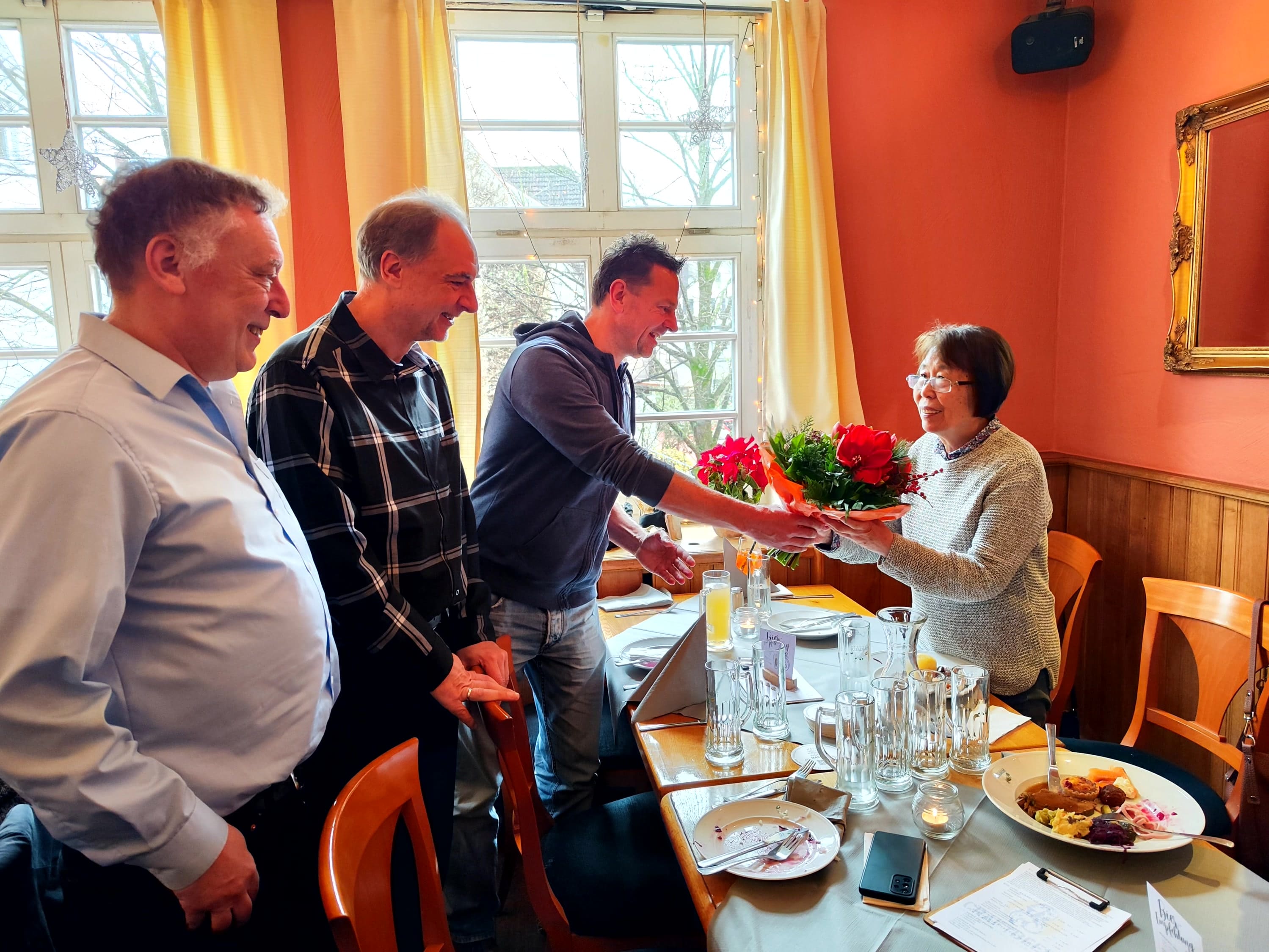}
\end{center}
\caption{Flowers for Jinying Zhi (Ning's wife) from the organizers of the memorial workshop.}
\end{figure}

Ning Cai's legacy continues to inspire, and his work remains a foundational element for ongoing and future research. The workshop not only honored his memory but also underscored the lasting impact of his contributions to the field.

\section*{Acknowledgments}
Huge thanks to Barbara Weitkamp from Babsi's Foto-Augenblicke for her amazing dedication and the fantastic shots she captured during the workshop. Her talent shines through in photos: Fig. 1-23, 26, 28, 32, 34, 38, 40, 54, and 57.
We extend our heartfelt thanks to Jens Stoye, Director of the ZiF at Bielefeld University, for his participation in the organisation of the workshop, which facilitated our access to the services and rooms of the ZiF.
The authors acknowledge the financial support by the Federal Ministry of Education and Research of Germany in the program of “Souverän. Digital. Vernetzt.” joint project 6G-life, project identification number: 16KISK002.{\color{black}{ Holger Boche and Christian Deppe further gratefully acknowledge
the financial support by the BMBF Quantum Programm QD-CamNetz, Grant
16KISQ077, QuaPhySI, Grant 16KIS1598K, and QUIET, Grant 16KISQ093.}} Christian Deppe was supported  by the Bundesministerium 
f\"ur Bildung und Forschung (BMBF) through Grant 16KIS1005. Rami Ezzine and Wafa Labidi were supported by the BMBF through Grant 16KIS1003K.

\printbibliography[title={Bibliography}]

@incollection{zif1994,
  title={Das Zentrum f{\"u}r interdisziplin{\"a}re Forschung},
  author={Sprenger, Gerhard and Weingart, Peter},
  booktitle={Reformuniversit{\"a}t Bielefeld: 1969-1994; zwischen Defensive und Innovation},
  year={1994}
}

@incollection{padberg2014center,
  title={The Center for Interdisciplinary Research (ZiF)—Epistemic and Institutional Considerations},
  author={Padberg, Britta},
  booktitle={University experiments in interdisciplinarity: Obstacles and opportunities},
  pages={95--116},
  year={2014},
  publisher={Transcript Verlag Bielefeld}
}

@inproceedings{MemorialNing,
  author       = {Ingo Althöfer and Holger Boche and Christian Deppe and Ulrich Tamm and Andreas Winter and Raymond Yeung},
title= {Ning Cai: A Tribute to a Pioneering Scholar in
Information Theory},
booktitle    = {Festschrift in Memory of Ning Cai
Information Theory and Related Fields, LNCS Vol. 14620} ,
year = {2025}
}

@inproceedings{PeterMemorialNing,
   title={On the Small Quasi-kernel conjecture}, 
      author={Péter L. Erdős and Ervin Győri and Tamás Róbert Mezei and Nika Salia and Mykhaylo Tyomkyn},
booktitle    = {Festschrift in Memory of Ning Cai
Information Theory and Related Fields, LNCS Vol. 14620} ,
year = {2025}
}

@inproceedings{UgoMemorialNing,
   title={Old and New Results on Alphabetic Codes}, 
      author={Roberto Bruno and Ugo Vaccaro},
booktitle    = {Festschrift in Memory of Ning Cai
Information Theory and Related Fields, LNCS Vol. 14620} ,
year = {2025}
}

@article{DBLP:journals/tit/AhlswedeCLY00,
  author       = {Rudolf Ahlswede and
                  Ning Cai and
                  Shuo{-}Yen Robert Li and
                  Raymond W. Yeung},
  title        = {Network information flow},
  journal      = {{IEEE} Trans. Inf. Theory},
  volume       = {46},
  number       = {4},
  pages        = {1204--1216},
  year         = {2000},
  url          = {https://doi.org/10.1109/18.850663},
  doi          = {10.1109/18.850663},
  bibsource    = {dblp computer science bibliography, https://dblp.org}
}

@INPROCEEDINGS{PosterAndrea,
  author={Boche, Holger and Grigorescu, Andrea and Schaefer, Rafael F. and Poor, H. Vincent},
  booktitle={GLOBECOM 2023 - 2023 IEEE Global Communications Conference}, 
  title={Algorithmic Computability of the Capacity of Additive Colored Gaussian Noise Channels}, 
  year={2023},
  volume={},
  number={},
  pages={4375-4380},
}

@ARTICLE{PosterArk,
       author = {{Viladomat Jasso}, Alonso and {Modi}, Ark and {Ferrara}, Roberto and {Deppe}, Christian and {N{\"o}tzel}, Janis and {Fung}, Fred and {Sch{\"a}dler}, Maximilian},
        title = "{Quantum and Quantum-Inspired Stereographic K Nearest-Neighbour Clustering}",
      journal = {Entropy},
         year = 2023,
        month = sep,
       volume = {25},
       number = {9},
          eid = {1361},
        pages = {1361},
          }

@INPROCEEDINGS{PosterEvagoras,
  author={Sidorenko, Vladimir and Stylianou, Evagoras and Deppe, Christian and Boche, Holger},
  booktitle={European Wireless 2023; 28th European Wireless Conference}, 
  title={Minimal Trellises for Decoding Quantum Stabilizer Codes}, 
  year={2023},
  volume={},
  number={},
  pages={364-369},
}

@inproceedings{Ning_Minglai,
	editor = {},
	author = {Cai, Minglai and Cai, Ning and Deppe, Christian},
	title = {Capacities of classical compound quantum wiretap and classical quantum compound wiretap channels},
	booktitle = {2012 IEEE International Symposium on Information Theory Proceedings},
	year = {2012},
	month = {},
	volume = {},
	publisher = {IEEE},
	organization = {},
	series = {},
	number = {},
	pages = {},
	isbn = {978146732579097814673258069781467325783},
	doi = {10.1109/isit.2012.6284548},
	language = {},
	note = {},
	url = {},
}

@article{Cai_Compound,
	author = {Boche, Holger and Cai, Minglai and Cai, Ning and Deppe, Christian},
	title = {Secrecy capacities of compound quantum wiretap channels and applications},
	journal = {Physical Review A},
	year = {2014},
	volume = {89},
	month = {},
	number = {5},
	pages = {},
	issn = {},
	doi = {10.1103/physreva.89.052320},
	language = {},
	note = {},
	url = {},
}

@incollection{ahlswede2006identification,
  title={Identification for sources},
  author={Ahlswede, Rudolf and Balkenhol, Bernhard and Kleinew{\"a}chter, Christian},
  booktitle={General theory of information transfer and combinatorics},
  pages={51--61},
  year={2006},
  publisher={Springer}
}

@article{ahlswede1989identification,
  title={Identification via channels},
  author={Ahlswede, Rudolf and Dueck, Gunter},
  journal={IEEE Transactions on Information Theory},
  volume={35},
  number={1},
  pages={15--29},
  year={1989},
  publisher={IEEE}
}

@article{ahlswede2008general,
  title={General theory of information transfer: Updated},
  author={Ahlswede, Rudolf},
  journal={Discrete Applied Mathematics},
  volume={156},
  number={9},
  pages={1348--1388},
  year={2008},
  publisher={Elsevier}
}

@article{Vrana2019,
  author    = {P\'{e}ter Vrana and Matthias Christandl},
  title     = {Asymptotic Spectrum of the Strassen Support Functionals},
  journal   = {IEEE Transactions on Information Theory},
  volume    = {65},
  number    = {9},
  pages     = {5945--5958},
  year      = {2019},
  month     = {Sep},
  doi       = {10.1109/TIT.2019.2929501},
}

@article{Devetak2008,
  author    = {Igor Devetak and Aram W. Harrow and Andreas Winter},
  title     = {A Resource Framework for Quantum Shannon Theory},
  journal   = {IEEE Transactions on Information Theory},
  volume    = {54},
  number    = {10},
  pages     = {4587--4618},
  year      = {2008},
  month     = {Oct},
  doi       = {10.1109/TIT.2008.929940},
}

@article{Csiszar2004,
  author    = {Imre Csisz{\'{a}}r and Prakash Narayan},
  title     = {Secrecy Capacities for Multiple Terminals},
  journal   = {IEEE Transactions on Information Theory},
  volume    = {50},
  number    = {12},
  pages     = {3047--3061},
  year      = {2004},
  month     = {Dec},
  doi       = {10.1109/TIT.2004.838089},
}

@article{Ning_BC,
	author = {Ahlswede, Rudolf and Cai, Ning and Deppe, Christian},
	title = {An Isoperimetric Theorem for Sequences Generated by Feedback and Feedback-Codes for Unequal Error Protection},
	journal = {Problems of Information Transmission},
	year = {2001},
	volume = {37},
	month = {},
	number = {4},
	pages = {332-338},
	issn = {},
	doi = {10.1023/a:1013823417501},
	language = {},
	note = {},
	url = {},
}

@inproceedings{LNCSQEC,
  author       = {Sidorenko, Vladimir and Stylianou, Evagoras and Deppe, Christian},
title= {Minimal Trellises for non-Degenerate and Degenerate Decoding of Quantum Stabilizer Codes},
booktitle    = {Festschrift in Memory of Ning Cai
Information Theory and Related Fields, LNCS Vol. 14620} ,
year = {2025}
}

@article{Salek2022,
  author={Salek, Farzin and Winter, Andreas},
  journal={IEEE Transactions on Information Theory},
  title={Multi-User Distillation of Common Randomness and Entanglement From Quantum States},
  year={2022},
  volume={68},
  number={2},
  pages={976-988},
  doi={10.1109/TIT.2021.3124965},
  month={Feb.}
}

@INPROCEEDINGS{PosterHolger,
  author={Boche, Holger and Pohl, Volker and Poor, H. Vincent},
  booktitle={2023 62nd IEEE Conference on Decision and Control (CDC)}, 
  title={The Wiener Theory of Causal Linear Prediction Is Not Effective}, 
  year={2023},
  volume={},
  number={},
  pages={8229-8234},
 }

@INPROCEEDINGS{PosterJuan,
  author={Hofmann, Pit and Cabrera, Juan A. and Bassoli, Riccardo and Fitzek, Frank H.P.},
  booktitle={GLOBECOM 2023 - 2023 IEEE Global Communications Conference}, 
  title={Analog Network Coding in Molecular Communications: A Practical Implementation}, 
  year={2023},
  volume={},
  number={},
  pages={571-576},
}

@ARTICLE{PosterMarcel,
       author = {{Wu}, Jyun-Sian and {Lin}, Pin-Hsun and {Mross}, Marcel A. and {Jorswieck}, Eduard A.},
        title = "{Worst-Case Per-User Error Bound for Asynchronous Unsourced Multiple Access}",
      journal = {arXiv e-prints},
         year = 2024,
        month = jan,
          eid = {arXiv:2401.14265},
        pages = {},
}

@ARTICLE{PosterMarcel1,
       author={Mross, Marcel A. and Lin, Pin-Hsun and Jorswieck, Eduard A.},
  journal={IEEE Transactions on Communications}, 
  title={Gaussian Broadcast Channels With Heterogeneous Finite Blocklength Constraints: Inner and Outer Bounds}, 
  year={2024},
  volume={72},
  number={5},
  pages={2731-2745},
  doi={10.1109/TCOMM.2024.3354203}}

@INPROCEEDINGS{PosterMoritz,
  author={Voichtleitner, Johannes and Wiese, Moritz and Frank, Anna and Boche, Holger},
  booktitle={GLOBECOM 2023 - 2023 IEEE Global Communications Conference}, 
  title={Improving Upper and Lower Bounds for the Security Performance of Wiretap Channels}, 
  year={2023},
  volume={},
  number={},
  pages={1-6},
 }

@ARTICLE{PosterPau,
       author = {{Colomer}, Pau and {Deppe}, Christian and {Boche}, Holger and {Winter}, Andreas},
        title = "{Zero-entropy encoders and simultaneous decoders in identification via quantum channels}",
      journal = {arXiv e-prints},
         year = 2024,
        month = feb,
          eid = {arXiv:2402.09116},
        pages = {arXiv:2402.09116}
}

@INPROCEEDINGS{PosterSifat,
  author={Rezwan, Sifat and Cabrera, Juan A. and Fitzek, Frank H. P.},
  booktitle={2022 International Conference on Electrical, Computer, Communications and Mechatronics Engineering (ICECCME)}, 
  title={Network Functional Compression for Control Applications}, 
  year={2022},
  volume={},
  number={},
  pages={1-6},
}

@ARTICLE{PosterTobias,
  author={Saeidian, Sara and Cervia, Giulia and Oechtering, Tobias J. and Skoglund, Mikael},
  journal={IEEE Transactions on Information Theory}, 
  title={Pointwise Maximal Leakage}, 
  year={2023},
  volume={69},
  number={12},
  pages={8054-8080},
  }

@ARTICLE{PosterVlad,
       author = {{Djuhera}, Aladin and {Andrei}, Vlad C. and {Li}, Xinyang and {M{\"o}nich}, Ullrich J. and {Boche}, Holger and {Saad}, Walid},
        title = "{R-SFLLM: Jamming Resilient Framework for Split Federated Learning with Large Language Models}",
      journal = {arXiv e-prints},
         year = 2024,
        month = jul,
          eid = {arXiv:2407.11654},
        pages = {},
         
}

@INPROCEEDINGS{PosterWafa,
  author={Labidi, Wafa and Deppe, Christian and Boche, Holger},
  booktitle={2023 IEEE International Symposium on Information Theory (ISIT)}, 
  title={Joint Identification and Sensing for Discrete Memoryless Channels}, 
  year={2023},
  volume={},
  number={},
  pages={442-447},
}

@ARTICLE{PosterXingyang,
       author = {{Li}, Xinyang and {Andrei}, Vlad C. and {Djuhera}, Aladin and {M{\"o}nich}, Ullrich J. and {Boche}, Holger},
        title = "{An Analysis of Capacity-Distortion Trade-Offs in Memoryless ISAC Systems}",
      journal = {arXiv e-prints},
         year = 2024,
        month = feb,
          eid = {arXiv:2402.17058},
        pages = {},
         
}

@INPROCEEDINGS{PosterYannick,
  author={Boche, Holger and Böck, Yannik N. and Speidel, Stefanie and Fitzek, Frank H. P.},
  booktitle={2023 62nd IEEE Conference on Decision and Control (CDC)}, 
  title={Turing Meets Machine Learning: Uncomputability of Zero-Error Classifiers}, 
  year={2023},
  volume={},
  number={},
  pages={8559-8566},
}

@INPROCEEDINGS{PosterZahra,
  author={Khanian, Zahra Baghali and Kuroiwa, Kohdai and Leung, Debbie},
  booktitle={2023 IEEE International Symposium on Information Theory (ISIT)}, 
  title={Rate-Distortion Theory for Mixed States}, 
  year={2023},
  volume={},
  number={},
  pages={749-754},
}

@INPROCEEDINGS{PosterZuhra,
  author={Amiri, Zuhra and Nötzel, Janis},
  booktitle={2023 IEEE 9th World Forum on Internet of Things (WF-IoT)}, 
  title={Comparing Latency and Power Consumption: Quantum vs. Classical Preprocessing}, 
  year={2023},
  volume={},
  number={},
  pages={01-06},
}

@ARTICLE{PosterCaspar,
  author={von Lengerke, Caspar and Hefele, Alexander and Cabrera, Juan A. and Reisslein, Martin and Fitzek, Frank H. P.},
  journal={IEEE Journal on Selected Areas in Communications}, 
  title={Beyond the Bound: A New Performance Perspective for Identification via Channels}, 
  year={2023},
  volume={41},
  number={8},
  pages={2687-2706},
  }

@techreport{PaperJohannes,
	author = {J. Rosenberger, H. Boche, J.A. Cabrera and C. Deppe},
	title = {Function Computation and Identification over Locally Homomorphic Multiple-Access Channels},
	year = {2024},
	month = {Apr},
	institution = {arXiv},
	address = {},
	number = {},
	doi = {},
	language = {en},
	note = {},
	url = {https://arxiv.org/abs/2404.14390},
}

@ARTICLE{CapacitycoloredNoise,
  author={Boche, Holger and Grigorescu, Andrea and Schaefer, Rafael F. and Poor, H. Vincent},
  journal={IEEE Transactions on Communications}, 
  title={Characterization of the Complexity of Computing the Capacity of Colored Noise Gaussian Channels}, 
  year={2024},
  volume={},
  number={},
  pages={1-1},
}

@ARTICLE{FiniteBL,
       author = {{Boche}, Holger and {Grigorescu}, Andrea and {Schaefer}, Rafael F. and {Poor}, H. Vincent},
        title = "{Finite Blocklength Performance of Capacity-achieving Codes in the Light of Complexity Theory}",
      journal = {arXiv e-prints},
         year = 2024,
        month = jul,
          eid = {arXiv:2407.07773},
        pages = {arXiv:2407.07773},
        
}

@INPROCEEDINGS{CRgaussian,
  author={Ezzine, Rami and Labidi, Wafa and Boche, Holger and Deppe, Christian},
  booktitle={GLOBECOM 2020 - 2020 IEEE Global Communications Conference}, 
  title={Common Randomness Generation and Identification over Gaussian Channels}, 
  year={2020},
  volume={},
  number={},
  pages={1-6},}

@INPROCEEDINGS{CRslowfading,
  author={Ezzine, Rami and Wiese, Moritz and Deppe, Christian and Boche, Holger},
  booktitle={2021 IEEE International Symposium on Information Theory (ISIT)}, 
  title={Common Randomness Generation over Slow Fading Channels}, 
  year={2021},
  volume={},
  number={},
  pages={1925-1930},
}

@INPROCEEDINGS{UCR,
  author={Ezzine, Rami and Wiese, Moritz and Deppe, Christian and Boche, Holger},
  booktitle={2022 IEEE Information Theory Workshop (ITW)}, 
  title={A General Formula for Uniform Common Randomness Capacity}, 
  year={2022},
  volume={},
  number={},
  pages={762-767},
}

@INPROCEEDINGS{epsilonUCR,
  author={Ezzine, Rami and Wiese, Moritz and Deppe, Christian and Boche, Holger},
  booktitle={2023 IEEE International Symposium on Information Theory (ISIT)}, 
  title={A Lower and Upper Bound on the Epsilon-Uniform Common Randomness Capacity}, 
  year={2023},
  volume={},
  number={},
  pages={240-245},
}

@article{Collisionflat,
title = "$\epsilon-$Almost collision-flat universal hash functions and mosaics of designs",
author = "Moritz Wiese and Holger Boche",
year = "2024",
month = apr,
language = "English",
volume = "92",
pages = "975--998",
journal = "Designs, Codes, and Cryptography",
publisher = "Springer Netherlands",
number = "4",
}

@ARTICLE{SmallMosaics,
       author = {{Kr{\v{c}}adinac}, Vedran},
        title = "{Small examples of mosaics of combinatorial designs}",
      journal = {arXiv e-prints},
         year = 2024,
        month = may,
          eid = {arXiv:2405.12672},
        pages = {},
         
archivePrefix = {},
       eprint = {},
 primaryClass = {},
}

@misc{PaperJulia,
      title={Multipartite multiplexing strategies for quantum routers}, 
      author={Julia A. Kunzelmann and Hermann Kampermann and Dagmar Bruß},
      year={2024},
      eprint={},
      archivePrefix={},
      primaryClass={},
      url={https://arxiv.org/abs/2406.13492}, 
}

@ARTICLE{HayashiTalk,
  author={Cai, Ning and Hayashi, Masahito},
  journal={IEEE Transactions on Information Theory}, 
  title={Secure Network Code for Adaptive and Active Attacks With No-Randomness in Intermediate Nodes}, 
  year={2020},
  volume={66},
  number={3},
  pages={1428-1448},
}

@misc{Hirchetalk,
      title={Quantum R\'enyi and $f$-divergences from integral representations}, 
      author={Christoph Hirche and Marco Tomamichel},
      year={2023},
      eprint={},
      archivePrefix={},
      primaryClass={},
      url={https://arxiv.org/abs/2306.12343}, 
}

@ARTICLE{Jorswiecktalk,
  author={Janda, Carsten Rudolf and Wiese, Moritz and Jorswieck, Eduard Axel and Boche, Holger},
  journal={IEEE Transactions on Information Theory}, 
  title={Arbitrarily Varying Wiretap Channels With Non-Causal Side Information at the Jammer}, 
  year={2023},
  volume={69},
  number={4},
  pages={2635-2663},}

@misc{Vincktalk,
      title={On the Binary Symmetric Channel with a Transition Probability Determined by a Poisson Distribution}, 
      author={A. J. Han Vinck and Fatma Rouissi},
      year={2023},
      eprint={2307.06073},
      archivePrefix={arXiv},
}

@INPROCEEDINGS{Abdallatalk,
  author={Rosenberger, Johannes and Ibrahim, Abdalla and Bash, Boulat A. and Deppe, Christian and Ferrara, Roberto and Pereg, Uzi},
  booktitle={2023 IEEE International Symposium on Information Theory (ISIT)}, 
  title={Capacity Bounds for Identification With Effective Secrecy}, 
  year={2023},
  volume={},
  number={},
  pages={1202-1207},}

@INPROCEEDINGS{ChenTalk,
  author={Chen, Yanling and Cai, Ning and Sezgin, Aydin},
  booktitle={2014 IEEE International Conference on Cloud Engineering}, 
  title={Wiretap Channel with Correlated Sources}, 
  year={2014},
  volume={},
  number={},
  pages={472-477},
}

@INPROCEEDINGS{BoulatTalk,
  author={Sheikholeslami, Azadeh and Bash, Boulat A. and Towsley, Don and Goeckel, Dennis and Guha, Saikat},
  booktitle={2016 IEEE International Symposium on Information Theory (ISIT)}, 
  title={Covert communication over classical-quantum channels}, 
  year={2016},
  volume={},
  number={},
  pages={2064-2068},
}

@INPROCEEDINGS{IlyaTalk,
  author={Vorobyev, Ilya and Lebedev, Vladimir and Lebedev, Alexey},
  booktitle={2023 IEEE International Symposium on Information Theory (ISIT)}, 
  title={Correcting One Error in Non-Binary Channels with Feedback}, 
  year={2023},
  volume={},
  number={},
  pages={1266-1270},
}

@article{Kramertalk,
author = {Kramer, Gerhard and Savari, Serap A.},
title = {Edge-Cut Bounds on Network Coding Rates},
year = {2006},
issue_date = {March 2006},
publisher = {Plenum Press},
address = {USA},
volume = {14},
number = {1},
issn = {1064-7570},
journal = {J. Netw. Syst. Manage.},
month = {mar},
pages = {49–67},
numpages = {19},
}

@INPROCEEDINGS{Janistalk,
  author={Sekavcnik, Simon and Noetzel, Janis},
  booktitle={European Wireless 2022; 27th European Wireless Conference}, 
  title={Entanglement Assisted Classical Communication Link for transmission of Pareto ON/OFF time series}, 
  year={2022},
  volume={},
  number={},
  pages={1-7},
  doi={}}

@book{Franktalk,
  title={Network Coding: From Theory to Practice},
  author={Medard, Muriel and Fitzek, Frank HP and R{\"o}tter, Daniel Enrique Lucani and Pedersen, Morten V and Heide, Janus},
  year={2013},
  publisher={Wiley}
}

@misc{Althoefertalk,
    author = {Ingo Althöfer},
    title = {Combinatorial Space Trajectories to Comets and Asteroids},
    year = {2019},
    url = {https://www.althofer.de/space-trajectories.html},
    note = {Updated: 2024-07-30}
}

\end{document}